\documentclass[3p]{scrartcl}%{article}%

\usepackage{titling}
\usepackage[hmarginratio=1:1,top=32mm,left=15mm,columnsep=1cm]{geometry} % 
\usepackage{array}
\usepackage{color} 
\usepackage{tabularx}
\usepackage{graphicx}
\usepackage[svgnames,table,xcdraw]{xcolor}
\usepackage{amsmath}
\usepackage{amssymb}
\usepackage{amsfonts}	
\usepackage{comment}
\usepackage{bm}
\usepackage{multirow}
\usepackage{relsize} % e.g. used for \mathsmaller
\usepackage{fourier} % for the norm comand \Vert (it has smaller space between the vertical bars 
\usepackage{epstopdf}
\usepackage[font=small,labelfont=bf]{caption}	% set font of caption text
\usepackage{multicol}

\usepackage{nomencl}	% create nomenclature
\usepackage{pbox} % figure in matrix cells
\makenomenclature

\usepackage[backend=bibtex,giveninits=true,sorting=nyt,url=false,doi=true,eprint=false,isbn=false,
backref,backrefstyle=none,maxbibnames=99]{biblatex}
\DefineBibliographyStrings{english}{%
	backrefpage = {Cited on p\adddot},%
	backrefpages = {Cited on pp\adddot}%
}

\bibliography{library}
\renewbibmacro{in:}{%
	\ifboolexpr{%
		test {\ifentrytype{article}}%
	}{}{\printtext{\bibstring{in}\intitlepunct}}%
}
\newcommand\BibTeX{{\rmfamily B\kern-.05em \textsc{i\kern-.025em b}\kern-.08em
T\kern-.1667em\lower.7ex\hbox{E}\kern-.125emX}}

\allowdisplaybreaks[1]

\usepackage{hyperref} 
\hypersetup{
    colorlinks=true,                          
    linkcolor=Blue, % Couleur des liens internes
    citecolor=DarkRed, % Couleur des num�ros de la biblio dans le corps
    urlcolor=blue  } % Couleur des url
\newcommand{\pd}{\mathcal{\partial}}

\newcommand{\x}{\mbf{x}}
\newcommand{\y}{\mbf{y}}
\newcommand{\mbf}[1]{\mathbf{#1}}			%

\newcommand{\cs}[1]{c_{\textrm{s},#1}}	% shear velocity in fluid
\newcommand{\cb}[1]{c_{\textrm{b},#1}}	% bulk velocity in fluid
\newcommand{\cp}[1]{c_{\textrm{p},#1}}	% compressional or p-wave velocity in pure phase

\newcommand{\Cfast}{C_\textrm{fast}}	% mixture fast characteristic velocity
\newcommand{\Cslow}{C_\textrm{slow}}	% mixture slow characteristic velocity
\newcommand{\Cshear}{C_\textrm{shear}}	% mixture shear characteristic velocity

\newcommand{\Vfast}{V_\textrm{fast}}	% mixture fast p-wave V(omega)
\newcommand{\Vslow}{V_\textrm{slow}}	% mixture slow p-wave V(omega)
\newcommand{\Vshear}{V_\textrm{shear}}	% mixture shear sound wave 

\newcommand{\bdm}{\begin{displaymath}}
\newcommand{\edm}{\end{displaymath}}

\newcommand{\bea}{\begin{eqnarray} }
\newcommand{\eea}{\end{eqnarray} }

\renewcommand{\AA}{{\tensor{A}}}

\newcommand{\GG}{{\tensor{G}}}
\newcommand{\g}{{\tensor{g}}}

\newcommand{\vv}{{\boldsymbol{v}}}

\newcommand{\II}{{\tensor{I}}}

\newcommand{\QQ}{{\mathbf{Q}}}
\renewcommand{\SS}{{\mathbf{S}}}

\newcommand{\ww}{{\bm{w}}}

\newcommand{\tr}{\textnormal{tr}}

\newcommand{\transpose}{{\textrm { T}}}

\newcommand*\samethanks[1][\value{footnote}]{\footnotemark[#1]}

\DeclareMathAlphabet{\mathsfbi}{OT1}{\sfdefault}{bx}{sl}
\DeclareMathVersion{sfletters}
\SetSymbolFont{letters}{sfletters}{OML}{ntxsfmi}{b}{it}

\makeatletter
\newcommand{\mathbfsbilow}[1]{%
	\text{\mathversion{sfletters}$\m@th#1$}%
}
\DeclareRobustCommand{\tensor}[1]{%
	\begingroup
	\ifcat\noexpand #1\relax
	\edef\greek@test{\detokenize{#1}}%
	\edef\greek@test{\expandafter\@cdr\greek@test\@nil}%
	\edef\greek@test{\expandafter\@car\greek@test\@nil}%
	\edef\x{\the\lccode\expandafter`\greek@test}%
	\edef\y{\number\expandafter`\greek@test}%
	\ifnum\x=\y\relax
	\mathbfsbilow{#1}%
	\else
	\mathsfbi{#1}%
	\fi
	\else
	\mathsfbi{#1}%
	\fi
	\endgroup
}
\makeatother

\newfont{\numerikEleven}{ecrm1000}
\newfont{\numerikTen}{cmss10}
\newfont{\numerikNine}{cmss9}
\newfont{\numerikEight}{cmss8}

\begin{document} 

% TITLE
\title{ 
	 \textbf{Two-phase hyperbolic model for porous media saturated with a viscous fluid and its application to wavefields simulation} 
%with a semi-implicit\\[1mm]
% staggered scheme and unified first order hyperbolic 
%formulation for\\[1mm]
% continuum mechanics 
}
%-------------------------------------------------------
%-------------------------------------------------------
% AUTHORS for revtex4-1 documentclass
%\author{Ilya Peshkov}
%\email{peshkov@math.nsc.ru}
%\affiliation{Department of Civil, Environmental and Mechanical Engineering, 
%	University of Trento, Via Mesiano 77, 38123 Trento, Italy}
%
%\author{Michael Dumbser}
%\email{michael.dumbser@unitn.it}
%\affiliation{Department of Civil, Environmental and Mechanical Engineering, 
%	University of Trento, Via Mesiano 77, 38123 Trento, Italy}
%
%\author{Evgeniy Romenski}
%\email{evrom@math.nsc.ru}
%\affiliation{Sobolev Institute of Mathematics, 4 Acad. Koptyug Avenue and 
%Novosibirsk State University, 2 Pirogova Str., Novosibirsk, Russia}

% AUTHORS for article documentclass
\author{
Evgeniy Romenski\thanks{Sobolev Institute of Mathematics, 4 Acad. Koptyug 
Avenue,	Novosibirsk, Russia, \href{mailto:evrom@math.nsc.ru}{evrom@math.nsc.ru}
},
\quad
Galina Reshetova\thanks{Institute of Computational Mathematics and Mathematical Geophysics, 6 Ac. Lavrentieva ave., 630090 Novosibirsk, Russia, 	
	\href{mailto:kgv@nmsf.sscc.ru}{kgv@nmsf.sscc.ru}}\ \,$ ^, $\samethanks[1],
\quad
Ilya Peshkov\thanks{Department of Civil, Environmental and Mechanical 
	Engineering, University of Trento, Via Mesiano 77, Trento, Italy, 
	\href{mailto:ilya.peshkov@unitn.it}{ilya.peshkov@unitn.it}}, 
}
\thanksmarkseries{arabic}
%\ead{michael.dumbser@unitn.it}

%\cortext[cor1]{{Ilya Peshkov is on leave from Sobolev Institute of Mathematics 
%, 4 Acad. Koptyug Avenue, 630090 Novosibirsk, Russia}}

% INSTITUTIONS
%\address[IMT]{{Institut de Math\'{e}matiques de Toulouse, Universit\'{e} 
%Toulouse III, F-31062 Toulouse, France.}}
%\address[UniTN]{Department of Civil, Environmental and Mechanical Engineering, 
%	University of Trento, Via Mesiano 77, 38123 Trento, Italy.} 
%\address[NSC]{{Sobolev Institute of Mathematics, 4 Acad. Koptyug Avenue, 
%630090 
%Novosibirsk, Russia}}
%\address[NSU]{{Novosibirsk State University, 2 Pirogova Str., 630090 
%Novosibirsk, Russia}}

%\pacs{}% insert suggested PACS numbers in braces on next line
%
%\keywords{Viscoplastic flows, non-Newtonian fluids, unified formulation of 
%continuum mechanics, semi-implicit staggered scheme}
\maketitle %\maketitle must follow title, authors, abstract and \pacs

\date{\today}

% ABSTRACT
\begin{abstract}
\noindent
We derive and study a new hyperbolic two-phase model of a porous deformable medium saturated by a 
viscous fluid. 
The governing equations of the model are derived in the framework of Symmetric Hyperbolic 
Thermodynamically Compatible (SHTC) systems and by generalizing the unified hyperbolic model of 
continuum 
fluid and solid mechanics. 
Similarly to the unified model, the presented model takes into account the viscosity of the 
saturating fluid through a hyperbolic reformulation.
The model accounts for such dissipative mechanisms as interfacial friction and viscous dissipation 
of the saturated fluid.  
Using the presented nonlinear finite-strain SHTC model, the governing equations for the propagation 
of small-amplitude waves in a porous medium saturated with a viscous fluid are derived. 
As in the conventional Biot theory of porous media, three types of waves can be found: fast and 
slow 
compression waves and shear waves.
It turns out that the shear wave attenuates rapidly due to the viscosity of the saturating fluid, 
and this wave is difficult to see in typical test cases.
However, some test cases are presented in which shear waves can be observed in the vicinity of 
interfaces between regions with different porosity.
\end{abstract}

\section{Introduction} \label{sec:introduction}

Modeling processes in saturated elastic porous media is of interest to many areas of industry and our daily life.  The research in this area 
began in the middle of the last century in the works of Biot \cite{Biot1956,Biot1956a}. In the subsequent development of Biot's theory, special 
attention was paid to seismic applications modeling, such as small amplitude wave propagation and energy loss, see, for example 
\cite{Carcione2010,Mavkoetal2009} and references therein. The main feature of wave propagation in Biot's model, in comparison with the 
theory of elastic waves, is that there are two compressional waves in it - fast and slow. Herewith, the slow compression wave attenuates quickly 
and therefore it is difficult to detect it (it can be clearly seen only for the high frequencies). The attenuation mechanism of this wave is an 
interfacial friction, which is also an important factor in the loss of wave energy in a porous saturated medium. 

In Biot's theory, there are no assumptions about the rheology of the saturating fluid and an interfacial friction is taken into account as a 
result of the average motion of the fluid relative to the solid skeleton \cite{Mavkoetal2009}. Fluid viscosity is indirectly presented in the 
friction coefficient, because its definition is based on the Darcy law (limiting case of Biot's theory). To the best of our knowledge only in a 
few papers the study of influence of saturating liquid rheology on wavefields can be found. Of particular note is the paper \cite{Gurevich2008}
that treats heavy oil as a viscoplastic saturating fluid, which is not of interest for the present paper.
There is an attempt to include the viscous stress directly into the model of the saturating fluid, which is assumed to be Newtonian 
\cite{Sahay2008,Muller2011,Gao2016} (see also the book \cite{Mavkoetal2009}). The effect of viscosity on the wavefield leads to the appearance of 
the so-called slow S-wave. As it is noted in \cite{Sahay2008}, this wave decays rapidly, therefore it is difficult to detect it even in numerical 
simulations. 
In \cite{Sahay2008} it is reported that S-wave can be found only in the vicinity of interfaces. % between fluid and porous medium. 
Thus, taking viscous stresses in a saturating fluid into account provides an additional mechanism for energy loss in a porous medium and should be taken into account in the interpretation of seismic data.

In fact, a saturated porous medium can be considered as a two-phase mixture of a saturating fluid and a solid skeleton. This means that
methods of the continuum mechanics theory of multiphase media can be used. In papers by Wilmanski
\cite{Wilmanski1998,Wilmanski2006}, the correspondence between Biot's theory and the theory of a two-phase fluid-solid mixture was investigated. From these papers, it can be concluded that Biot's model in its governing equations can be interpreted as a continuous two-phase model under certain assumptions.

Recently, a new Symmetric Hyperbolic Thermodynamically Compatible (SHTC) model of a two-phase continuum has been proposed for an elastic porous 
medium saturated with a compressible fluid \cite{Romenski2020}. The SHTC system theory allows formulating a class of symmetric hyperbolic systems 
of partial differential equations that satisfy the laws of thermodynamics and have good mathematical properties 
\cite{GodRom2003,Godunov:1995a,Godunov1996,Rom1998,Romenski2001,Peshkov2018}. 
This class includes the governing equations of many known models of continuum mechanics, and moreover SHTC systems theory can be a powerful tool for developing new models of complex processes. 
The finite-strain poroelastic model presented in \cite{Romenski2020} is based on the unified model of continuum \cite{HPR2016,DPRZ2016,DPRZ2017} 
which is capable to describe a medium in elastic, elastoplastic and viscous fluid states using a single system of governing differential 
equations. 
It extends the unified model by considering a saturated porous medium as a two-phase solid-fluid mixture, and its governing equations combine the 
equations of the unified model and equations of the SHTC model for compressible two-phase fluid flows 
\cite{Romenski2007,RomDrikToro2010,Romenski2016}. 

In \cite{Romenski2020}, the rheology of the saturating fluid is not taken into account, and the only mechanism of energy loss in wave processes 
is interfacial friction.
In the present paper we extend the model \cite{Romenski2020} by introducing the viscosity of the saturating fluid in the model. This extension is 
based on the use of a hyperbolic viscous fluid model, which is essentially part of the unified model \cite{HPR2016,DPRZ2016} and represents a 
viscous fluid as a solid medium with very small shear stress relaxation time. Thus, the two-phase continuum model of porous medium discussed here 
is considered 
as a mixture of two viscoelastic constituente, one of which corresponds to the fluid. The closing 
relations (elastic moduli and mixture shear stress relaxation time) in the model are taken as functions of the volume and mass fractions of 
constituents and determined by some mixture rules. This allows one to obtain the governing equations for a pure solid and a pure fluid as 
limiting cases of the model system with vanishing volume fractions of the constituents. 
It is important that the presented model is hyperbolic and guaranties finite velocities for all types of propagating waves. 

Limiting the full non-linear finite-strain model to the case of small amplitude wave propagation, we derive a linear poroelastic model that can 
be used to study 
wavefields in a porous medium saturated with a viscous fluid with inclusions of pure solid and pure fluid regions. This is very advantageous 
property because it allows one to simulate waves in regions with a complex geometry of the internal structure by the single partial differential 
equation (PDE) system. 
For example, in \cite{Reshetova2021} the so-called diffuse interface approach is used to simulate wavefields in porous media with pure solid and pure fluid inclusions, which demonstrates the effectiveness of the model.

The results of the analysis of the propagation of one-dimensional waves show that there are three types of waves in the model: fast and slow compressional waves and shear waves
(as in Biot's theory). The appropriate choice of closing relations gives a good qualitative agreement of the frequency-dependence of both 
compressional waves with those obtained by Biot's theory. 
The behavior of the shear wave strongly depends on the parameters that determine the effective viscosity of the medium: the velocity of the shear wave tends to zero at low frequencies, and the corresponding attenuation factor increases with increasing viscosity. What concerns the observation of the so-called slow S-wave, in the model there is only one shear wave, but in some circumstances its velocity can be small. 
This results in that, in numerical simulations, one can see some smeared waves at the vicinity of the interfaces between media with different 
porosities, which is similar to the observations given in \cite{Sahay2008}.
Thus, the presented model is capable of simulating wave phenomena in a porous medium saturated with a viscous fluid. Yet, for a correct 
quantitative description of shear wavefields, reliable experimental data on their propagation are needed.

The rest of the paper is organized as follows. In Section\,\ref{sec.general.model}, the general two-phase solid-fluid SHTC model is formulated in 
the finite-strain 
settings and its properties are studied. Section\,\ref{sec.single.pressure} contains a description of a simplified isentropic single pressure 
model and its limiting 
versions for vanishing phase volume fractions. In Section\,\ref{sec.small.waves}, a derivation of the equations for small amplitude wave 
propagation in a porous 
medium saturated with a viscous fluid is presented. In Sections\,\ref{sec.charact} and \ref{sec.dispers}, we analyze characteristic velocities 
and dispersion relations (frequency-dependence of the sound speeds) of the model. Finally, in Section\,\ref{sec.numerics}, a series of 
two-dimensional test 
problems is solved using a 
finite difference scheme on staggered grids. In particular, an observation of the slow S-wave is discussed. 

\section{Two-phase thermodynamically compatible model of solid-fluid mixture applicable for the description of saturated porous 
medium}\label{sec.general.model}
In this section, we consider the two-phase solid-fluid mixture model \cite{Romenski2020}, also called master model here, that later will be used 
for the derivation 
of PDEs for small amplitude wave propagation is based on the model for deformed porous medium saturated with the compressible fluid. 
If solid and fluid constituents are characterized by their volume fractions $\alpha_1$ and $\alpha_2$, ($\alpha_1+\alpha_2=1$), then the
master system reads as
\begin{subequations}\label{eqn.MS}
	\begin{eqnarray}
&&\displaystyle\frac{\partial \rho v^i}{\partial t}+\frac{\partial 
	(\rho v^i v^k + \rho^2 E_\rho \delta_{ik} + w^iE_{w^k} +\rho 
A_{ki} E_{A_{kj}} )}{\partial x_k}=0, 
\label{eqn.momentumMS}\\[2mm]
	&&\displaystyle\frac{\partial A_{i k}}{\partial t}+\frac{\partial A_{im} 
	v^m}{\partial x_k}+v^j\left(\frac{\partial A_{ik}}{\partial 
	x_j}-\frac{\partial A_{ij}}{\partial x_k}\right)
=-\dfrac{ \psi_{ik} }{\theta},\label{eqn.deformationMS}\\[2mm]
&& \frac{\partial \rho}{\partial t}+\frac{\partial \rho v^k}{\partial 
	x_k}=0,\label{eqn.contiMS}\\[2mm]
&& \frac{\partial \rho c_1}{\partial t}+\frac{\partial (\rho c_1 v^k+\rho 
E_{w^k})}{\partial 
	x_k}=0,\label{eqn.contiMS1}\\[2mm]
&&\displaystyle\frac{\partial w^k}{\partial t}+\frac{\partial (w^lv^l+E_{c_1})}{\partial 
x_k}
+v^l\left(\frac{\partial w^k}{\partial x_l}-
\frac{\partial w^l}{\partial x_k}\right)
=-\dfrac{ \lambda_{k} }{\theta_2},\label{eqn.relvelMS}\\[2mm]
&& \frac{\partial \rho \alpha_1}{\partial t}+\frac{\partial \rho \alpha_1 v^k 
}{\partial 
	x_k}=-\frac{\rho \varphi}{\theta_1},\label{eqn.alphaMS}\\[2mm]
&&\displaystyle\frac{\partial \rho s}{\partial t}+\frac{\partial \rho 
	s v^k }{\partial x_k}=\dfrac{\rho}{\theta E_s}\psi_{ik} \psi_{ik}+
\dfrac{\rho}{\theta_1 E_s}\varphi^2 +
\dfrac{\rho}{\theta_2 E_s}\lambda_k \lambda_k \geq0. 
\label{eqn.entropyMS}
\end{eqnarray}
\end{subequations}
Here, $\rho=\alpha_1\rho_1+\alpha_2\rho_2$ is the mixture density, $\rho_1, \rho_2$ are the fluid and solid mass densities, 
$c_1=\alpha_1\rho_1/\rho$ is the fluid mass fraction ($c_2=1-c_1=\alpha_2\rho_2/\rho$ is the solid phase mass fraction), 
$v^i=c_1v^i_1+c_2v^i_2$ is the mixture velocity, $v^i_1, v^i_2$ are fluid and solid phase velocities,
$w^i=v^i_1-v^i_2$ is the relative velocity of phase motion, $s$ is the entropy of the mixture, $A_{ik}$ is the distortion matrix characterizing 
local elastic deformation of the mixture. 
Equation \eqref{eqn.momentumMS} is the mixture momentum equation, \eqref{eqn.deformationMS} is the evolution equation for the elastic distortion 
of the mixture $A_{ik}$, \eqref{eqn.contiMS} is the mixture mass conservation law, \eqref{eqn.contiMS1} is the fluid mass conservation law, 
\eqref{eqn.relvelMS} is the balance equation for the relative velocity, \eqref{eqn.alphaMS} is the balance equation for the liquid volume 
fraction, and \eqref{eqn.entropyMS} is the mixture entropy balance law. The source term in the latter equation provides the growth of the mixture 
entropy due to the dissipative processes as the second law of thermodynamics requires.

The most important closing relation for system \eqref{eqn.MS} is the generalized energy $E$, depending on the mixture parameters $\rho, c_1, 
\alpha_1, w^k, s$ and $A_{ik}$ and we discuss its definition below. As soon as the generalized energy is defined, one can compute the 
thermodynamic forces $\frac{\pd E}{\pd \rho} = E_\rho$, $\frac{\pd E}{\pd A_{kj}} = E_{A_{kj}}$, $\frac{\pd E}{\pd w^k} = E_{w^k}$, $\frac{\pd 
E}{\pd c_1} = E_{c_1}$, $\frac{\pd E}{\pd s} = E_s$ in \eqref{eqn.momentumMS}, 
\eqref{eqn.contiMS1}, \eqref{eqn.relvelMS}, \eqref{eqn.entropyMS} and source terms in \eqref{eqn.deformationMS}, \eqref{eqn.relvelMS}, 
\eqref{eqn.alphaMS}, \eqref{eqn.entropyMS}:
\begin{equation}
\psi_{ik} = E_{A_{ik}}, \quad \lambda_k=E_{w^k}, \quad \varphi = E_{\alpha_1}.     
\end{equation}
Parameters $\theta$, $\theta_1$, and $\theta_2$ characterize the rate of the mixture shear stress relaxation, and the rates of relative velocity 
relaxation and pressure relaxation respectively and can depend on state variables.
Thus, the definition of generalized energy $E$ gives us a governing PDEs for the solid-fluid mixture written in terms of phase parameters.  
The pressure $p$, the shear stress tensor $\sigma_{ij}$ and the temperature $T$ are also computed via the generalized energy as: 
\begin{equation}
p = \rho^2 E_\rho, \quad  \sigma_{ij} = -\rho A_{ki} E_{A_{kj}}, \quad T = E_s. 
\end{equation}

Note that the solution to system \eqref{eqn.MS} satisfies the energy conservation law 
\begin{equation} \label{eqn.energyMS}
\displaystyle\frac{\partial \rho E}{\partial t}+
\frac{\partial \left(\rho  v^k E +v^i(p \delta_{ik}+\rho w^i E_{w^k}-\sigma_{ik}) +\rho E_{c_1}E_{w^k} \right)}{\partial x_k}=0
\end{equation}
in accordance with the first law of thermodynamics. Also note that in numerical simulations, it is necessary to use the energy conservation law 
\eqref{eqn.energyMS} instead of entropy balance law \eqref{eqn.entropyMS}.

As stated in \cite{Romenski2020}, system \eqref{eqn.MS}, \eqref{eqn.energyMS} belongs to the class of SHTC (Symmetric Hyperbolic 
Thermodynamically Compatible) systems and describes dynamic processes in the solid-fluid mixture at finite-strains. 

Our goal is to apply governing equations \eqref{eqn.MS} to the description of processes in the deformed medium saturated by the compressible 
(viscous or inviscid) fluid. We shall do it assuming that the fluid volume fraction $\alpha_1$ is identical to the porosity $\phi$ of the 
medium, $\alpha_1=\phi$. Before discussing the definition of the generalized energy it is important to emphasize that we take the phase mass 
densities and mixture distortion  as parameters characterizing the deformation of the element of the medium.
This choice of parameters is made due to the fact that the parameters of the medium associated with volumetric deformation, such as density, energy and pressure, are additive values with respect to mass or volume fractions.
As for the general strain, one can determine the strain tensor of each phase separately, but it is unclear how to formulate the equations for 
these individual strains and how to take into account their interaction. That is why we use the distortion of the entire mixture as the strain 
measure of the medium. Its evolution is determined by the velocity of the mixture via equation \eqref{eqn.deformationMS}.

To define the generalized energy density $\rho E$, we, as in \cite{Romenski2020}, assume that it is a sum of the kinetic energy $\rho E_0$ 
of the center of mass of the mixture element, the kinematic energy of the relative motion $\rho E_1$, the energy of volumetric 
deformation $\rho 
E_2$, and the 
energy of shear deformations $\rho E_3$:
\begin{equation}
\rho E=\rho E_0(\vv)+\rho E_1(c_1,\ww)+
\rho E_2(\alpha_1, c_1, \rho,s)+\rho E_3(c_1,\rho,s,\AA). 
\label{energy.SF}    
\end{equation}
The kinetic energy of the center of mass of the mixture element and kinematic energy of the relative motion are defined as
\begin{equation}
\rho E_0(\vv)= \frac{1}{2}\rho v^jv^j, \quad   
\rho E_1(c_1,\ww) =  \frac{1}{2}c_1(1-c_1)\rho  w^jw^j.
\label{kineticEn}
\end{equation} 
Not that due to the definition of $v^j, w^j$, 
\begin{equation}
    \rho E_0+\rho E_1=\alpha_1\rho_1\frac{v^j_1v^j_1}{2}+
    \alpha_1\rho_1\frac{v^j_2v^j_2}{2}.
\end{equation}
The energy of volumetric deformation is defined as
\begin{equation} \label{energy.mix}
\rho E_2(\alpha_1, c_1, \rho,s)=\alpha_1 \rho_1e_1(\rho_1,s)+\alpha_2 \rho_2 e_2(\rho_2,s) \quad
\text{or} \quad E_2(\alpha_1, c_1, \rho,s)=c_1e_1\left(\frac{\rho c_1}{\alpha_1},s\right)+c_2e_2\left(\frac{\rho c_2}{\alpha_2},s\right).
\end{equation}
The energy of shear deformation depends on the distortion of the entire solid-fluid mixture element and we define it as
\begin{equation}
E_3=\frac18 \cs{m}^2\left(\tr({\g^2})-3\right), 
\label{ShearEnergy}
\end{equation} 
where $\cs{m}$ is the shear sound speed in the mixture to be determined, and $\g$ is the 
normalized Finger (or metric) strain tensor: $\g={\GG} (\det\hspace{-0.4mm}{\GG})^{-1/3}$, $\GG=\AA^\transpose\AA$.

We assume now that the saturating fluid is viscous and its rheology is described by the unified SHTC (Symmetric Hyperbolic Thermodynamically 
Compatible) model of continuum \cite{HPR2016,DPRZ2016}. This unified model considers a fluid as a visco-elastic medium with a small shear strain 
relaxation time. In such an approach, the solution to the Navier-Stokes equations of Newtonian fluids is approximated well by the solutions to 
the unified solid-fluid model \cite{DPRZ2016} in the relaxation limit. Later, we will see that in the limiting cases of a
pure fluid $(\alpha_1=1, 
\alpha_2=0)$ or a
pure solid $(\alpha_1=0, \alpha_2=1)$, governing equations of the two-phase solid-fluid mixture \eqref{eqn.MS} reduce to the unified SHTC 
governing equations 
for pure fluid and pure 
solid. 

Because \eqref{ShearEnergy} is the shear energy of the mixture, the shear mixture sound speed $ \cs{m} $ should depend on the porosity (fluid 
volume fraction).
We assume that $\cs{m}$ is computed by the simple 
mixture rule via the shear sound speeds of fluid and solid phases $\cs{1}$ and $\cs{2}$:
\begin{equation} \label{cs_mix}
\cs{m}^2=c_1\cs{1}^2+c_2\cs{2}^2.   
\end{equation}

Note that in \cite{Romenski2020}, the two-phase model for deformed saturated porous medium was formulated under the assumption $\mu_1=0$ that 
means that the saturating fluid is inviscid and has no resistance to shear.

Using the definition of the generalized energy \eqref{energy.SF}, \eqref{kineticEn}, \eqref{energy.mix}, \eqref{ShearEnergy} and relations 
between mixture's and phase parameters', one can find thermodynamic forces $E_\alpha$, 
$E_\rho$, $E_{A_{kj}}$, $E_{w^k}$, $E_{c_1}$ and other thermodynamic parameters $p=\rho^2E_\rho$,   $\sigma_{ij} = -\rho A_{ki} E_{A_{kj}}$,  $T 
= E_s$:
\begin{subequations}\label{Thermod.forcesSF}
\begin{eqnarray}
& E_{\alpha_1}=\frac{p_2-p_1}{\rho}, \quad p=\rho^2E_\rho={\alpha_1p_1+\alpha_2p_2}, 
 \\ 
&\frac{\partial E}{\partial \AA}=
\frac{\cs{m}^2}{2} \AA^{-\transpose} \left(\g^2-\frac{\tr({\g^2})}{3} \II \right),
\quad 
\sigma_{ij}=-\frac{\rho \cs{m}^2}{2}\left({g_{ik}g_{kj}-
	\frac{1}{3}{g_{lk}g_{kl}}\delta_{ij}}\right),
\quad   \\[2mm]
&E_{w^i}=c_1c_2 w^i, \quad 
E_{c_1}=e_1+\frac{p_1}{\rho_1}-e_2-\frac{p_2}{\rho_2}
+\frac{\left(\cs{1}^2-\cs{2}^2\right)}{\cs{m}^2}E_3+(1-2c_1)\frac{\Vert\ww\Vert^2}{2}, \\[2mm]
& E_s=T=c_1\frac{\partial e_1}{\partial s}+c_2\frac{\partial e_2}{\partial s}.
\end{eqnarray}  
\end{subequations}

To close the model completely, one needs to define the parameters 
$\theta$, $\theta_1$, $\theta_2$ characterizing the rates of the mixture shear stress relaxation, relative velocity relaxation and pressure relaxation.

For further considerations it is convenient to formulate governing equations in terms of phase parameters of state, which can be obtained with 
the use of relationships between the mixture and phase parameters and formulae for 
thermodynamic forces \eqref{Thermod.forcesSF}
\begin{subequations}\label{eqn.PV}
	\begin{eqnarray}
&&\displaystyle\frac{\partial  
(\alpha_1\rho_1v^i_1+\alpha_2\rho_2v^i_2)}{\partial t}+
\frac{\partial(\alpha_1\rho_1v^i_1v^k_1+\alpha_2\rho_2v^i_2v^k_2 
+p\delta_{ik}-\sigma_{ik})}{\partial x_k}=0, 
\label{eqn.momentumPV}\\[2mm]
&&\displaystyle\frac{\partial A_{i k}}{\partial t}+\frac{\partial A_{ij} 
	v^j}{\partial x_k}+v^j\left(\frac{\partial A_{ik}}{\partial 
	x_j}-\frac{\partial A_{ij}}{\partial x_k}\right)
=-\dfrac{ \psi_{ik} }{\theta},\label{eqn.deformationPV}\\[2mm]
&& \frac{\partial \alpha_1\rho_1}{\partial t}+\frac{\partial \alpha_1\rho_1 
v^k_1}{\partial 
	x_k}=0,\label{eqn.contiPV1}\\[2mm]
&& \frac{\partial \alpha_2\rho_2}{\partial t}+\frac{\partial \alpha_2\rho_2 
v^k_2}{\partial 
	x_k}=0,\label{eqn.contiPV2}\\[2mm]
&&\displaystyle\frac{\partial w^k}{\partial 
t}+\frac{\partial\left((v_1^j v_1^j-v_2^j v_2^j)/2+e_1+p_1/\rho_1-e_2-p_2/\rho_2+
{\left(\cs{1}^2-\cs{2}^2\right)}E_3/{\cs{m}^2}\right)}
{\partial x_k}
+v^l\left(\frac{\partial w^k}{\partial x_l}-
\frac{\partial w^l}{\partial x_k}\right)
=-\dfrac{ \lambda_{k}
}{\theta_2}, \label{eqn.relvelPV}\\[2mm]
&& \frac{\partial \rho \alpha_1}{\partial t}+\frac{\partial \rho \alpha_1 v^k 
}{\partial 
	x_k}=-\frac{\rho \varphi}{\theta_1},\label{eqn.alphaPV}\\[2mm]
&&\displaystyle\frac{\partial \rho s}{\partial t}+\frac{\partial \rho 
	s v^k }{\partial x_k}=\dfrac{\rho}{\theta T}\psi_{ik} \psi_{ik}+
\dfrac{\rho}{\theta_1 T}\varphi^2 +
\dfrac{\rho}{\theta_2 T}\lambda_k \lambda_k \geq0.
\label{eqn.entropyPV}
\end{eqnarray}
\end{subequations}

\section{Simplified isentropic single pressure model}\label{sec.single.pressure}
In this section, we formulate a simplified model which is applicable to studying processes in porous medium with small temperature variations and 
instantaneous relaxation of interphasial pressures.
If the saturated porous medium is initially in a thermal equilibrium, its deformations are small and there is no intensive thermal sources, then 
we can assume that the temperature variations in the process under consideration are small. This means that we can neglect thermal processes and 
can use the isentropic version of equations \eqref{eqn.PV}, neglecting equation \eqref{eqn.entropyPV} and assuming that the phase energies and 
pressures do not depend on entropy.

Furthermore, we assume that the characteristic pore space in the deformed porous medium is small.
This assumption allows us to reduce the set of state parameters and simplify system \eqref{eqn.PV}. 
Indeed, in this case  we can assume that the pressure relaxation is instantaneous. The reason for this is that the phase pressures are equalizing 
due to the pressure wave propagating in the fluid in pores and skeleton. 
The pressure equalization time can be estimated as the time of several runs of the pressure wave in a pore of a characteristic size, and it is 
obvious that it is small in comparison with the characteristic time of interest to us in macroscopic samples. That is why we can assume that in 
model \eqref{eqn.PV} the fluid and solid pressures are equal, $p_1=p_2$. 
Note that this assumption allows us to use the single pressure $p=p_1=p_2$ in \eqref{eqn.PV}, and thus it is necessary to replace the equation 
for the volume fraction \eqref{eqn.alphaPV} by the algebraic equation $p_1(\rho_1)=p_2(\rho_2)$. We have to do this replacement because the 
relation $p_1(\rho_1)=p_2(\rho_2)$ can be obtained as the relaxation limit of the model if $\theta_1 \to 0$. Taking into account the 
algebraic equation $p_1=p_2$, and using phase mass conservation equations \eqref{eqn.contiPV1}, \eqref{eqn.contiPV2} one can derive the equation 
for the volume fraction of the single-pressure model:
\begin{equation}
({\alpha_1K_2+\alpha_2K_1})\frac{\partial \alpha_1}{\partial t}+
(\alpha_2 K_1 v_1^k+\alpha_1 K_2v_2^k)\frac{\partial\alpha_1}{\partial x_k}
+\frac{\alpha_1 \alpha_2K_1}{\rho_1}(v_1^k-v_2^k)\frac{\partial \rho_1}{\partial x_k}+
{\alpha_1 \alpha_2 K_1}\frac{\partial v_1^k}{\partial x_k}-
{\alpha_1 \alpha_2 K_2}\frac{\partial v_2^k}{\partial x_k}=0, \label{eqn.alphaFFK}     
\end{equation}
where $K_1=\rho_1 c_1^2 $ and $ K_2=\rho_2 c_2^2$ are the phase bulk moduli and $c_1$ and $ c_2$ are the phase bulk sound velocities 
($c_i^2=\partial 
p_i/\partial 
\rho_i$).

Thus, the simplified system takes the following form:
\begin{subequations}\label{eqn.Isentropic}
	\begin{eqnarray}
&&\displaystyle\frac{\partial 
(\alpha_1\rho_1v^i_1+\alpha_2\rho_2v^i_2)}{\partial t}+
\frac{\partial(\alpha_1\rho_1v^i_1v^k_1+\alpha_2\rho_2v^i_2v^k_2 
+p\delta_{ik}-\sigma_{ik})}{\partial x_k}=0, 
\label{eqn.momentumPVs}\\[2mm]
&&\displaystyle\frac{\partial A_{i k}}{\partial t}+\frac{\partial A_{ij} 
	v^j}{\partial x_k}+v^j\left(\frac{\partial A_{ik}}{\partial 
	x_j}-\frac{\partial A_{ij}}{\partial x_k}\right)
=-\dfrac{ \psi_{ik} }{\theta},\label{eqn.deformationPVs}\\[2mm]
&& \frac{\partial \alpha_1\rho_1}{\partial t}+\frac{\partial \alpha_1\rho_1 
v^k_1}{\partial 
	x_k}=0,\label{eqn.contiPV1s}\\[2mm]
&& \frac{\partial \alpha_2\rho_2}{\partial t}+\frac{\partial \alpha_2\rho_2 
v^k_2}{\partial 
	x_k}=0,\label{eqn.contiPV2s}\\[2mm]
&&\displaystyle\frac{\partial (v_1^k-v_2^k)}{\partial 
t}+\frac{\partial\left((v_1^j v_1^j-v_2^j v_2^j)/2+e_1+p_1/\rho_1-e_2-p_2/\rho_2+
{\left(\cs{1}^2-\cs{2}^2\right)}E_3/{\cs{m}^2}\right)}
{\partial x_k}
+v^l\left(\frac{\partial w^k}{\partial x_l}-
\frac{\partial w^l}{\partial x_k}\right)
=-\dfrac{ \lambda_{k}
}{\theta_2}, \label{eqn.relvelPVs}\\[2mm]
&& ({\alpha_1K_2+\alpha_2K_1})\frac{\partial \alpha_1}{\partial t}+
(\alpha_2 K_1 v_1^k+\alpha_1 K_2v_2^k)\frac{\partial\alpha_1}{\partial x_k}
+\frac{\alpha_1 \alpha_2K_1}{\rho_1}(v_1^k-v_2^k)\frac{\partial \rho_1}{\partial x_k}+
{\alpha_1 \alpha_2 K_1}\frac{\partial v_1^k}{\partial x_k}-
{\alpha_1 \alpha_2 K_2}\frac{\partial v_2^k}{\partial x_k}=0,\label{eqn.alphaPVs}
\end{eqnarray}
\end{subequations}

The derivation of the governing equations for small amplitude wave propagation will be based on the above system, but first we note that in the 
limiting cases $\alpha_1=0$, $\alpha_2=1$ (pure solid) and $\alpha_1=1$, $\alpha_2=0$ (pure fluid) this system reduces to the unified solid-fluid 
model of finite-strain elastoplastic solid and viscous fluids respectively \cite{HPR2016,DPRZ2016}.  
In fact, if we put $\alpha_1=0$, $\alpha_2=1$ in \eqref{eqn.Isentropic} and recall that 
$v^j=c_1v^j_1+c_2v^j_2$, we then obtain the following system
\begin{subequations}\label{eqn.IsentropicSolid}
	\begin{eqnarray}
&&\displaystyle\frac{\partial 
\rho_2v^i_2}{\partial t}+
\frac{\partial(\rho_2v^i_2v^k_2 
+p\delta_{ik}-\sigma_{ik})}{\partial x_k}=0, 
\label{eqn.momentumPVsSolid}\\[2mm]
&&\displaystyle\frac{\partial A_{i k}}{\partial t}+\frac{\partial A_{ij} 
	v^j_2}{\partial x_k}+v^j_2\left(\frac{\partial A_{ik}}{\partial 
	x_j}-\frac{\partial A_{ij}}{\partial x_k}\right)
=-\dfrac{ \psi_{ik} }{\theta},\label{eqn.deformationPVsSolid}\\[2mm]
&& \frac{\partial \rho_2}{\partial t}+\frac{\partial \rho_2 
v^k_2}{\partial x_k}=0,\label{eqn.contiPV2Solid}\\[2mm]
&&\displaystyle\frac{\partial (v_1^k-v_2^k)}{\partial 
t}+\frac{\partial\left((v_1^j v_1^j-v_2^j v_2^j)/2+e_1+p_1/\rho_1-e_2-p_2/\rho_2+
{\left(\cs{1}^2-\cs{2}^2\right)}E_3/{\cs{m}^2}\right)}
{\partial x_k}
+v^l_2\left(\frac{\partial w^k}{\partial x_l}-
\frac{\partial w^l}{\partial x_k}\right)
=-\dfrac{ \lambda_{k}
}{\theta_2}, \label{eqn.relvelPVsSolid} \\[2mm]
&& \frac{\partial \alpha_1}{\partial t}+v_2^k\frac{\partial\alpha_1}{\partial x_k}=0,\label{eqn.alphaPVsSolid}
\end{eqnarray}
\end{subequations}
where $p=p_2=\rho_2^2\frac{\partial e_2}{\partial \rho_2}$ is the pressure in the solid component, 
$\sigma_{ij} = 
-\rho 
A_{ki} E_{A_{kj}}=-\frac{\rho_2 \cs{2}^2}{2}\left({g_{ik}g_{kj}-
	\frac{1}{3}{g_{lk}g_{kl}}\delta_{ij}}\right)$ is the shear stress in the solid.
System \eqref{eqn.IsentropicSolid} does not include the limiting case of mass conservation equation 
for fluid phase \eqref{eqn.contiPV2s} because of the absence of the fluid ($\alpha_1=0$). We see 
also that equations \eqref{eqn.relvelPVsSolid}, \eqref{eqn.alphaPVsSolid} have no influence on the 
solid behavior, which is fully described by equations 
\eqref{eqn.momentumPVsSolid}--\eqref{eqn.contiPV2Solid} because \eqref{eqn.alphaPVsSolid} states 
the constancy of the phase 
volume fractions ($\alpha_1=0$, $\alpha_2=1$) and the relative velocity is not presented in all 
other equations. Thus the reduced model for the solid phase is governed by equations 
\eqref{eqn.momentumPVsSolid}--\eqref{eqn.contiPV2Solid}
which are exactly the equations of elastoplastic isentropic medium written in terms of velocities, pressure and shear stress, the full version of which, taking into account entropy, can be found in \cite{DPRZ2016}.

The similar system can be obtained in the limiting case of pure fluid. Thus, assuming $\alpha_1=1$, 
$\alpha_2=0$ and neglecting again the equation for the relative velocity, equation for the mass 
conservation of the solid phase,  and for the volume fraction of the solid phase, we arrive to the 
following system
\begin{subequations}\label{eqn.IsentropicFluid}
	\begin{eqnarray}
&&\displaystyle\frac{\partial 
\rho_1v^i_1}{\partial t}+
\frac{\partial(\rho_1v^i_1v^k_1 
+p\delta_{ik}-\sigma_{ik})}{\partial x_k}=0, 
\label{eqn.momentumPVsFluid}\\[2mm]
&&\displaystyle\frac{\partial A_{i k}}{\partial t}+\frac{\partial A_{ij} 
	v^j_1}{\partial x_k}+v^j_1\left(\frac{\partial A_{ik}}{\partial 
	x_j}-\frac{\partial A_{ij}}{\partial x_k}\right)
=-\dfrac{ \psi_{ik} }{\theta},\label{eqn.deformationPVsFluid}\\[2mm]
&& \frac{\partial \rho_1}{\partial t}+\frac{\partial \rho_1 
v^k_1}{\partial x_k}=0,\label{eqn.contiPV2Fluid}
\end{eqnarray}
\end{subequations}
where $p=p_1=\rho_1^2\frac{\partial e_1}{\partial \rho_1}$ is the fluid pressure, $\sigma_{ij} = 
-\rho 
A_{ki} E_{A_{kj}}=-\frac{\rho_1 \cs{1}^2}{2}\left({g_{ik}g_{kj}-
	\frac{1}{3}{g_{kl}g_{lk}}\delta_{ij}}\right)$ is the the shear stress and the source term in 
	\eqref{eqn.deformationPVsFluid} contains 
$\Psi=[\psi_{ik}]=\frac{\partial E}{\partial \AA}=
\frac{\cs{1}^2}{2} \AA^{-\transpose} \left(\g^2-\frac{\tr({\g^2})}{3} \II \right)$.
The system \eqref{eqn.IsentropicFluid} is similar to the isentropic version of the unified model of 
continuum presented in \cite{HPR2016,DPRZ2016} and describes Newtonian viscous fluid flows as 
an asymptotic limit for small shear strain relaxation time $\tau$ if to choose 
$$\theta= \theta_0 \tau, \quad \theta_0= \frac{2 \cs{1}^2}{\rho_1 |\GG|^{1/3}},$$
that for small $\tau$ gives us 
$$
\sigma_{ik} = \rho \tau \cs{1}^2 \left(\left(\frac{\partial v_1^i}{\partial x_k}+\frac{\partial 
v_1^k}{\partial x_i}\right) - \frac{2}{3}\frac{\partial v_1^j}{\partial x_j}\delta_{ik}\right).$$
This formula is exactly the definition of the Navier-Stokes shear stress and can be written 
as 
$$
\sigma_{ik} = 2\eta (\dot \varepsilon_{ik}-\delta_{ik}{\dot\varepsilon_{jj}}/{3}),
$$
where $\dot \varepsilon_{ik}=\frac{1}{2}\left(\frac{\partial v_1^i}{\partial x_k}+\frac{\partial 
v_1^k}{\partial x_i}\right)$ is the strain rate tensor and $\eta=\rho \tau \cs{1}^2$ is the 
effective dynamic viscosity.
The details of the asymptotic analysis and derivation of the Navier-Stokes stress for small $\tau$ 
can be found in Appendix\,\ref{app.asympt}.

Thus, the presented hyperbolic thermodynamically compatible model describes a two-phase porous 
medium with a viscoelastic skeleton and a viscoelastic saturating fluid. 
This model allows us to derive differential equations for the propagation of small amplitude waves 
in a porous medium saturated with a viscoelastic fluid in the next section.

\section{Governing equations for small amplitude wave propagation in an elastic porous medium saturated with a viscous 
fluid}\label{sec.small.waves}

In this section, a system of partial differential equations is derived that can be used to model 
small-amplitude waves in an initially unstressed medium. The derivation can be done by the standard 
linearization procedure and is similar to that presented in \cite{Romenski2020}.

Consider a stationary unstressed medium and denote the parameters of state of this medium by the 
symbol "0". Assume that the initial value of the fluid and solid volume fractions $\alpha_1^0, 
\alpha_2^0, (\alpha_1^0+\alpha_2^0=1)$ are known.The immovability of the medium means that the 
mixture velocity and relative velocity in this initial state are equal to zero $v^{i}_{0}=0, 
w^{k}_{0}=0$ and therefore individual phase velocities are also equal to zero $v_{10}^i=0, 
v_{20}^i=0$. Furthermore, in the unstressed state, the pressure and shear stress of the mixture are 
also equal to zero, that means that $p_1^0=p_2^0=p^0=0$, $\sigma_{ik}^0=0$. Zero values of phase 
pressures correspond to the reference density values $\rho_{10},\rho_{20}$ and zero values of shear 
stress correspond to the distortion $A^0_{ij}=\delta_{ij}$.

Now, assume that the solution has the form
\begin{equation} \label{Perturb}
v_1^k=v_{10}^{k}+\Delta v_1^k=\Delta v_1^k, v_2^k=v_{20}^{k}+\Delta v_2^k=\Delta v_2^k, A_{ij}=A_{ij}^0 + \Delta A_{ij}=\delta_{ij} + \Delta A_{ij}, \rho_1=\rho_{1}^{0}+\Delta \rho_1, \rho_2=\rho_{2}^{0}+\Delta \rho_2, 
\alpha_1=\alpha_1^0+\Delta \alpha_1, 
\end{equation}
where the small perturbations of corresponding state variables are denoted by the symbol $\Delta$.
If to substitute \eqref{Perturb} into system \eqref{eqn.Isentropic} and then 
neglect second and higher-order terms with $\Delta$, we arrive to the following system of linear 
PDEs
\begin{subequations}\label{eqn.Isentropic_L}
	\begin{eqnarray}
&&\displaystyle\frac{\partial 
(\alpha_1^0\rho_1^0 \Delta v^i_1+\alpha_2^0\rho_2^0 \Delta v^i_2)}{\partial t}+
\frac{\partial(\Delta p\delta_{ik}-\Delta \sigma_{ik})}{\partial x_k}=0, 
\label{eqn.momentumPVsL}\\[2mm]
&&\displaystyle\frac{\partial (\Delta A_{i k})}{\partial t}+\frac{\partial 
(\Delta v^i)}{\partial x_k}
= - \dfrac{ \Delta \sigma_{ik} }{\rho_0 \theta},\label{eqn.deformationPVsL}\\[2mm]
&& \frac{\partial (\alpha_1^0 \Delta \rho_1+\Delta \alpha_1\rho_1^0)}{\partial t}+\frac{\partial (\alpha_1^0\rho_1^0 \Delta v^k_1)}{\partial x_k}=0,\label{eqn.contiPV1sL}\\[2mm]
&& \frac{\partial (\alpha_2^0 \Delta \rho_2+\Delta \alpha_2\rho_2^0)}{\partial t}+\frac{\partial (\alpha_2^0\rho_2^0 \Delta v^k_2)}{\partial 
	x_k}=0,\label{eqn.contiPV2sL}\\[2mm]
&&\displaystyle\frac{\partial (\Delta v_1^k- \Delta v_2^k)}{\partial 
t}+\frac{\partial\left(\Delta p/\rho_1^0- \Delta p/\rho_2^0\right)}
{\partial x_k}
=-\dfrac{c_1^0 c_2^0 (\Delta v_1^k- \Delta v_2^k)
}{\theta_2}, \label{eqn.relvelPVsL}\\[2mm]
&& ({\alpha_1^0 K_2^0+\alpha_2^0 K_1^0})\frac{\partial \Delta \alpha_1}{\partial t}+
{\alpha_1^0 \alpha_2^0 K_1^0}\frac{\partial \Delta v_1^k}{\partial x_k}-
{\alpha_1^0 \alpha_2^0 K_2^0}\frac{\partial \Delta v_2^k}{\partial x_k}=0,\label{eqn.alphaPVsL}
\end{eqnarray}
\end{subequations}
Here, $\Delta p =\frac{K_1^0}{\rho_1^0} \Delta \rho_1 = \frac{K_2^0}{\rho_2^0} \Delta \rho_2$, 
where 
$K_1^0=\left. \rho_1^0\frac{\partial p_1}{\partial \rho_1}\right|_{\rho_1=\rho_1^0},
\quad
K_2^0=\left. \rho_2^0\frac{\partial p_2}{\partial \rho_2}\right|_{\rho_2=\rho_2^0}$, 
$\Delta \sigma_{ik}=2\rho_0 (\cs{m}^0)^2 \left(\varepsilon_{ik}-\frac{1}{3}
(\varepsilon_{11}+\varepsilon_{22}+\varepsilon_{33})
\right)$, $(\cs{m}^0)^2=c_1^0 \cs{1}^2+c_2^0\cs{2}^2$, and 
$\varepsilon_{ij}=-(\Delta A_{ij}+\Delta A_{ji})/2$ is the small deformation tensor.

System \eqref{eqn.Isentropic_L} can be simplified by introducing new variables and reducing the 
number of equations. Let us take as a set of independent state variables the mixture velocity 
$U^i=(\alpha_1^0\rho_1^0 \Delta v^i_1+\alpha_2^0\rho_2^0 \Delta v^i_2)/\rho_0$, $(\rho_0=c_1^0 
\rho_1^0+c_2^0 \rho_2^0)$, the relative velocity $W^i=\Delta v_1^i-\Delta v_2^i$, the pressure of 
the mixture $P=\Delta p = \frac{K_1^0}{\rho_1^0} \Delta \rho_1= \frac{K_2^0}{\rho_2^0} \Delta 
\rho_2$,
the shear stress tensor 
$\Sigma_{ik}=\Delta \sigma_{ik}=2\rho_0 (\cs{m}^0)^2 \left(\varepsilon_{ik}-\frac{1}{3}
(\varepsilon_{11}+\varepsilon_{22}+\varepsilon_{33})\right)$.
The coefficient $\theta$ characterizing the rate of stress relaxation is taken as $\theta = 
\theta_0 \tau$, where $\theta_0=2 (\cs{m}^0)^2/\rho_0$, and $\tau$ is the relaxation time.
The complete PDE system obtained from \eqref{eqn.Isentropic_L} and written in terms of the listed 
above variables reads as
\begin{subequations}\label{stress.velocity}
	\begin{eqnarray} 
	&& \rho_0 \frac{\pd V^i}{\pd t}+\frac{\pd P}{\pd x_i}-
	\frac{\pd \Sigma_{ik}}{\pd x_k} = 0,  \\
	&& \frac{\pd W^k}{\pd t}+ \left(\frac{1}{\rho_1^0} - 
	\frac{1}{\rho_2^0}\right)\frac{\pd P}{\pd x_k}=
	-\frac{c_1^0c_2^0}{\theta_2}W^k,  \label{stress.velocity.w}\\
	&& \frac{\pd P}{\pd t} +K\frac{\pd V^k}{\pd x_k}+
	\frac{\alpha_1^0\alpha_2^0}{\rho^0}\left(\rho^0_2-\rho^0_1 \right)
	K\frac{\pd W^k}{\pd x_k} =0, \label{stress.velocity.P}   \\   \label{sij}
	&&\frac{\pd \Sigma_{ik}}{\pd t} - 
	\mu\left(\frac{\pd V^i}{\pd x_k} +\frac{\pd V^k}{\pd x_i}-
	\frac{2}{3}\delta_{ik}\frac{\pd V^j}{\pd x_j} \right) = 
	- \frac{\Sigma_{ik}}{\tau},
	\end{eqnarray}
\end{subequations}
Here $K$ is the bulk modulus of the mixture defined as 
\begin{equation}\label{eqn.Ki}
K = \left( \frac{\alpha_1^0}{K_1^0} + \frac{\alpha_2^0}{K_2^0} \right)^{-1},
\quad
K_1^0 = \left. \rho_1^0 \frac{\pd p_1}{\pd \rho_1} \right\vert_{\rho_1 = \rho_1^0},
\quad
K_2^0 = \left. \rho_2^0 \frac{\pd p_2}{\pd \rho_2} \right\vert_{\rho_2 = \rho_2^0},
\end{equation}
$\mu$ is the shear modulus of the mixture which is defined according to \eqref{cs_mix} 
\begin{equation}\label{eqn.mu}
\mu =\rho_0 \cs{m}^2= \alpha^0_1 \mu_1 + \alpha^0_2 \mu_2, \quad \mu_1=\rho_1^0 \cs{1}^2, \quad
\mu_2=\rho_2^0 \cs{2}^2.
\end{equation}

Note that in the complete system we do not need to include equation \eqref{eqn.alphaPVsL} for the  volume fraction (porosity) because its perturbation does not affect wavefields and, if necessary, $\Delta \alpha_1$ can be computed by solving this equation.

Thus, system of linear partial differential equations \eqref{stress.velocity} has been formulated 
and it 
is applicable to modeling the propagation of small-amplitude waves in a stationary unstressed 
porous medium saturated with a fluid. It covers wave processes for the entire porosity range $\phi 
= 
\alpha_1^0 \in [0,1]$. For the modeling of real media, it is necessary to determine the material 
constants in \eqref{stress.velocity}. In addition to the initially specified densities 
$\rho_1^0$, $\rho_1^0$ and porosity $\phi=\alpha_1^0$, we should define the bulk and shear modulus 
of  the  solid and fluid constituents $K_1^0$, $K_2^0$ and $\mu_1$, $\mu_2$, which give us the
mixture bulk $K$ and shear modulus $\mu$ defined by \eqref{eqn.Ki}, \eqref{eqn.mu}. It is also 
necessary to define 
coefficients $\theta_2$ responsible for the interface friction and the shear stress relaxation time 
$\tau$.
In our study we take the the interfacial friction coefficient $\theta_2$ constant, as in \cite{Romenski2020}. What concerns the shear stress relaxation time $\tau$, we take it in the following form  
\begin{equation}\label{eqn.tau}
\tau =  \left( \frac{\left (\alpha_1^0\right )^n}{\tau_1} + \frac{\left (\alpha_2^0\right )^n}{ 
\tau_2}\right)^{-1}.
\end{equation}
It is clear that the relaxation time should be a function of the porosity $\phi=\alpha_1 ^ 0$, and this functional dependence is the subject of further theoretical studies. 
We take the empirical form of the relaxation time \eqref{eqn.tau} due to the fact that
for the limiting cases of porosity $\phi=0$ or $\phi=1$ it should correspond to the relaxation 
times of pure solid and pure fluid and the nonlinearity on $\phi$ gives us a reasonable values of 
$\tau$ in the vicinity of pure phases. 
If the skeleton is pure elastic, then we take $ \tau_2 = \infty $  and the relaxation time $ \tau $ 
is computed from the formula 
\begin{equation}\label{eqn.tau.simplified}
\tau =  \frac{ \tau_1}{\left (\alpha_1^0\right )^n}.
\end{equation}

\begin{table}[]
	\begin{center}
		\begin{tabular}{llll}
			\rowcolor[HTML]{EFEFEF} 
			& $ \cs{1} = 10 \, m/s$ & $ \cs{1} = 100,m/s $ & $ \cs{1} = 750\, m/s $ \\
			$ n=1 $		& $ \tau = 4.808\cdot10^{-7} $ & $ \tau = 4.808\cdot10^{-9} $ & $ \tau = 
			8.547\cdot10^{-11} $ \\
			\rowcolor[HTML]{EFEFEF} 
			$ n=5 $		& $ \tau = 3.005\cdot10^{-4} $ & $ \tau = 3.005\cdot10^{-6} $ & $ \tau = 
			5.342\cdot10^{-8} $ \\
			$ n=8 $	    & $ \tau = 0.038 $ 			 & $ \tau = 3.756\cdot10^{-4} $ &  $ \tau = 
			6.677\cdot10^{-6} $\\
			\rowcolor[HTML]{EFEFEF} 
			$ n=10 $	& $ \tau = 0.939 $ 			 & $ \tau = 9.390\cdot10^{-3} $ &  $ \tau = 
			1.669\cdot10^{-4} $
		\end{tabular}
		\caption{Values of $ \tau $ (in seconds) for porosity $ \phi = 0.2 $ and different values 
		of $ n $ and fluid shear characteristic speed $ \cs{1} = \sqrt{\mu_1/\rho^0_1} $.}
		\label{tab.tau.n.cs}
	\end{center}
\end{table}

Fluid shear modulus $ \mu_1 $, relaxation time $ \tau_1 $ and dynamic viscosity $ \eta_1 $  are 
connected by the relation
\begin{equation}\label{eta}
\eta = \mu_1 \tau_1 = \rho^0_1 \cs{1}^2 \tau_1.
\end{equation}
Table\,\ref{tab.tau.n.cs} shows some values of the relaxation time \eqref{eqn.tau.simplified} for 
several values of $ n $ and shear sound speed $ \cs{1} $ of the fluid.

All of the above-mentioned material constants should be chosen in such a way that the behavior of 
the medium under consideration corresponds to the available experimental data or the results known 
in the literature. In \cite{Romenski2020}, a comparison of the two-phase thermodynamically 
compatible model of a porous medium saturated with an inviscid fluid and Biot's model is carried 
out. It is shown that the appropriate choice of material constants in the two-phase model allows 
one to obtain velocities of the fast and slow longitudinal waves and their frequency dependencies 
close to those obtained by Biot's model. To our knowledge, there are only a few research works in 
which the influence of the viscosity of a saturating liquid on wave fields in saturated porous 
medium is studied \cite{Sahay2008,Gao2016}. The main goal of these studies is the analysis of shear 
wave propagation and attenuation. In \cite{Sahay2008}, it is discovered that the accounting of 
viscosity of the saturating liquid generates the so-called {\em slow shear wave}. 
Indeed, such a wave appears instead of a typical elastic shear wave and looks like a certain 
diffuse wave, the smeared structure of which arises due to the viscosity of the saturating liquid. 
In the above-mentioned papers, it is reported that a slow shear wave is rather difficult to 
observe, and it can be seen when elastic waves interact with interfaces between media with 
different porosities. In Section\,\ref{sec.numerics}, shear waves in a porous medium saturated with 
a viscous fluid can be seen in some test cases, despite the strong attenuation caused by viscosity.
As far as we know, there is no available experimental data concerning the shear wave propagation 
that 
is why we present qualitative study of such kind of waves. Since our goal is to see the main 
features of shear waves in a porous medium, we will choose material constants in such a way that 
clearly demonstrates these features. Then, as soon the experimental data will be available, one can 
vary material constants in order to achieve an agreement between theoretical and experiments 
studies.

\begin{figure}[t]
	\begin{center}
		\begin{tabular}{c}
			\includegraphics[draft=false,trim=0 0 0 0,clip,scale=0.6]{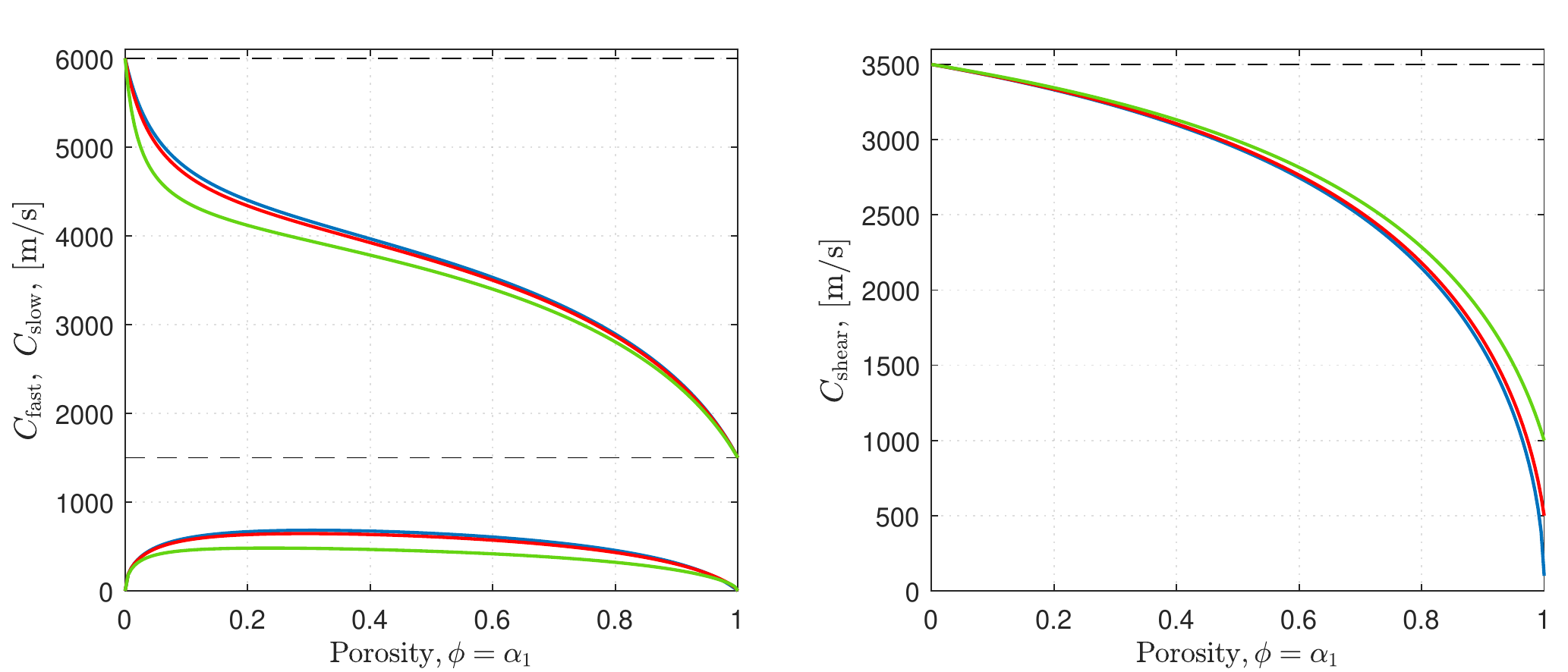}
		\end{tabular}
		\caption{Dependence of the 
		mixture 
		characteristic speeds \eqref{speeds.SHTC} of the fast $ \Cfast(\phi) $ and slow $ 
		\Cslow(\phi) $ pressure modes 
		(left), and shear mode $ \Cshear(\phi) $ 
		(right) on porosity $ \phi = \alpha_1 $ for different values of the shear characteristic 
		speed 
		of the 
		liquid $ \cs{1} $: $ \cs{1}=100 $\,m/s 
		(blue), $ \cs{1}=500 $\,m/s (red), and $ \cs{1}=1000 $\,m/s (green). 
		The top dashed line in left figure corresponds to the compressional velocity in the pure 
		solid $ \cp{2} $
		(i.e. $ \phi = 0 $), the bottom dashed line is the bulk fluid characteristic speed $ \cb{1} 
		$. 
		The dashed line in the right figure is the shear characteristic speed in the pure solid 
		$ \cs{2} $.}	
		\label{fig.characteristics}
	\end{center}
\end{figure}

\section{Characteristic velocities}\label{sec.charact}
We consider system \eqref{stress.velocity} in one space dimension along the coordinate $ x_1 $. 
Collecting all the variables into the single vector $ \QQ $:
\begin{equation}
\QQ = (V^1,V^2,V^3,W^1,W^2,W^3,P~,~\Sigma_{11},\Sigma_{21},\Sigma_{31})^\transpose,
\end{equation}
system \eqref{stress.velocity} can be written in a matrix form as follows
\begin{equation}\label{eqn.1Dfull}
\QQ_t + \mathbb{A} \QQ_x = \SS,
\end{equation}
where 
\begin{equation}\label{eqn.1Dfull.mat}
\mathbb{A} = \left( \begin{array}{cccccccccc}
0      &  0   &  0   & 0  & 0 & 0 & \rho_0^{-1} & - \rho_0^{-1} &          0          
&          0          \\
0      &  0   &  0   & 0  & 0 & 0 &     0     &          0          & - \rho_0^{-1} 
&          0          \\
0      &  0   &  0   & 0  & 0 & 0 &     0     &          0          &          0          
& - \rho_0^{-1} \\
0      &  0   &  0   & 0  & 0 & 0 &     R     &          0          &          0          
&          0          \\
0      &  0   &  0   & 0  & 0 & 0 &     0     &          0          &          0          
&          0          \\
0      &  0   &  0   & 0  & 0 & 0 &     0     &          0          &          0          
&          0          \\
K      &  0   &  0   & K' & 0 & 0 &     0     &          0          &          0          
&          0          \\
-\frac43\mu &  0   &  0   & 0  & 0 & 0 &     0     &          0          &          0          
&          0          \\
0      & -\mu &  0   & 0  & 0 & 0 &     0     &          0          &          0          
&          0          \\
0      &  0   & -\mu & 0  & 0 & 0 &     0     &          0          &          0          
&          0
\end{array} 
\right), \qquad \SS = \left(\begin{array}{c}
0 \\ 
0 \\ 
0 \\ 
-\frac{1}{\theta_2'} W^1 \\ 
-\frac{1}{\theta_2'} W^2 \\ 
-\frac{1}{\theta_2'} W^3 \\ 
0 \\ 
-\frac{1}{\tau} \Sigma_{11}\\ 
-\frac{1}{\tau} \Sigma_{21}\\ 
-\frac{1}{\tau} \Sigma_{31}
\end{array} \right) 
\end{equation}
with $ R = 1/\rho_1^0 - 1/\rho_2^0 $, $ K' = 
\frac{\alpha_1^0\alpha_2^0}{\rho^0}\left(\rho^0_2-\rho^0_1 
\right) K $, and $ \theta_2' = \theta_2/(c_1^0 c_2^0) $. The characteristic 
speeds  $ \lambda_i $ of system 
\eqref{eqn.1Dfull}  are the roots of the characteristic polynomial $ 
\det(\mathbb{A} - \lambda \mathbb{I}) = 0 $, where $ \mathbb{I} $ is the identity matrix. In the 
case of vanishing of all dissipative source 
terms ($ \tau = \infty $, $ \theta'_2 = \infty $), $ \lambda_i $ coincide with the speeds of small 
amplitude waves (sound waves) in the medium. 
If some dissipation mechanisms are presented ($ \tau < \infty $, $ \theta'_2 < \infty $), the 
sound speeds are different from the characteristic speeds $ \lambda_i $ and depend on the wave 
frequency, see the next section. 

The roots of the characteristic polynomial $\det(\mathbb{A}-\lambda \mathbb{I})$ are given by the 
formulas
\begin{subequations}\label{speeds.SHTC}
	\begin{equation}\label{SHTC.charar.pressure}
	\lambda_{1,2,3,4} = 
	\sqrt{\frac{X+Y+Z \pm 
			\sqrt{(X+Y+Z)^2 - 4(X Y)}}{2}},
	\qquad
	\lambda_{5,6,7,8} = \sqrt{\frac{\mu}{\rho_0}},
	\qquad
	\lambda_{9,10} = 0,
	\end{equation}
	\begin{equation}
	X = R K', 
	\qquad 
	Y = \frac{4}{3} \frac{\mu}{\rho_0} , 
	\qquad
	Z = \frac{K}{\rho_0} .
	\end{equation}
\end{subequations}
The speeds $ \lambda_{1,2,3,4} $ correspond to the left-propagating and right-propagating so-called fast $ \Cfast $ and slow
 $ \Cslow $ characteristic speeds of compression waves (P-waves), while
$ \lambda_{5,6,7,8} $ are the shear wave characteristic speeds $ \Cshear $.
Fig.\,\ref{fig.characteristics} shows the dependence of $ \Cfast $, $ \Cslow $, and $ \Cshear 
$ on the porosity $ \phi=\alpha_1 $.

\section{Dispersion relations}\label{sec.dispers}

If the dissipation mechanisms (fluid viscosity and interfacial friction) are presented in 
\eqref{stress.velocity}, i.e. if the relaxation parameters $ \tau < \infty $ or $ \theta'_2 < 
\infty$ are finite, then propagating waves exhibit dispersion, that is the speed of a wave 
(including small amplitude waves) 
depends on the wave frequency $ \omega $. For example, as shown in \cite{Ruggeri1992}, the speeds 
of small amplitude waves $ \Vfast(\omega) $, $ \Vslow(\omega) $, and $ \Vshear(\omega) $ in a 
medium with relaxation processes are smaller than the characteristic speeds \eqref{speeds.SHTC} in 
the same medium but without dissipation mechanisms. In the high-frequency limit ($ \omega \to 
\infty $), the sound speeds tend  
towards the characteristic speeds \eqref{speeds.SHTC}.

To find the dependencies $ \Vfast(\omega) $, $ \Vslow(\omega) $, and $ \Vshear(\omega) $, one needs to find the roots of the following equation 
\cite{Ruggeri1992,Romenski2020}
\begin{equation}\label{eqn.1Dfull.disp}
	\det\left( \mathbb{I} - \frac{k}{\omega} \mathbb{A} + \frac{i}{\omega} \mathbb{S}\right) = 0 ,
\end{equation}
where $ \mathbb{S} = \pd \SS/\pd\QQ $, $ k $ is the complex wave number, $ \omega $ is the real 
frequency, and $ i $ is the imaginary unit.
The phase velocity $ V (\omega)$  and the attenuation factor $ a 
$ are then given by
\begin{equation}
	V = \frac{\omega}{{\textrm{Re}}(k)}, \qquad a = -{\textrm{Im}}(k).
\end{equation}
In addition, it is convenient to use the attenuation per wavelength 
\cite{Ruggeri2015}
\begin{equation}\label{atten.wavelength}
	a_\lambda = a \lambda = \frac{2 \pi V a }{\omega} = -2 \pi \frac{{\textrm{Im}}(k)}{{\textrm{Re}}(k)},
\end{equation}
where $ \lambda $ is the wavelength.

Denoting $ \zeta = k/\omega $, the roots $ \zeta_\textrm{fast} $, $ \zeta_{\textrm{slow}} $, and $ 
\zeta_{\textrm{shear}} $ of \eqref{eqn.1Dfull.disp} are given by the following formulas
\begin{subequations}
	\begin{align}
		\zeta_{\textrm{fast}} &= 
		\sqrt{\frac{\Omega_\gamma (X - B^2) + \Omega_1 \Omega_\gamma B^2 + \Omega_1 Y \gamma - 
		\sqrt{-4 \Omega_1 \Omega_\gamma Y (X - 
		B^2) \gamma + \left(\Omega_\gamma (X - B^2) + 
		\Omega_1 ((Y + B^2) \gamma - i B^2/\Omega)\right)^2}}{2Y (X - B^2) \gamma}},
		\\
		\zeta_{\textrm{slow}} &= 
		\sqrt{\frac{\Omega_\gamma (X - B^2) + \Omega_1 \Omega_\gamma B^2 + \Omega_1 Y \gamma + 
				\sqrt{-4 \Omega_1 \Omega_\gamma Y (X - 
					B^2) \gamma + \left(\Omega_\gamma (X - B^2) + 
					\Omega_1 ((Y + B^2) \gamma - i B^2/\Omega)\right)^2}}{2Y (X - B^2) \gamma}},
		\\
		\zeta_{\textrm{shear}} &= \sqrt{\frac{\rho_0 \Omega_\gamma}{\mu \gamma}},
	\end{align}
\end{subequations}
where $ B^2 = K/\rho_0 $ is the bulk characteristic speed of the saturated porous medium, $ Y = 
\frac{4}{3} \mu/\rho_0$, $ R = 1/\rho_1^0 - 1/\rho_2^0$, $ X = B^2 + K' R $. Furthermore,  $ \gamma 
= 
\theta'_2/\tau$ is the ratio of relaxation parameters, 
$ \Omega_\gamma = \gamma - i/\Omega $, 
$ \Omega_1 = 1- i/\Omega $, and $ \Omega = \tau \omega $ is the non-dimensional frequency.

\begin{figure}[t]
	\begin{center}
		\begin{tabular}{c}
			\includegraphics[draft=false,trim=0 0 0 
			0,clip,scale=0.7]{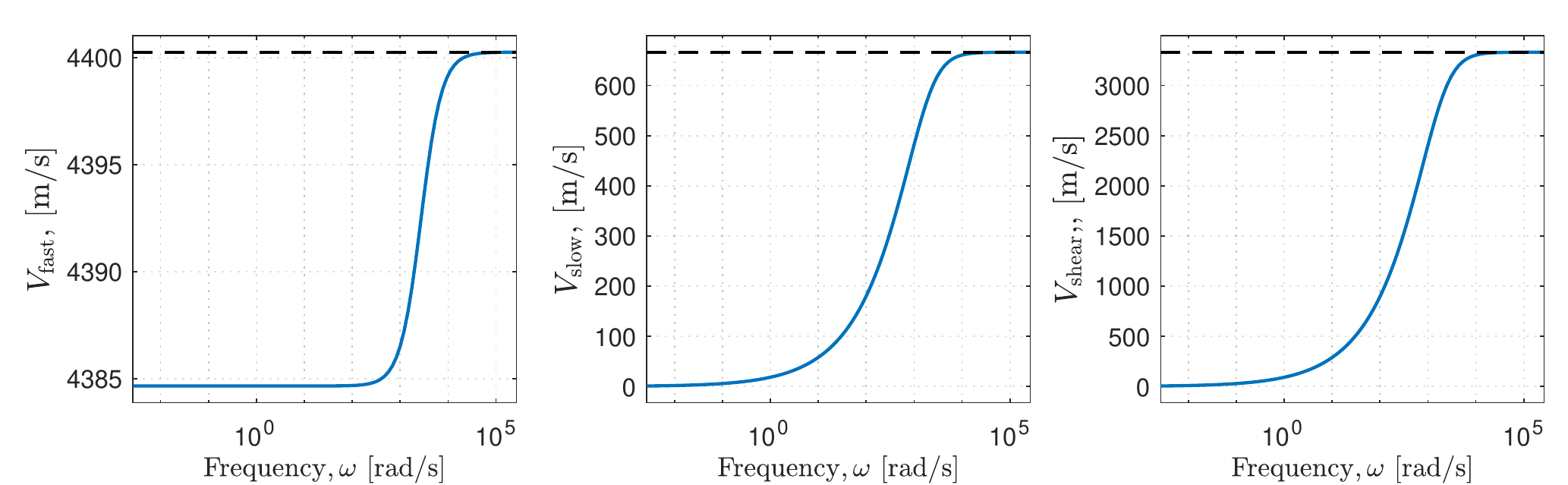}\\
			\includegraphics[draft=false,trim=0 0 0 
			0,clip,scale=0.7]{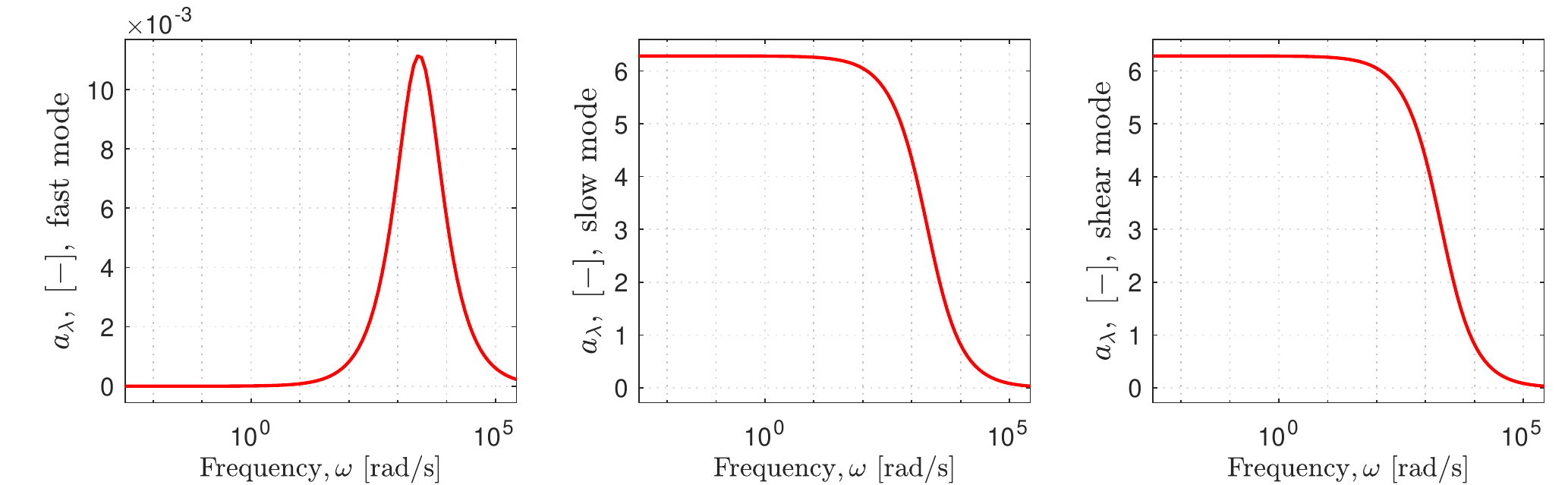}
		\end{tabular}
		\caption{ Sound wave dispersion and 
			attenuation factor per wave length $ a_\lambda $ for all three modes for the case $ \theta'_2 = 
			\infty $, $ \tau < \infty $: fast P-mode (left column) 
			slow P-mode 
			(middle column), and shear mode (right column). Porosity $ \phi = \alpha_1 = 0.2$. Other material parameters 
			are given in Table\,\ref{tab.param}.
				}  
		\label{fig.disp1}
	\end{center}
\end{figure}

\begin{figure}[t]
	\begin{center}
		\begin{tabular}{c}
			\includegraphics[draft=false,trim=0 0 0 
			0,clip,scale=0.7]{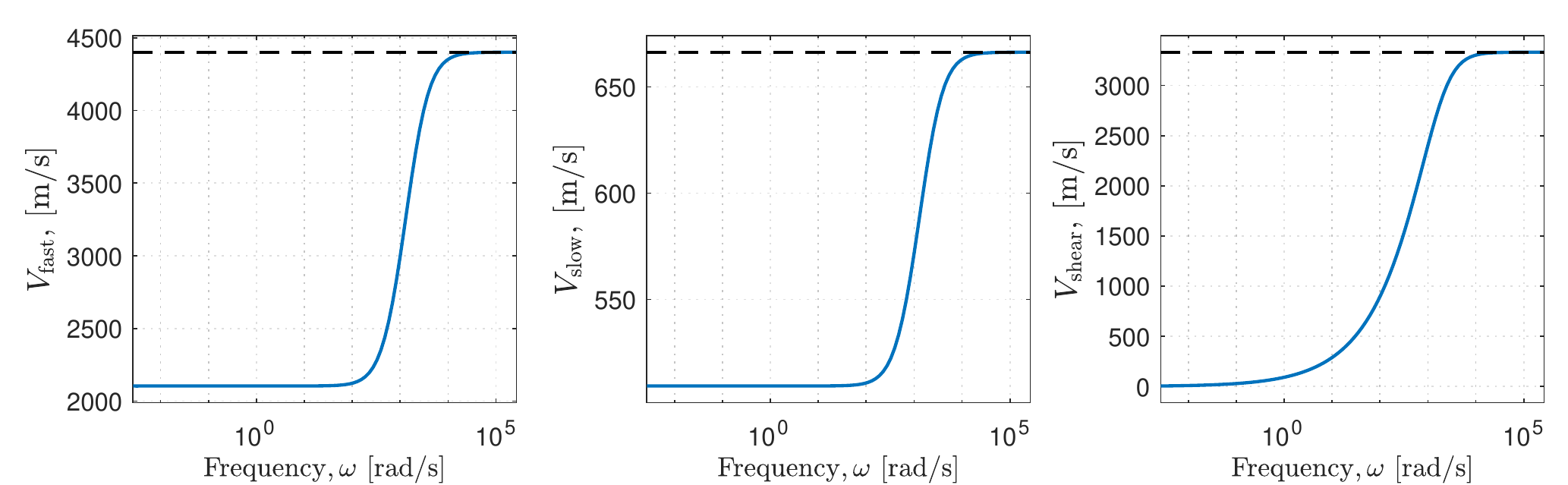}\\
			\includegraphics[draft=false,trim=0 0 0 
			0,clip,scale=0.7]{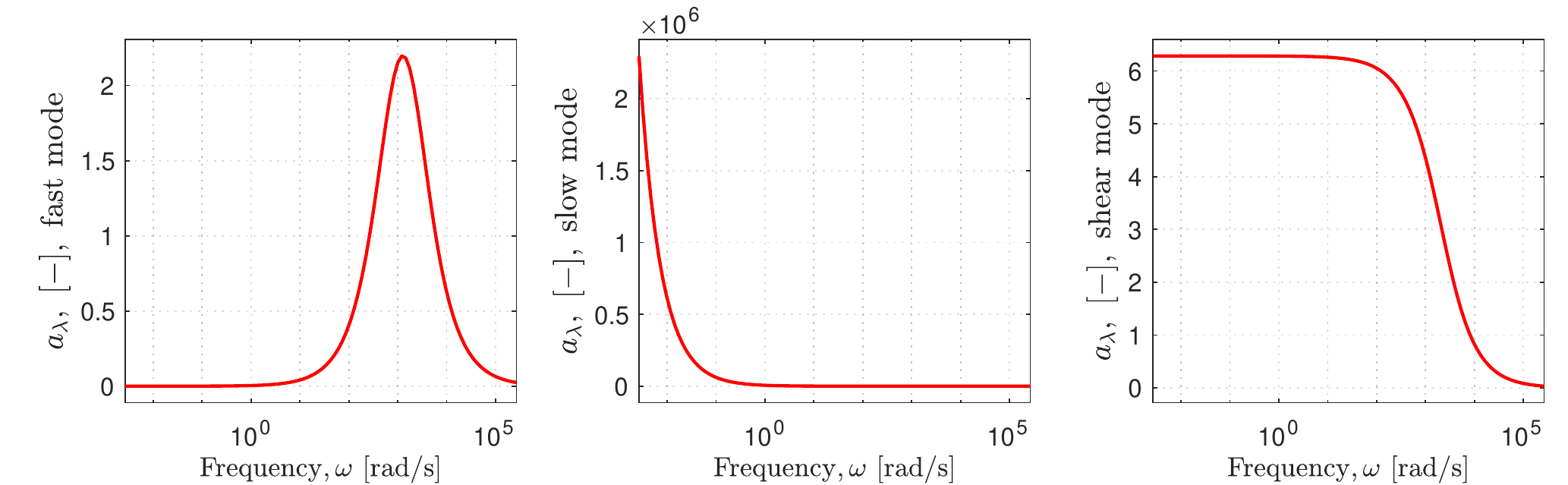}
		\end{tabular}
		\caption{ Sound wave dispersion and 
			attenuation factor per wave length $ a_\lambda $ for all three modes for the case $ \tau = \theta_2 
			< \infty $: fast P-mode (left column) 
			slow P-mode 
			(middle column), and shear mode (right column). Porosity $ \phi = \alpha_1 = 0.2$. Other material parameters 
			are given in Table\,\ref{tab.param}.
		}  
		\label{fig.disp1.theta2}
	\end{center}
\end{figure}

Fig.\,\ref{fig.disp1} shows the dispersion curves $ \Vfast(\omega) $, $ \Vslow(\omega) $, $ \Vshear(\omega) $ and 
attenuation factors for $ \tau < \infty $, $ \theta'_2=\infty $ (the fluid is viscous but the interfacial friction is ignored) and 
Fig.\,\ref{fig.disp1.theta2} depicts the dispersion curves when both 
dissipative mechanisms are presented, $ \tau < \infty $ and $ \theta'_2 < \infty $. The 
material parameters are: $ \phi = \alpha_1 =0.2 $, $ \cs{1} =750 $\,m/s, liquid viscosity $ \eta = 
10^{-2} $\,Pa$ \cdot $s, the skeleton relaxation 
time $ \tau_2 = \infty $ (pure elastic medium), saturating liquid relaxation time $ \tau_1 = \eta/\mu_1 = 
1.709\cdot10^{-11}$\,s. The poroelastic medium relaxation time $ \tau $ is computed from \eqref{eqn.tau.simplified}. When $ \theta'_2 < \infty $ 
we take $ \theta'_2 = \tau $. 

One can notice the following differences between two cases presented in Fig.\,\ref{fig.disp1} and 
Fig.\,\ref{fig.disp1.theta2}. First, the shape 
of the 
phase velocity $ \Vslow(\omega) $ changes from a longitudinal-like shape in Fig.\,\ref{fig.disp1} 
to a shear-like shape in 
Fig.\,\ref{fig.disp1.theta2}. The low frequency limits ($ \omega \to 0 $) of $ \Vslow(\omega) $ are 
different as well, 
509.2\,m/s in Fig.\,\ref{fig.disp1.theta2} versus 0\,m/s in Fig.\,\ref{fig.disp1}. Second, the low 
frequency limits of $ \Vfast(\omega) $ are also very different. Third, the 
attenuation factor $ a_\lambda(\omega) $ of the fast mode increases over 3 orders of magnitude from 
Fig.\,\ref{fig.disp1} to Fig.\,\ref{fig.disp1.theta2}. And finally, the attenuation factor $ 
a_\lambda(\omega) $ of the slow mode increases even 
more dramatically over 6 
orders of magnitude from 
Fig.\,\ref{fig.disp1} to Fig.\,\ref{fig.disp1.theta2}.

 \begin{table}[h]
	\begin{center}
	\begin{tabular}{lccccccc}
			\rowcolor[HTML]{EFEFEF} 
	\hline %inserts one horizontal lines
			  State & $ i $ & $c_{\textrm{p},i}$, [m/s] & $ c_{\textrm{s},i} $, [m/s] & $ \rho_i $, 
			  [kg/m$ ^3$] & $ \eta $,[Pa$\cdot$s] & $\tau_i $, [s] & $ n $, [-]\\
			 \hline %inserts one horizontal lines
		Fluid & 1 & 1500 & 100 & 1040 & $ 10^{-2} $ & $ \tau_1 = \eta/\mu_1 $ & 8\\
 	\rowcolor[HTML]{EFEFEF} 
		Solid & 2 & 6000 & 3500 & 2500 & -- & $ \tau_2 = \infty $ & 8\\
		\end{tabular}
	\caption{Material parameters used in the numerical simulation.}
	\label{tab.param}
	\end{center}
\vspace{-0.75cm}
\end{table}

%_______________________________
\newpage
\section{Numerical tests}\label{sec.numerics}

\subsection{General descriptions} 

In this section, we numerically verify the correctness of the obtained theoretical results and compare computed wave velocities for different frequencies with the velocities corresponds to dispersion curves. 

To generate seismic waves, it is necessary to specify the source term in the right-hand side of system  \eqref{stress.velocity}. On can define a volumetric-type source exciting only compressional waves in a homogeneous media by inserting source term in the pressure equation \eqref{stress.velocity.P} or in equations for the diagonal components $\Sigma_{ii} (i=1,2,3)$ of the deviatoric stress \eqref{sij}. To generate all types of waves propagating from the source in a homogeneous medium, including S wave,  it is necessary to define oriented-type source by adding source term only in one of the equations for  $\Sigma_{ii}$. In this case, we obtain vertical- or horizontal-type sources with maximum seismic energy propagation in the corresponding direction.
We define the source term as 
\begin{equation}
F(x,y,t)=\delta(x_0,y_0)f(t),
\end{equation}
where $\delta$-Dirac's delta function, localizing the source in point $(x_0,y_0)$ and $f(t)$-Ricker's wavelet in time
\begin{equation}
f(t)=(1-\omega^2(t-t_{0})^{2}/2)exp[-\omega^2(t-t_{0})^{2}/4],
\end{equation}
 where $\omega=2\pi f_0$ is the angular frequency, $f_0$ is peak frequency in hertz  and $t_{0}$ is the time wavelet delay, selected as $t_{0}=2/f_0$ in our further consideration.
 
As a computational method, we choose finite difference schemes on staggered grids. The equations \eqref{stress.velocity} form a symmetric hyperbolic system which is well suited to using staggered grids. The method is widely used due to computational efficiency and easy implementation. We use the classic second-order accurate staggered-grid stencil designed on the Cartesian coordinate system. The derivation of finite-difference approximation for system \eqref{stress.velocity} can be found in \cite{Romenski2020}. To suppress artificial reflections from non-physical computational boundaries, we apply the perfectly matched layer technique (PML) in the original split-field formulation \cite{Collino2001}.

%____________
\subsection{Homogeneous medium} 

We consider a set of two-dimensional test cases, in which the wave propagation is studied in a 
square domain $\Omega$ covered by a numerical grid with $N$ grid points in each spatial direction 
and grid space interval $\Delta l$ so that $\Omega=[N\Delta l]^2$ with centering at the origin of 
the Cartesian coordinate system $(X,Y)$. To simulate an unbounded medium, the domain $\Omega$ is 
surrounded by PML layers to avoid non-physical reflections from computational boundaries. 

We consider three typical angular frequencies $\omega=10^{5}, 10^{3}$ and  $10$ [rad/s],
covering all cases of the characteristic behaviour of the dispersion curves presented in the previous section. For each selected angular frequency $\omega$, we scale the spatial step $\Delta l$ and recording time $T$ by some scaling parameter $\varepsilon$, leaving the number of nodes $N$ constant. 
This is done because the seismic wavelength is proportional to the frequency and hence the size of 
the computational domain and the recording time must change with the frequency scaling.
The parameter $\varepsilon$ is selected for the convenience of comparison: if the wavefields properties do not depend on the frequency, then for all three cases the seismograms should be identical in scales proportional to this parameter.

\begin{figure}[h]
\centering
\begin{tabular}{cc}
%  \hline
%  control: length you want (using vspace) and the width, using pbox 
   \pbox{8cm}{\vspace{0.0ex} \includegraphics[draft=false,width=8cm]{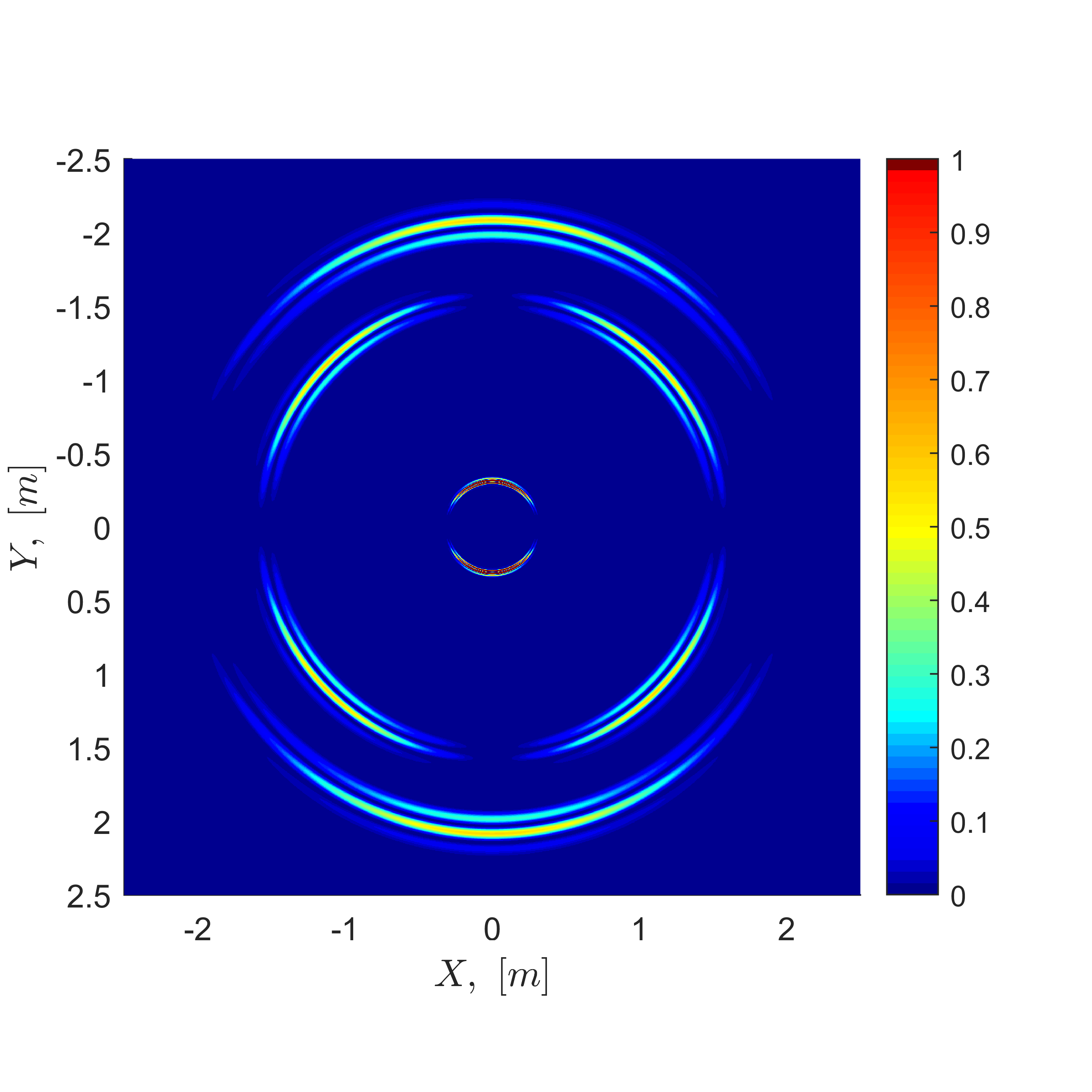}} &
   \pbox{8cm}{\vspace{0.0ex} \includegraphics[draft=false,width=7.5cm]{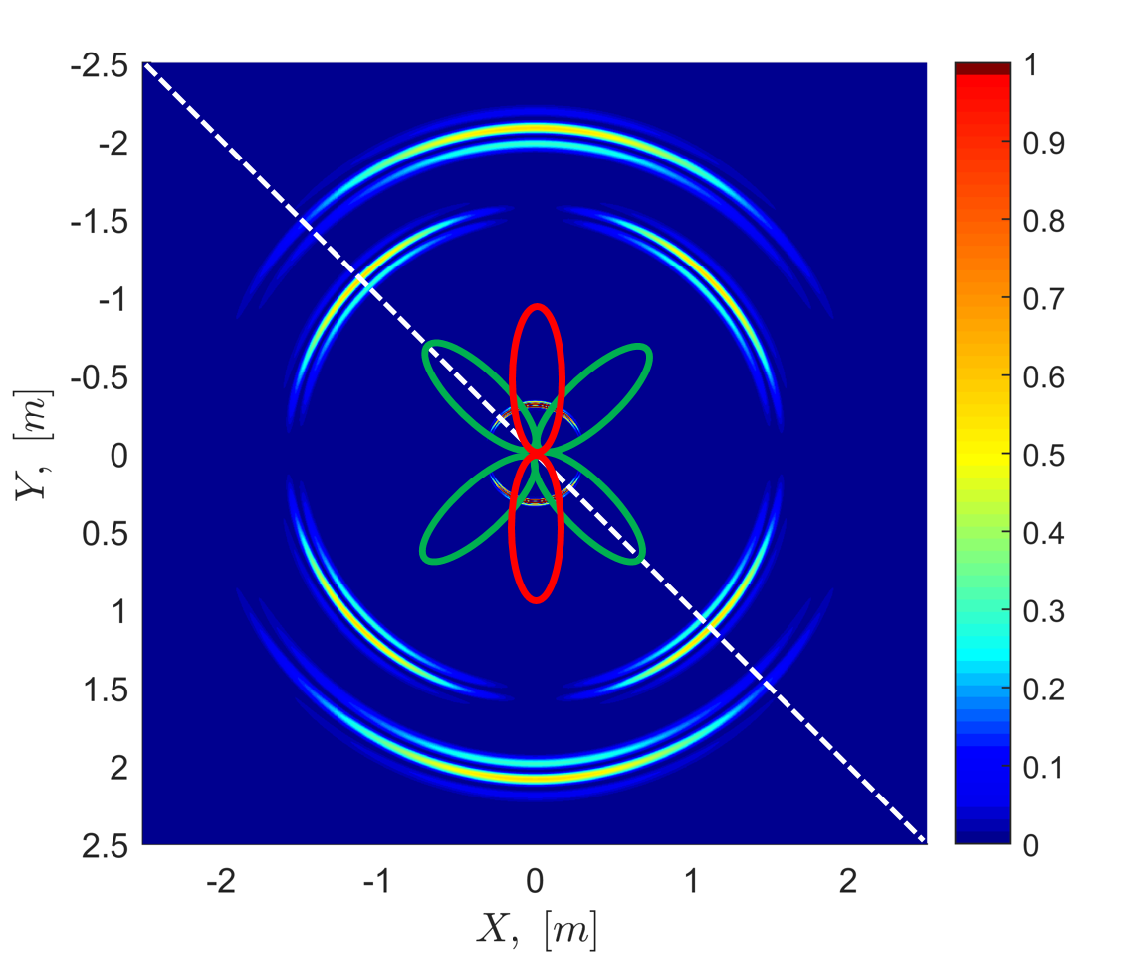}} \\
 % {\pbox{9cm}{\vspace{2ex} $\tau = \theta_2 = \infty$            \vspace{1ex}}}     &       
 % {\pbox{9cm}{\vspace{2ex} $\tau = \theta_2 = 3.75\cdot10^{-4}$  \vspace{1ex}}} \\
% \hline
\end{tabular}
          \caption{ Snapshot of the total velocity vector for $\omega=10^{5}$  and  time $t = 5\cdot10^{-4} $\,s computed with parameters $\tau = \theta_2 = 3.75\cdot10^{-4}$ for the vertically-type source(on the left picture). Radiation patterns for the P (red line) and the S (green line) waves and receivers location (dotted white line) (on the right picture).}
          \label{fig:Snap_omega_10_5}
\end{figure}

First, let us consider the case $\omega=10^{5}$ for which the velocity dispersion curve reaches the 
upper asymptotic constant. For simulation we take grid parameters $N=10^{4}$, $\Delta l=\Delta 
l_0=5\cdot10^{-4} m$, medium parameters from Table \ref{tab.param}, porosity $\phi = 
\alpha_1^0=0.2$, power index $n=8$, vertically-type source location $(x_0,y_0)=(0,0)$ and 
seismogram recording time $T=T_0=10^{-3} s$.  Grid spacing $\Delta l_0$ provides approximately an 
amount of 10 points per slow compressional wavelength that is needed to avoid the numerical 
dispersion. Fig.\,\ref{fig:Snap_omega_10_5} (left) shows a snapshot of the normalized total 
velocity vector for  time $t = 5\cdot10^{-4} $\,s. The appearance of fast and slow compressional P 
waves and shear S wave exited by a vertically-type source radiating in the $ Y $-direction can be 
observed. 
It is clearly can be seen the difference in radiation patterns, denoting the angular dependence of 
the strength of the  waves radiated by a source for P and S waves: the maximum amplitude of P wave 
corresponds to the $ Y $-direction, while for S wave this direction is $ \pm45^{\circ}$. 
Fig.\,\ref{fig:Snap_omega_10_5} (right) plot the radiation patterns for the P (red line) and the S 
(green line) waves. The amplitude of the S wave is totally vanishing in directions that coincide 
with the $X$ and $Y$ grid axis. This remark explains why we choose the diagonal of the 
computational domain as a seismogram recording line. For all subsequent tests in this section, we 
assume that the receivers locate on the diagonal of $\Omega$ with a uniform spacing (dotted white 
line in Fig.\,\ref{fig:Snap_omega_10_5}). 

For each selected angular frequency $\omega=10^{5}, 10^{3},10$ we simulate wavefields with 
different values of interfacial friction (relative velocity relaxation) coefficient and shear 
stress relaxation time to study their influence on the wavefield formation. In case of 
$\omega=10^{3}$ we set $T=\varepsilon T_0$ $s$ and $\Delta l=\varepsilon\Delta l_0$ $m$ with 
$\varepsilon=10^{2}$ and for the case of $\omega=10$, we set $\varepsilon=10^{4}$.

\begin{figure}[h]
\centering
\begin{tabular}{cc}
%  \hline
%  control: length you want (using vspace) and the width, using pbox 
% after \\: % \hline or \cline{col1-col2} \cline{col3-col4} ...
   \pbox{9cm}{\vspace{0.01ex} \includegraphics[draft=false,width=9cm]{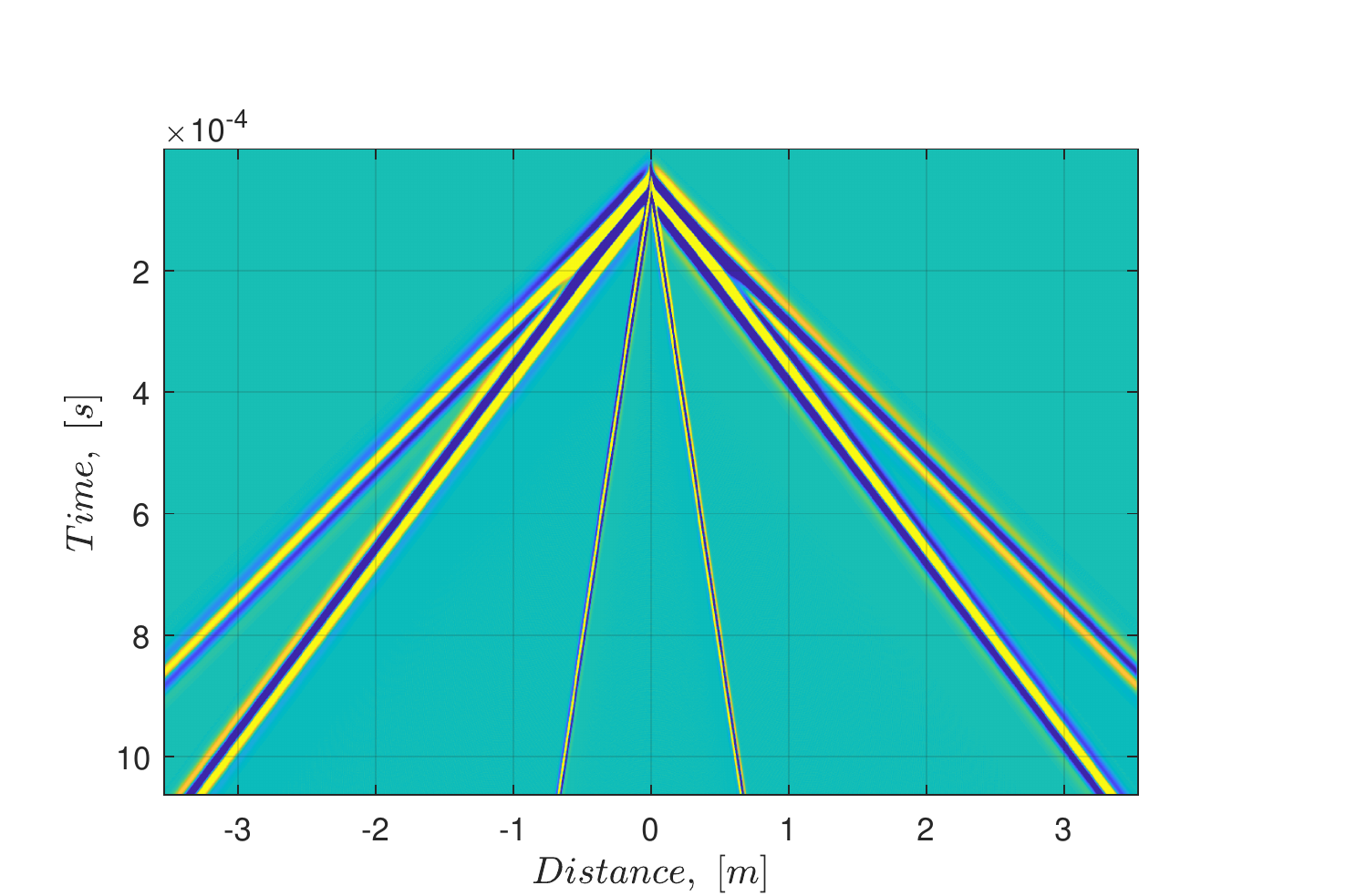}} &
   \pbox{9cm}{\vspace{0.01ex} \includegraphics[draft=false,width=9cm]{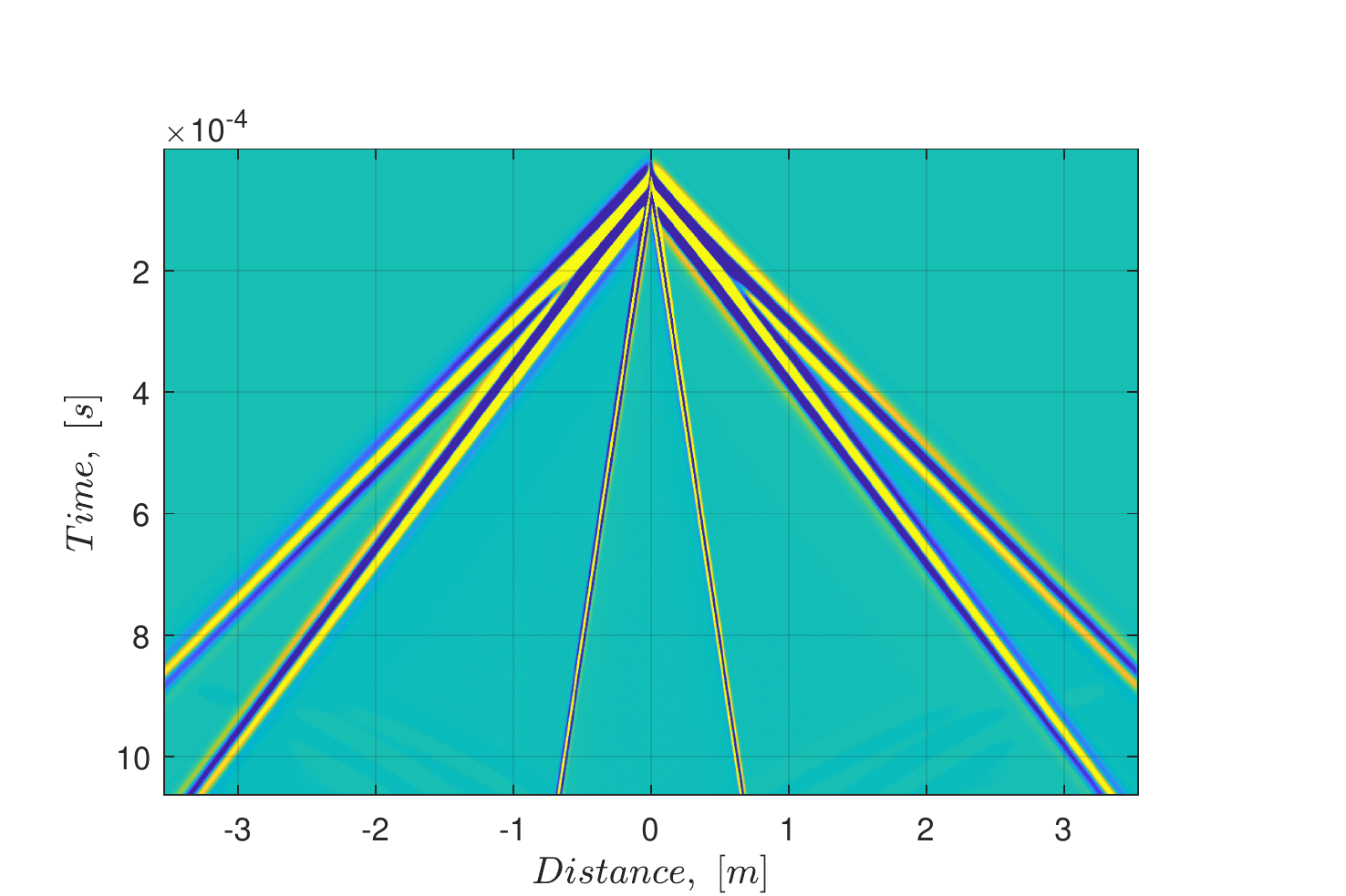}} \\
  {\pbox{9cm}{\vspace{2ex} $\tau = \theta_2 = \infty$            \vspace{1ex}}}     &       
  {\pbox{9cm}{\vspace{2ex} $\tau = \theta_2 = 3.75\cdot10^{-4}$  \vspace{1ex}}} \\
 &  \\
  % \hline
  
   \pbox{9cm}{\vspace{0.01ex} \includegraphics[draft=false,width=9cm]{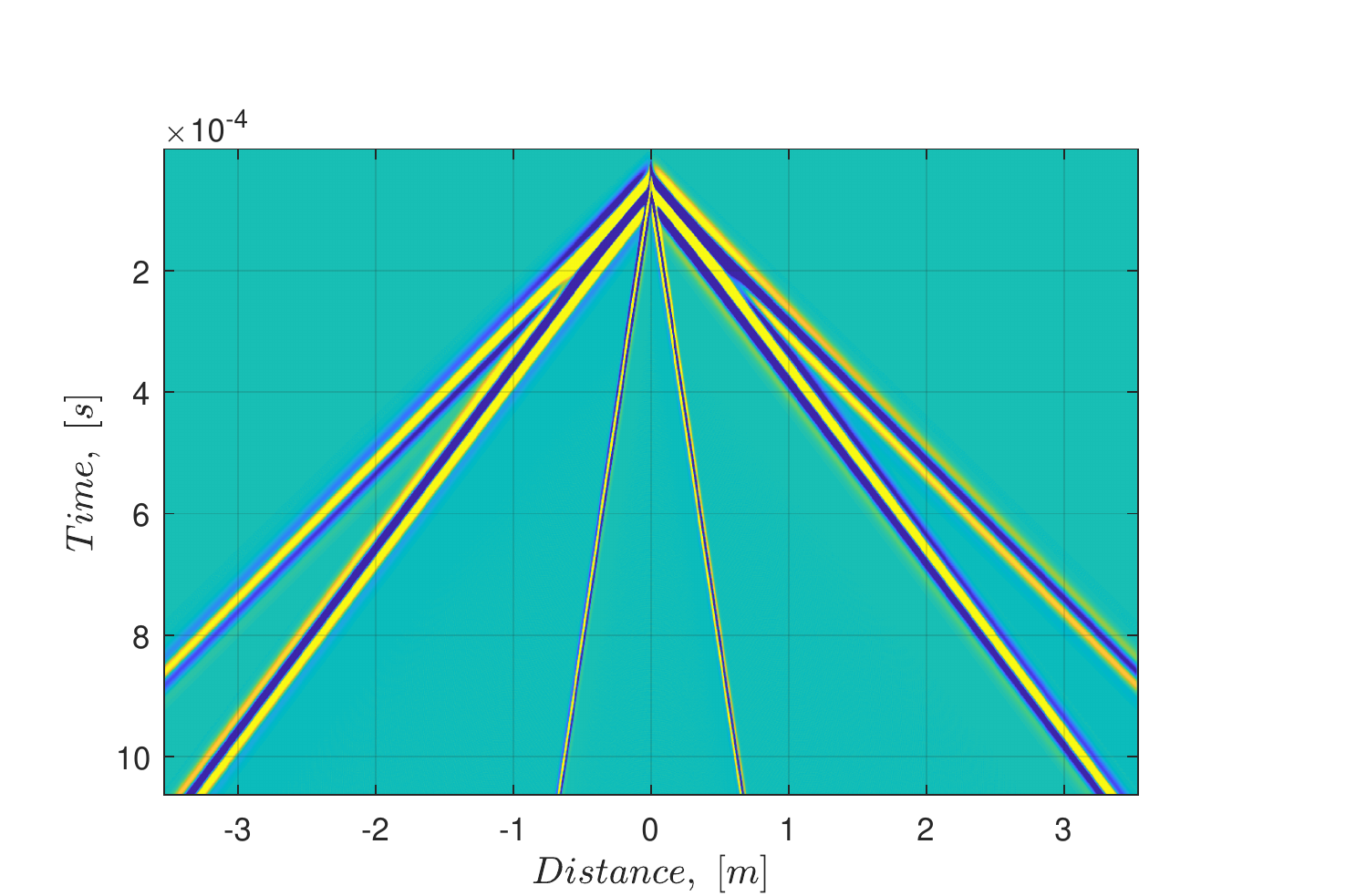}} &
   \pbox{9cm}{\vspace{0.01ex} \includegraphics[draft=false,width=9cm]{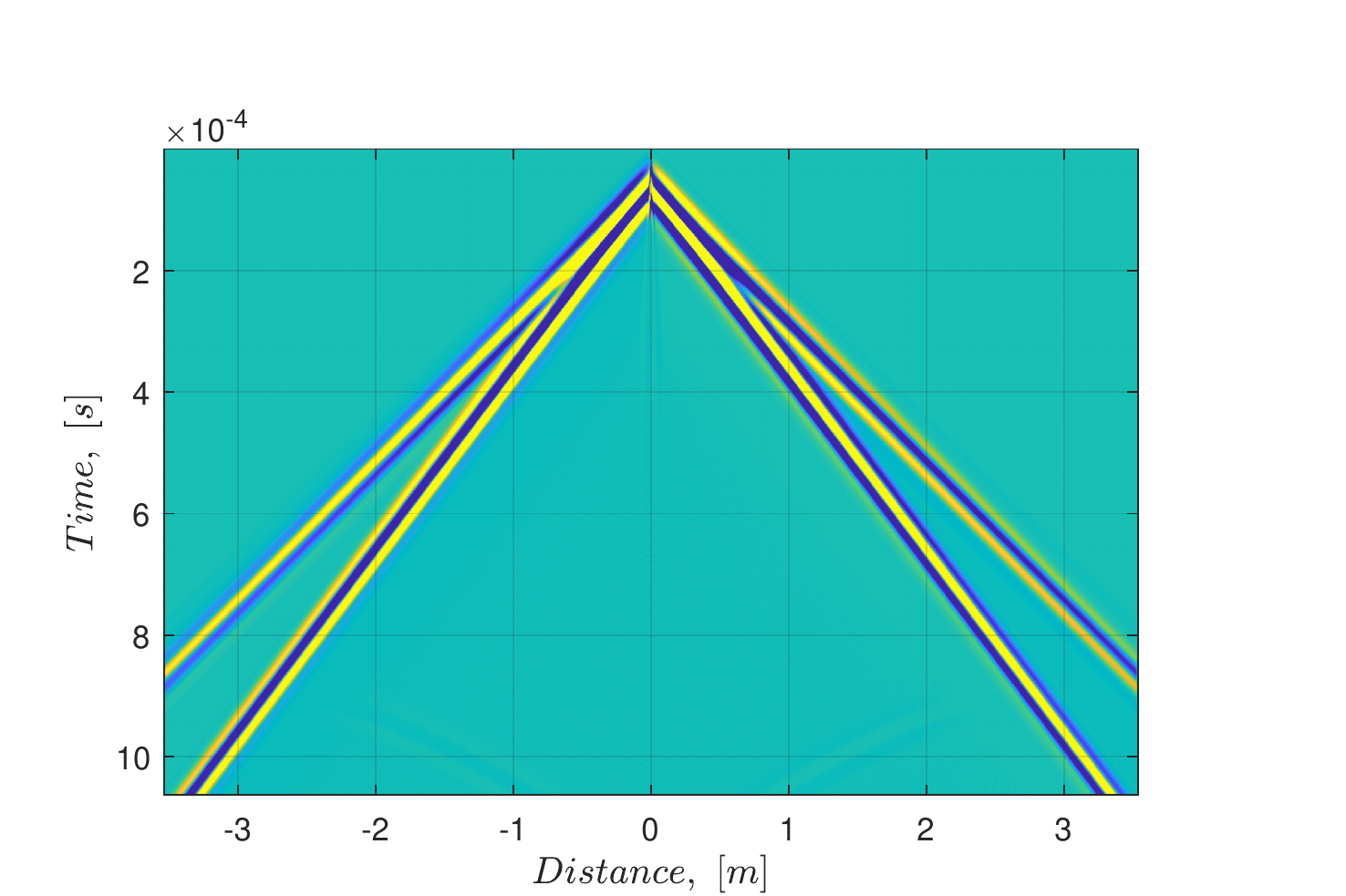}} \\
  {\pbox{9cm}{\vspace{2ex} $\tau = \infty, \;\;\;\theta_2 = 3.75\cdot10^{-4}$  \vspace{4ex}}}     &   
  {\pbox{9cm}{\vspace{2ex} $\tau = \infty, \;\;\;\theta_2 = 3.36\cdot10^{-7}$  \vspace{4ex}}} \\

% \hline
\end{tabular}
          \caption{ Seismograms for the mixture vertical velocity $v^2$ for $\omega=10^{5}$ and 
          different values of the relative velocity relaxation time $\theta_2$ and shear stress 
          relaxation time $\tau$.}
	      \label{fig:Joint_Seismogr_omega10_5}
\end{figure}

\begin{figure}[h]
\centering
\begin{tabular}{cc}
  % \hline
%  control: length you want (using vspace) and the width, using pbox 
% after \\: % \hline or \cline{col1-col2} \cline{col3-col4} ...
   \pbox{9cm}{\vspace{0.01ex} \includegraphics[draft=false,width=9cm]{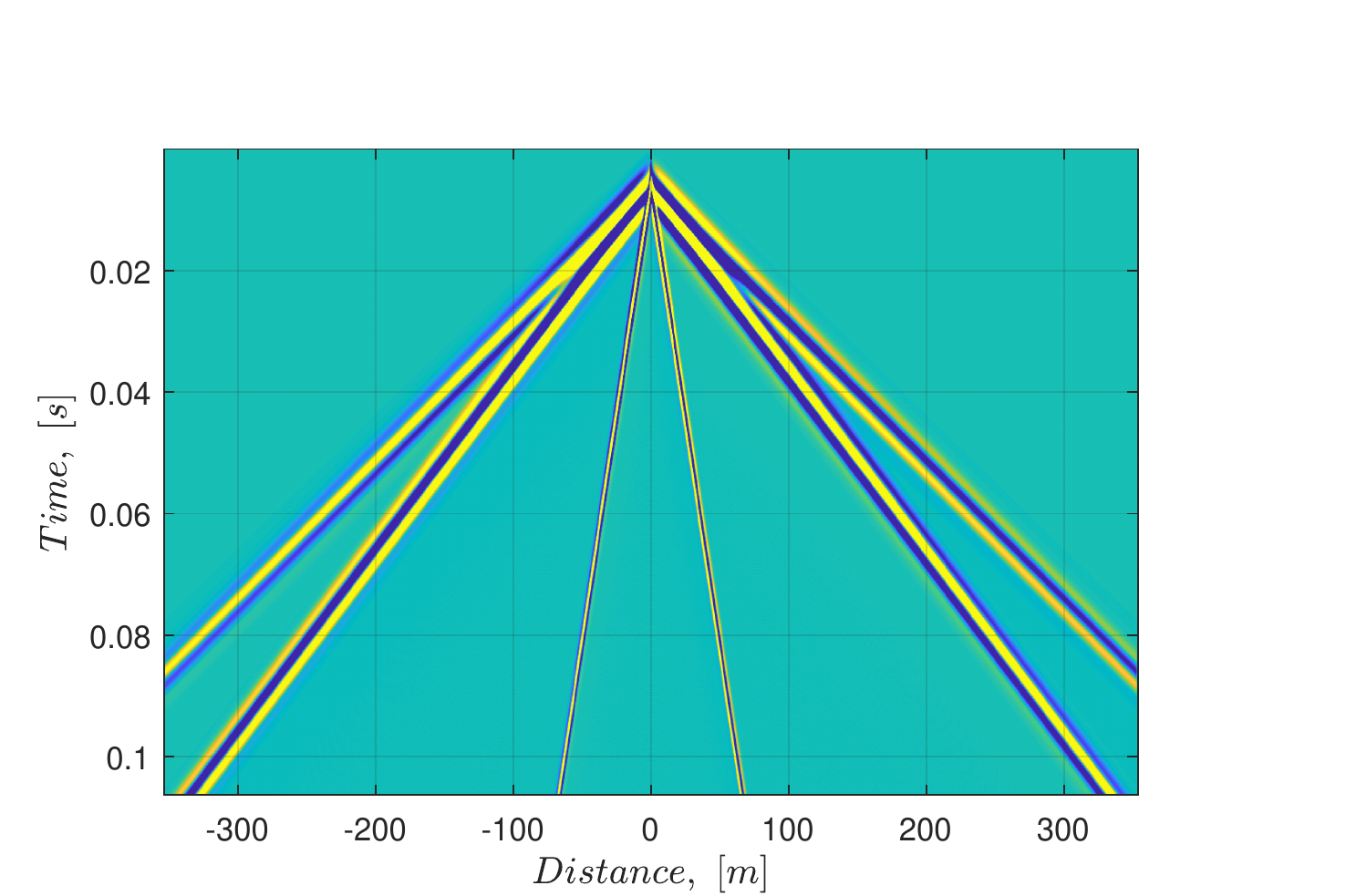}} &
   \pbox{9cm}{\vspace{0.01ex} \includegraphics[draft=false,width=9cm]{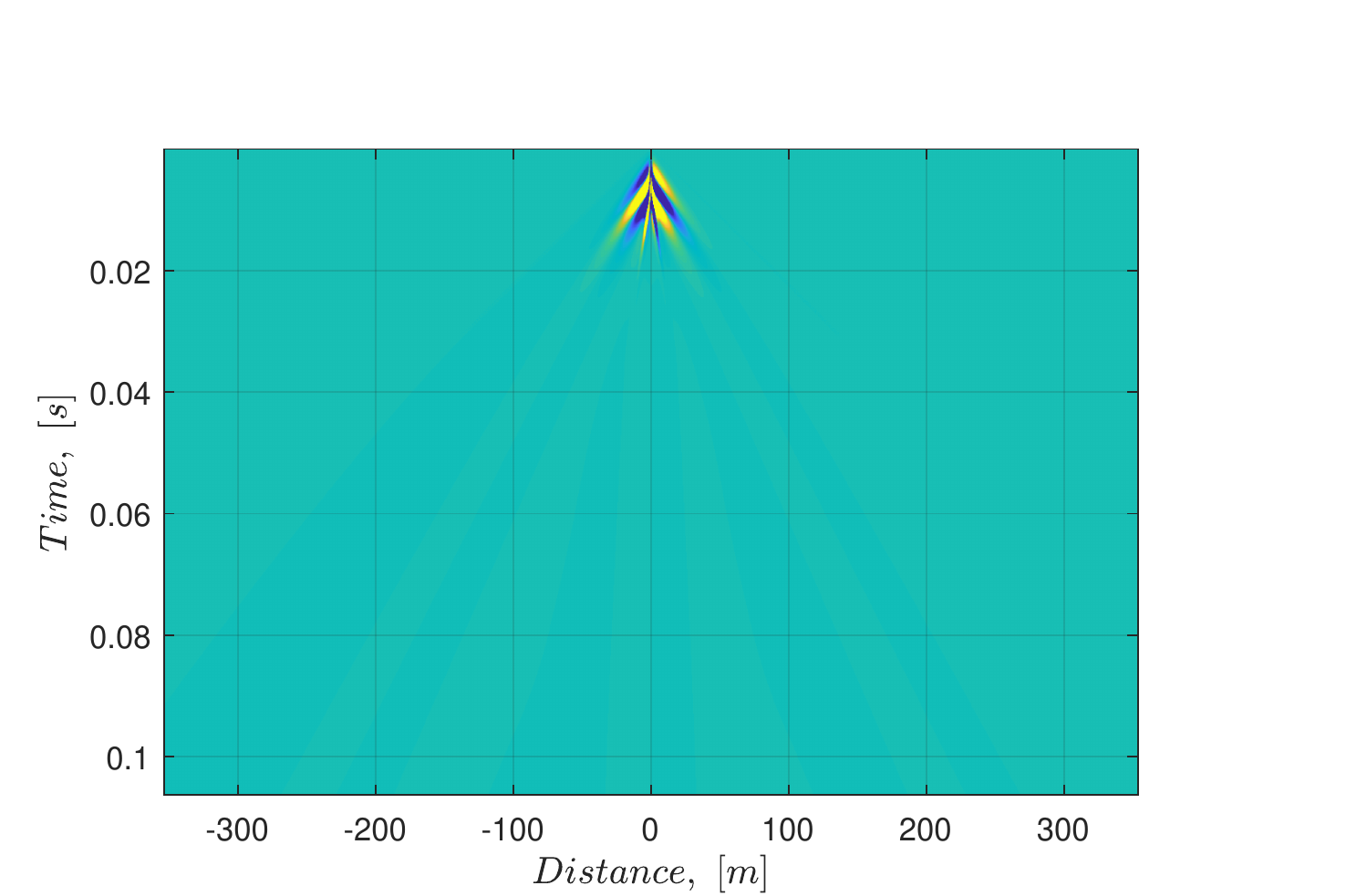}} \\
  {\pbox{9cm}{\vspace{2ex} $\tau = \theta_2 = \infty$            \vspace{1ex}}}     &       
  {\pbox{9cm}{\vspace{2ex} $\tau = \theta_2 = 3.75\cdot10^{-4}$  \vspace{1ex}}} \\
 &  \\
  % \hline
  
   \pbox{9cm}{\vspace{0.01ex} \includegraphics[draft=false,width=9cm]{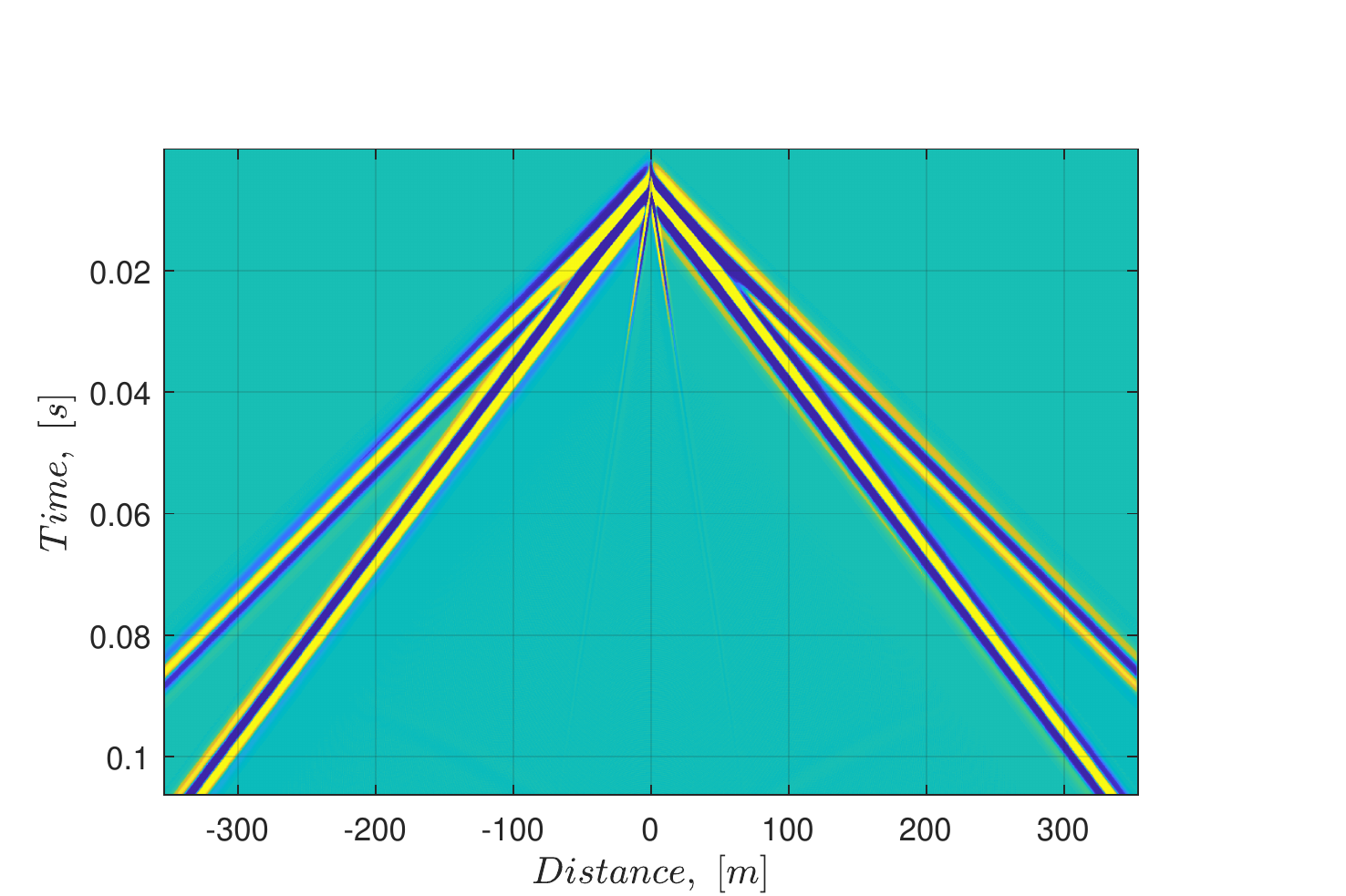}} &
   \pbox{9cm}{\vspace{0.01ex} \includegraphics[draft=false,width=9cm]{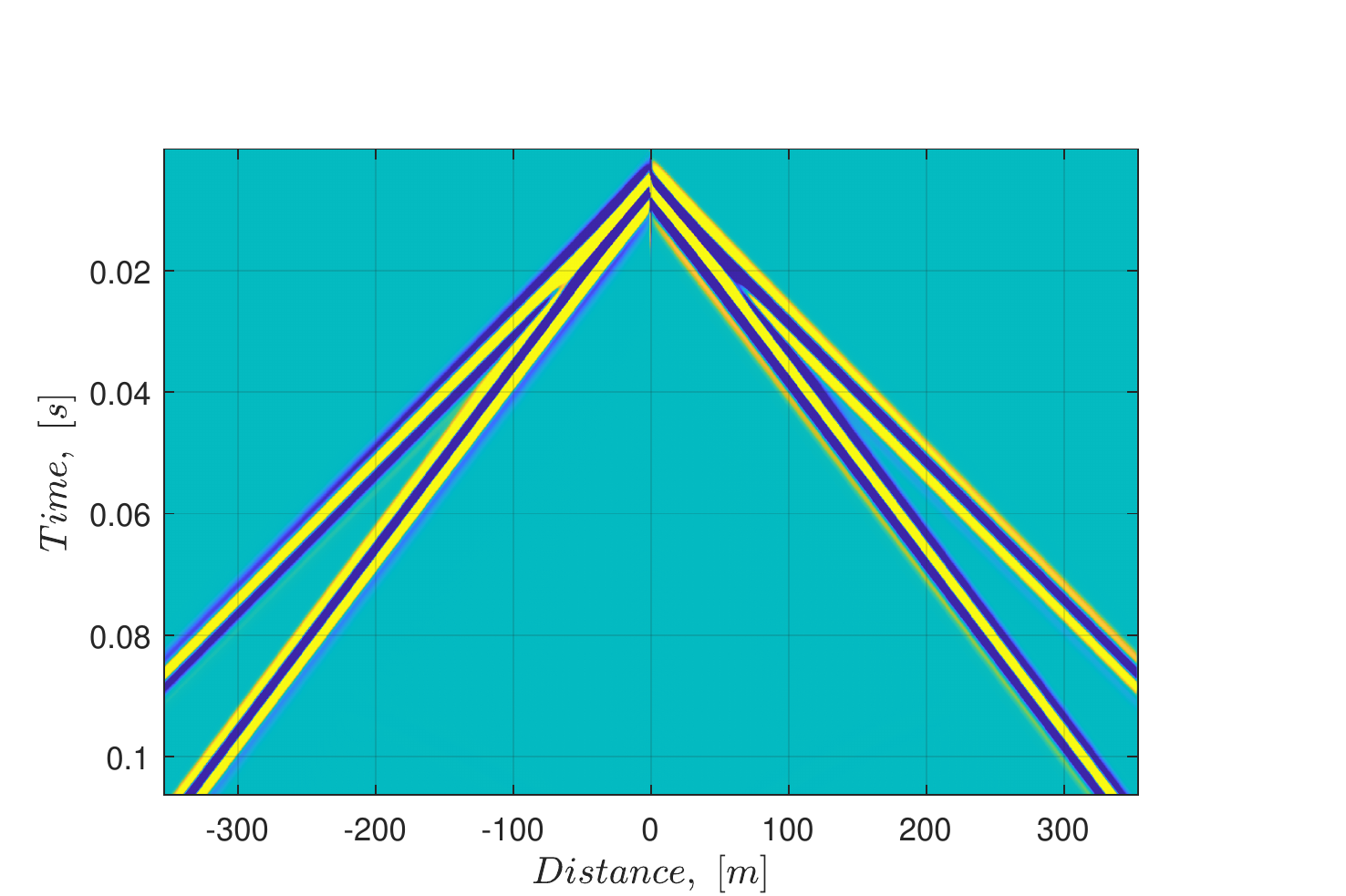}} \\
  {\pbox{9cm}{\vspace{2ex} $\tau = \infty, \;\;\;\theta_2 = 3.75\cdot10^{-4}$  \vspace{4ex}}}     &   
  {\pbox{9cm}{\vspace{2ex} $\tau = \infty, \;\;\;\theta_2 = 3.36\cdot10^{-7}$  \vspace{4ex}}} \\

% \hline
\end{tabular}
          \caption{ Seismograms for the mixture velocity $v^2$ for $\omega=10^{3}$ and different values of the relative velocity relaxation time $\theta_2$ and shear stress relaxation time $\tau$.}
	      \label{fig:Joint_Seismogr_omega10_3}
\end{figure}

\begin{figure}[h]
\centering
\begin{tabular}{cc}
  % \hline
%  control: length you want (using vspace) and the width, using pbox 
% after \\: % \hline or \cline{col1-col2} \cline{col3-col4} ...
   \pbox{9cm}{\vspace{0.01ex} \includegraphics[draft=false,width=9cm]{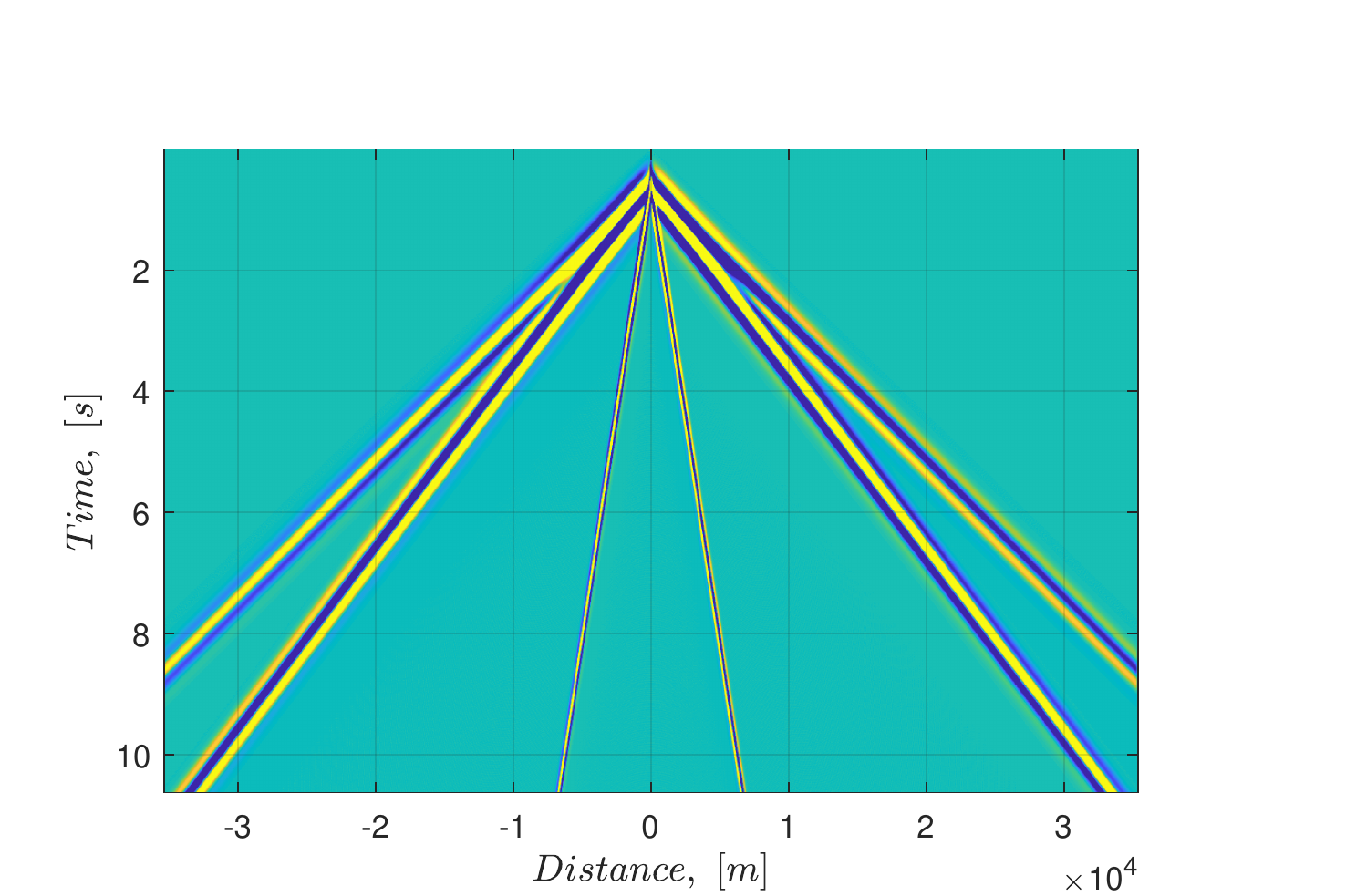}} &
   \pbox{9cm}{\vspace{0.01ex} \includegraphics[draft=false,width=9cm]{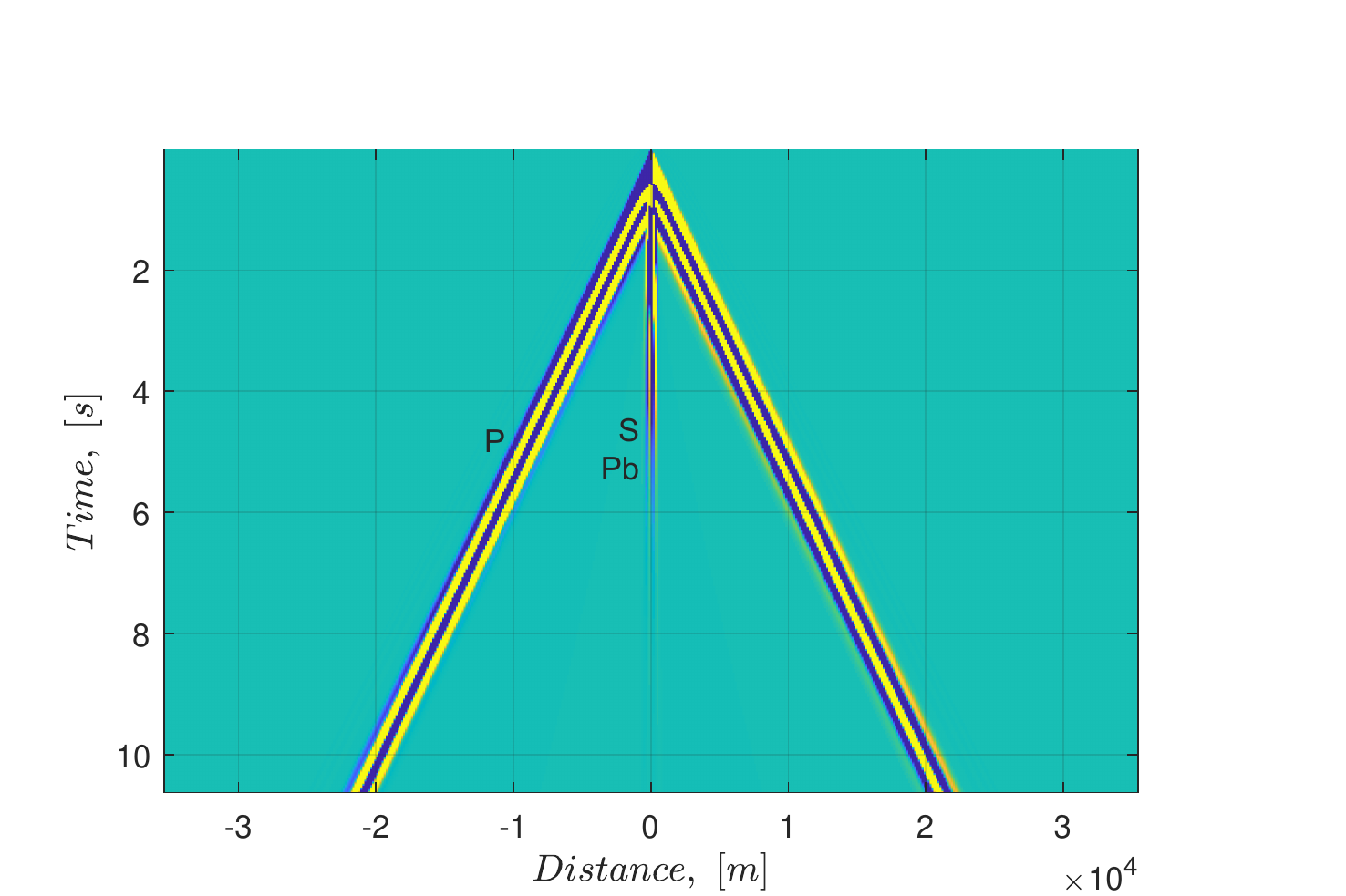}} \\
  {\pbox{9cm}{\vspace{2ex} $\tau = \theta_2 = \infty$            \vspace{1ex}}}     &       
  {\pbox{9cm}{\vspace{2ex} $\tau = \theta_2 = 3.75\cdot10^{-4}$  \vspace{1ex}}} \\
 &  \\
  % \hline
  
   \pbox{9cm}{\vspace{0.01ex} \includegraphics[draft=false,width=9cm]{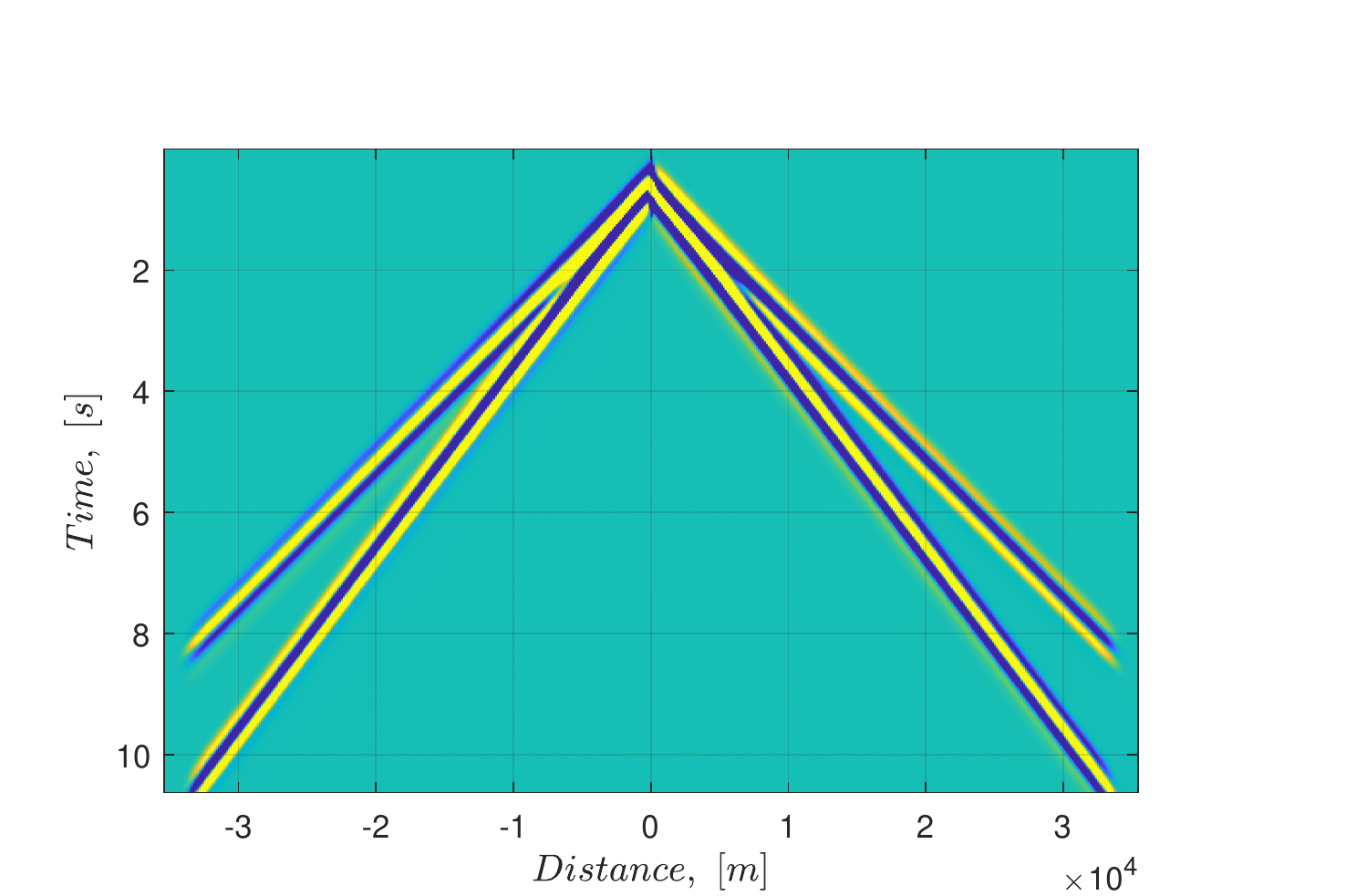}} &
   \pbox{9cm}{\vspace{0.01ex} \includegraphics[draft=false,width=9cm]{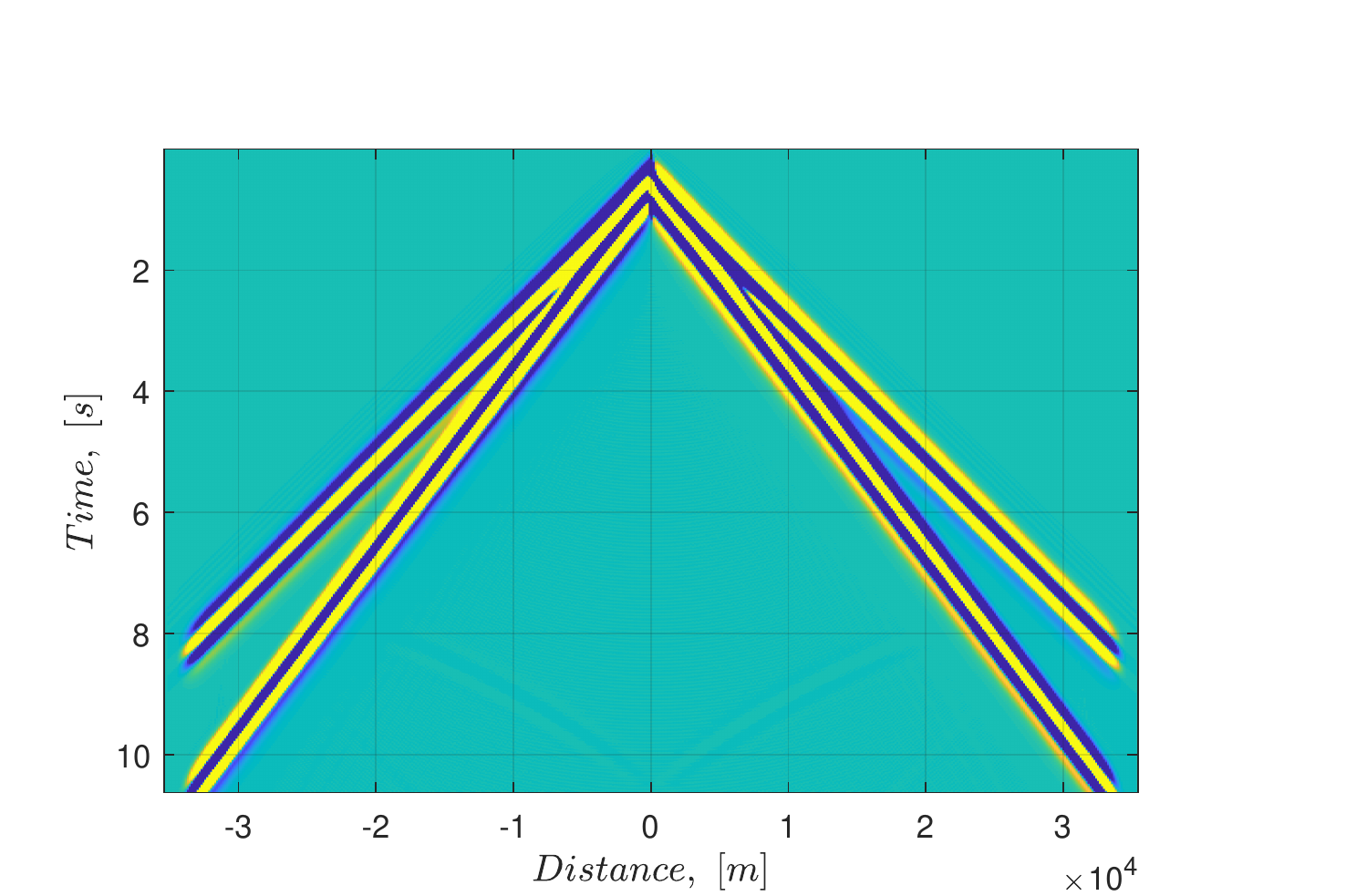}} \\
  {\pbox{9cm}{\vspace{2ex} $\tau = \infty, \;\;\;\theta_2 = 3.75\cdot10^{-4}$  \vspace{4ex}}}     &   
  {\pbox{9cm}{\vspace{2ex} $\tau = \infty, \;\;\;\theta_2 = 3.36\cdot10^{-7}$  \vspace{4ex}}} \\

% \hline
\end{tabular}
          \caption{ Seismograms for the mixture velocity $v^2$ for $\omega=10$ and different values of the relative velocity relaxation time $\theta_2$ and shear stress relaxation time $\tau$.}
	      \label{fig:Joint_Seismogr_omega10_1}
\end{figure}

Fig.\,\ref{fig:Joint_Seismogr_omega10_5}--\ref{fig:Joint_Seismogr_omega10_1} show seismograms for 
the mixture vertical velocity $v^2$ for different $\omega$ and different values of relaxation 
parameters: relative velocity relaxation time $\theta_2$ and shear stress relaxation time $\tau$. 
First we examine the case when the medium does not have any dissipation ($\tau = \theta_2 = 
\infty$). Then we consider the case with two dissipation mechanisms having the same finite values 
of shear stress and relative velocity relaxation times ($\tau = \theta_2 = 3.75\cdot10^{-4}$). The 
value of 
shear stress relaxation time is obtained by formula \eqref{eta} for the viscosity $ \eta_1 = 
10^{-2} $\,Pa$ \cdot $s and parameters $n=8$, $ \cs{1} = 100$ $m/s$ (Table\,\ref{tab.tau.n.cs}). 
Additionally, in order to study the effect of the relative velocity relaxation only, we examine 
cases with  $\tau = \infty$ and two different parameters of $\theta_2$ .

The comparison of seismograms in 
Fig.\,\ref{fig:Joint_Seismogr_omega10_5}--\ref{fig:Joint_Seismogr_omega10_1} allows us to formulate 
a conclusion about the dependence of wavefields on the two dissipation mechanisms. It is clearly 
seen that in the absence of dissipation, the velocities of the compressional and shear waves does 
not change for all frequencies. The situation is different when we switch on both dissipation 
mechanisms (case $\tau = \theta_2 = 3.75\cdot10^{-4}$). The dependence of velocity on frequency 
that coincides with the theoretical one predicted by the dispersion curves in 
Fig.\,\ref{fig.characteristics} can be observed. In addition, for frequency $\omega= 10^{3}$ we see 
rather strong wave attenuation  due to the predicted attenuation factor depicted in 
Fig.\,\ref{fig.disp1}. Comparing seismograms for all three cases ($\tau = \theta_2 = 
3.75\cdot10^{-4}$), ($\tau = \infty, \;\;\;\theta_2 = 3.75\cdot10^{-4}$) and  ($\tau = \infty, 
\;\;\;\theta_2 = 3.36\cdot10^{-7}$) we can see that the wavefield attenuation mostly depends on 
parameter $\tau$ and the parameter $\theta_2$ is responsible for the appearance and attenuation of 
the Biot mode.

%____________
\subsection{Two-layered  medium} 

In this section, we consider a two-layered elastic/poroelastic medium with a horizontal interface located in the origin $x=0$. The purpose of this experiment is to study the manifestation of a disperse shear S wave  near the layer interface. 

\begin{figure} [hbt!]
\centering
\begin{tabular}{cc}
  % \hline
   \pbox{9cm}{\vspace{0.0ex} \includegraphics[draft=false,width=8cm]{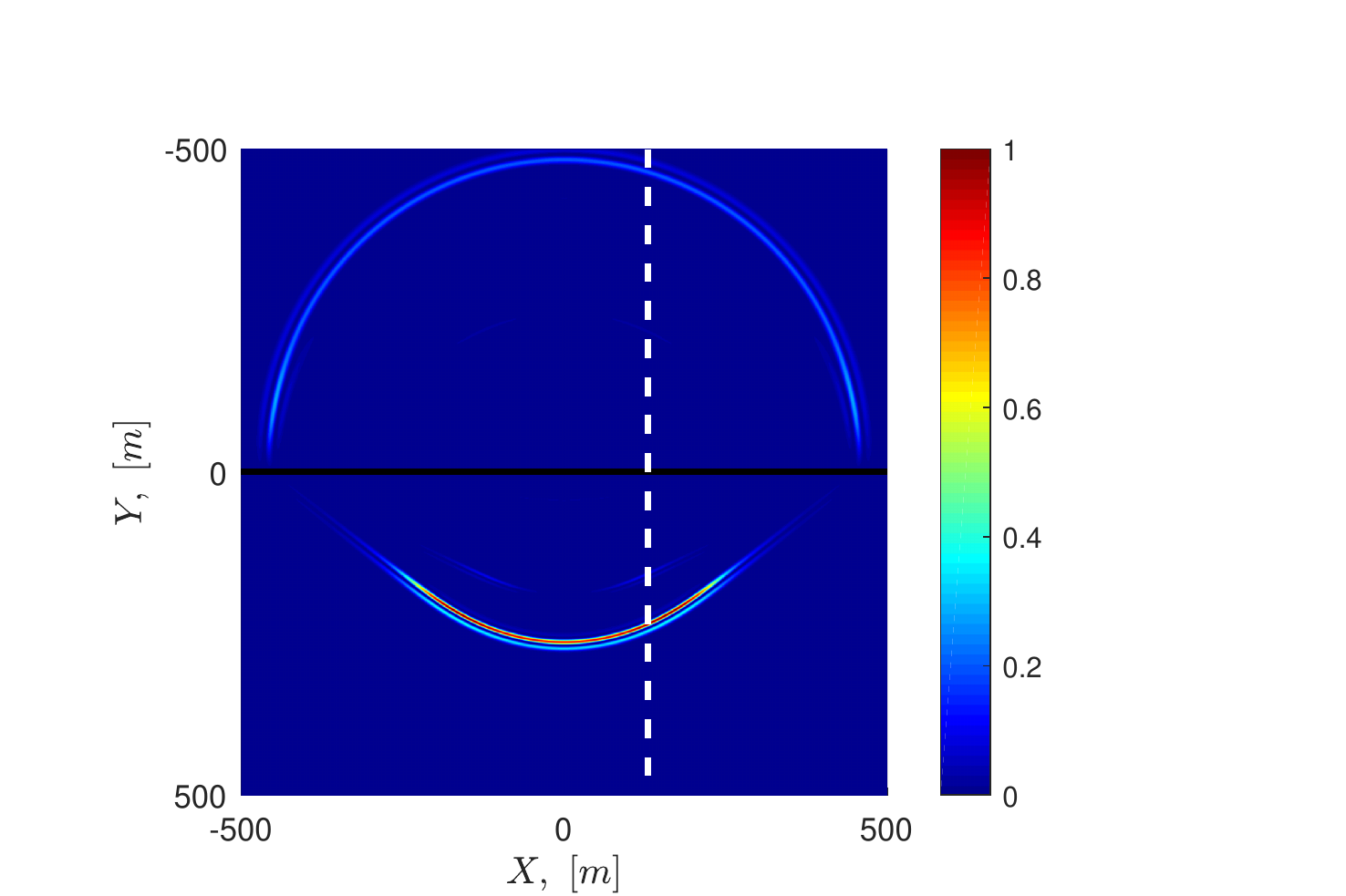}} &
   \pbox{9cm}{\vspace{0.0ex} \includegraphics[draft=false,width=8cm]{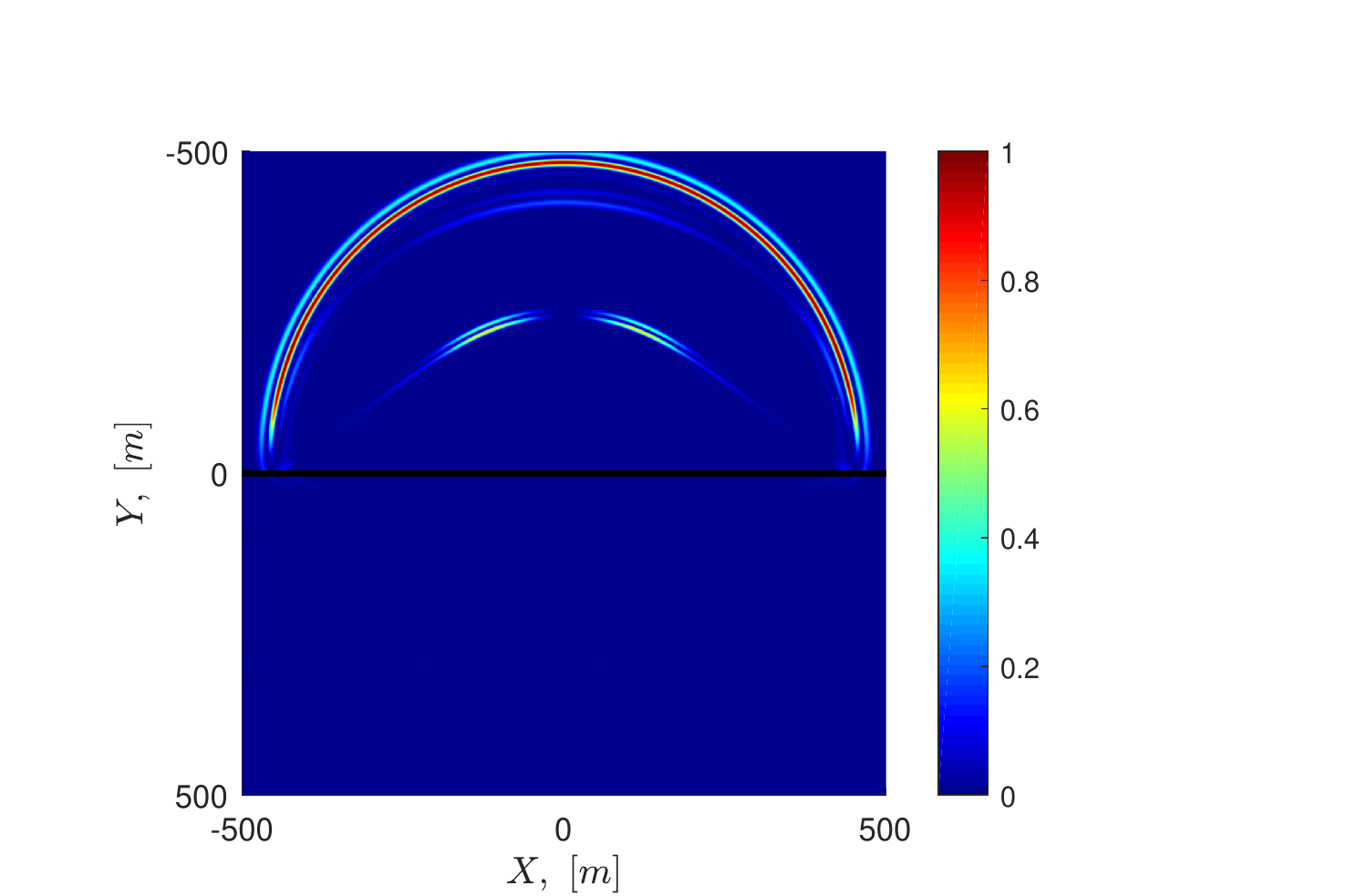}} \\
  {\pbox{9cm}{\vspace{2ex} $\tau = \theta_2 = \infty$            \vspace{1ex}}}     &       
  {\pbox{9cm}{\vspace{2ex} $\tau = \theta_2 = 3.75\cdot10^{-4}$  \vspace{1ex}}} \\
  &  \\
    % \hline
   \pbox{9cm}{\vspace{0.0ex} \includegraphics[draft=false,width=8cm]{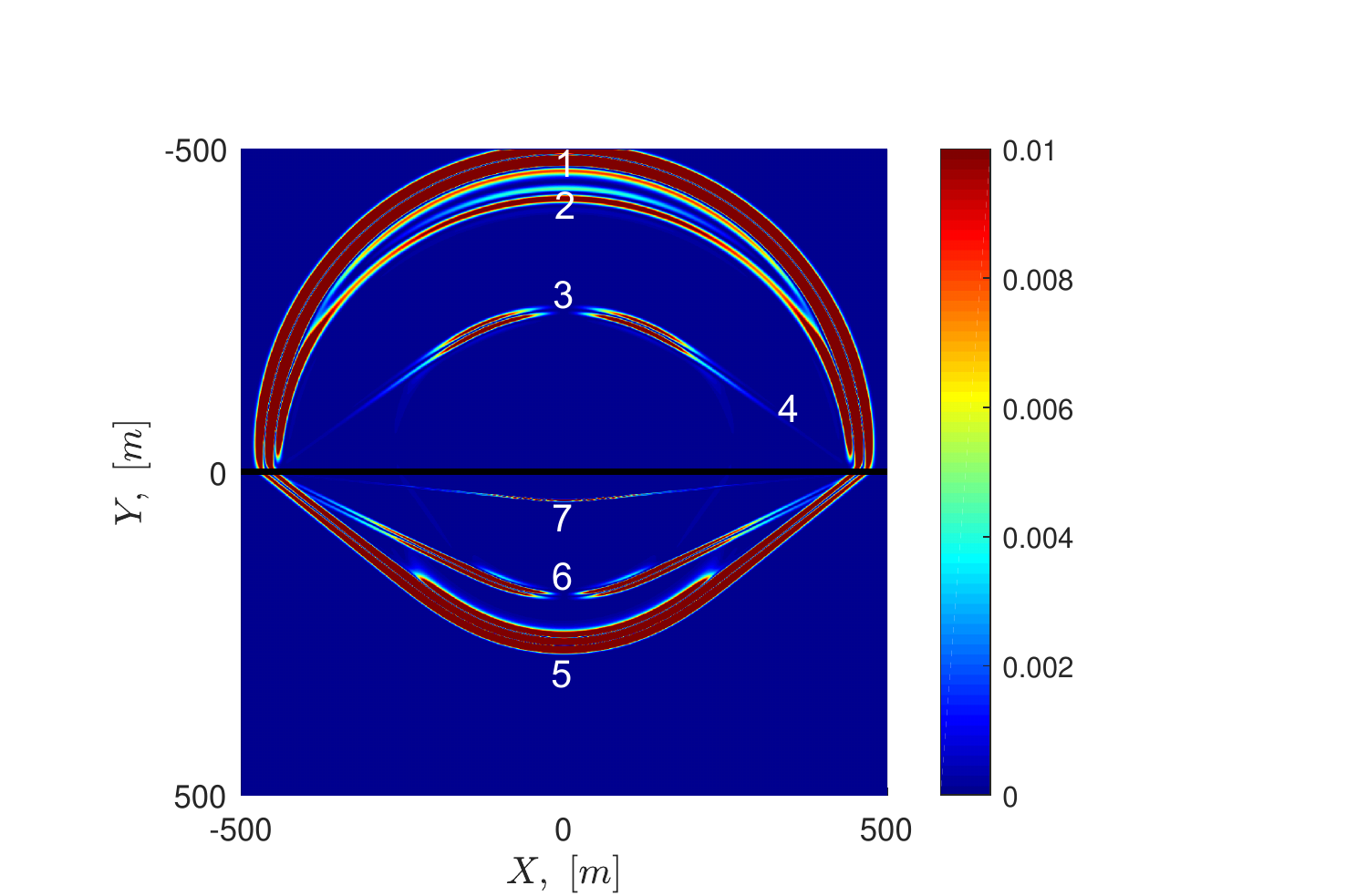}} &
   \pbox{9cm}{\vspace{0.0ex}\includegraphics[draft=false,width=8cm]{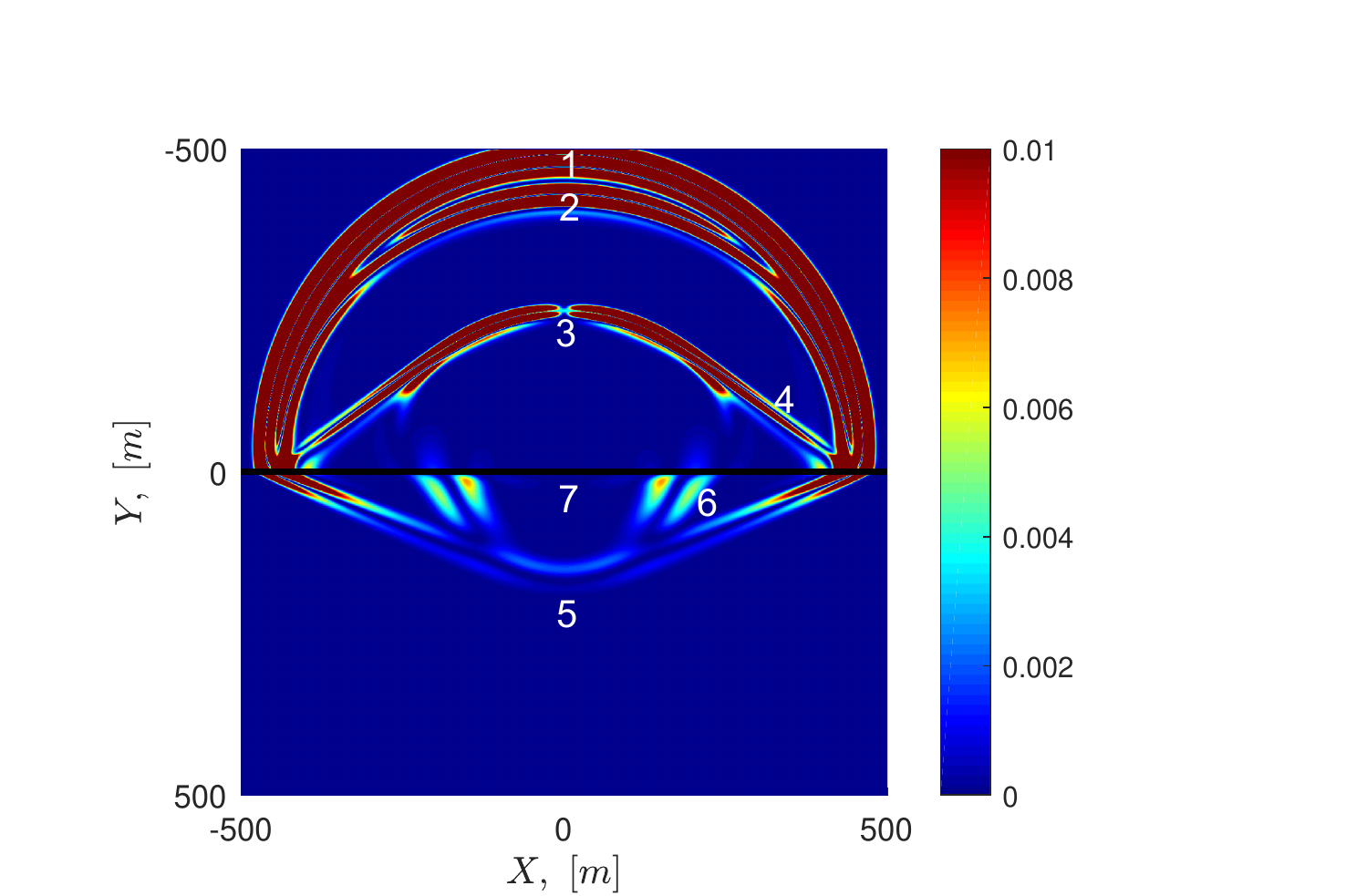}} \\
  {\pbox{9cm}{\vspace{2ex} $\tau = \theta_2 = \infty$, magnification           \vspace{1ex}}}     &       
  {\pbox{9cm}{\vspace{2ex} $\tau = \theta_2 = 3.75\cdot10^{-4}$, magnification  \vspace{1ex}}} \\
 % \hline
\end{tabular}
          \caption{ Snapshot of the normalized total velocity vector for $\omega=10^{3}$  at  time $t = 8\cdot10^{-2} $\,s computed with parameters  $\tau = \theta_2 = \infty$ (left column) and $\tau = \theta_2 = 3.75\cdot10^{-4}$ (right column) for two-layered medium. Notations: black line - layer interface, white dotted line - receivers location,  
           1 = direct P-wave generated by the source, 
           2 = reflected P-wave, 
           3 =  reflected S-wave converted from the P-wave,
           4 =  P-to-S head wave,
           5 = transmitted fast P-wave, 
           6 = converted and transmitted S-wave,
           7 = converted and transmitted slow P-wave.}
          \label{fig:Layered_compare_snap}
\end{figure}

\begin{figure}
\centering
\begin{tabular}{cc}
  % \hline
   \pbox{8cm}{\vspace{0.0ex} \includegraphics[draft=false,width=8cm]{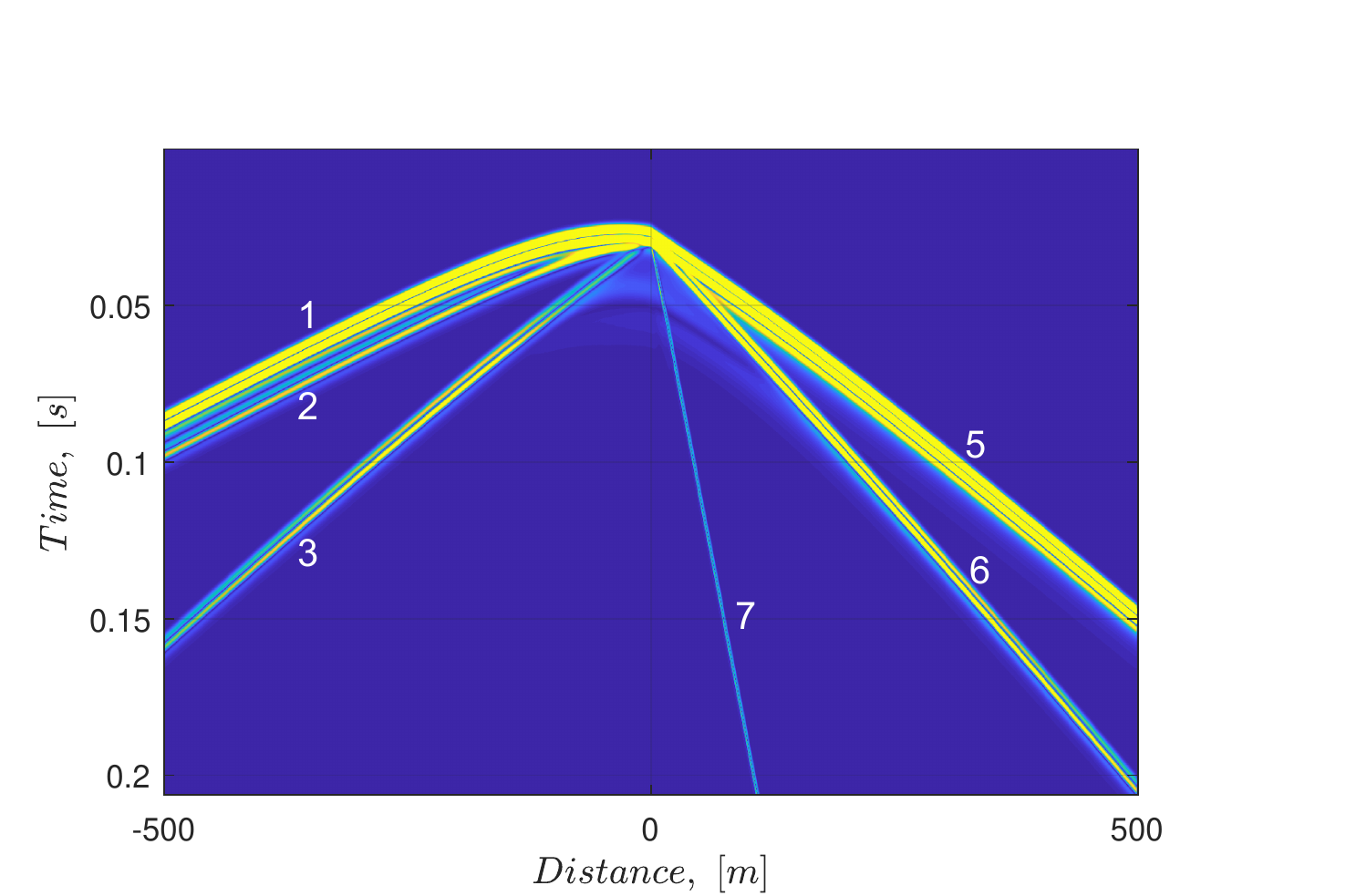}} &
   \pbox{8cm}{\vspace{0.0ex} \includegraphics[draft=false,width=8cm]{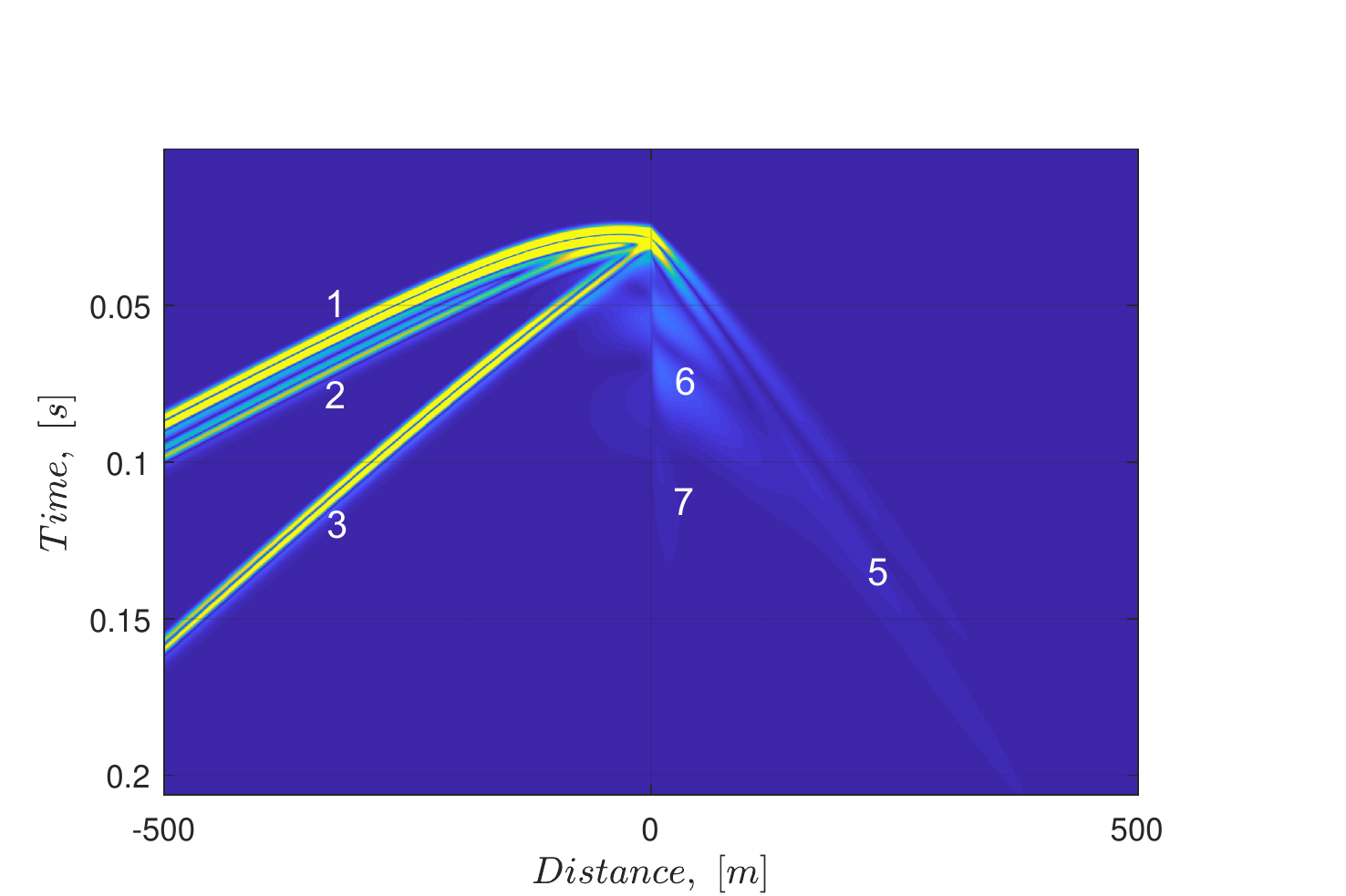}} \\
  {\pbox{9cm}{\vspace{2ex} $\tau = \theta_2 = \infty$            \vspace{1ex}}}     &       
  {\pbox{9cm}{\vspace{2ex} $\tau = \theta_2 = 3.75\cdot10^{-4}$  \vspace{1ex}}} \\
 % \hline
\end{tabular}
          \caption{ Seismograms of the total velocity vector for $\omega=10^{3}$  at  time $t = 
          8\cdot10^{-2} $\,s computed with parameters  $\tau = \theta_2 = \infty$ (left) and $\tau 
          = \theta_2 = 3.75\cdot10^{-4}$ (right) for two-layered medium. Notations:  
           1 = direct P-wave generated by the source, 
           2 = reflected P-wave, 
           3 =  reflected S-wave converted from the P-wave,
           5 = transmitted fast P-wave, 
           6 = converted and transmitted S-wave,
           7 = converted and transmitted slow P-wave.}
          \label{fig:Layered_compare_seismog}
\end{figure}

Let us suppose that the upper part of the computational domain $\Omega$ is a pure elastic medium 
($\phi =0$) with physical parameters Solid1 from Table \ref{tab.param.layered} whereas the lower 
part of $\Omega$ is a poroelastic medium  with the porosity $\phi=\alpha_1^0 = 0.2$ and material 
parameters Solid2, Viscous Fluid2 from Table \ref{tab.param.layered}. The size and numerical grid 
characteristics of $\Omega$  are the same as in the previous homogeneous case for $\omega=10^{3}$. 
The volumetric-type source with frequency $\omega=10^{3}$ $Hz$ is located in the upper half-plane 
at point $(x_0,y_0)=(0,-30)$ $m$ and it excites only one compressional  wave in an elastic medium. 
The choice of volumetric-type source is explained by the desire to avoid the complex wavefield 
mixture and to focus on the analysis of reflected and transmitted waves generated by only one 
incident P-wave.
For comparison we consider two cases of poroelastic layer with $\tau = \theta_2 = \infty$ and $\tau 
= \theta_2 = 3.75\cdot10^{-4}$. Fig.\,\ref{fig:Layered_compare_snap} and 
Fig.\,\ref{fig:Layered_compare_seismog} present the results of  numerical simulations. 
Fig.\,\ref{fig:Layered_compare_snap} shows wavefield snapshots of the normalized total velocity 
vector at  time $t = 8\cdot10^{-2} $\,s   and  Fig.\,\ref{fig:Layered_compare_seismog} shows 
seismograms for time $T=10^{-3}$\,$ s $ recorded in uniform spacing receivers located on the 
vertical line shifted $130$\,$ m $ right from the source position (dotted white line in 
Fig.\,\ref{fig:Layered_compare_snap}).  It clearly can be seen the wavefield difference for this 
two tests. Although the parameters of the upper elastic layer do not change, the wavefield in the 
upper layer changes due to reflection from the boundary of the poroelastic layer with different 
physical characteristics. If to take into account the shear stress and velocity relaxation times 
$\tau = 
\theta_2 = 3.75\cdot10^{-4}$, we can observe the occurrence of the disperse shear S wave with low 
velocity in the poroelastic layer. The amplitude of this wave  attenuates rather quickly, so we can 
observe it near the interface only, where the wave appears and then attenuates.
The shear wave can be seen more clearly if the amplitude of the wavefield is increased by two orders (see Fig.\,\ref{fig:Layered_compare_snap}, bottom row).
 \begin{table}[h]
	\begin{center}
	\begin{tabular}{lcccccc}
			\rowcolor[HTML]{EFEFEF} 
	% \hline %inserts one horizontal lines
			  State & $c_{\textrm{p},i}$, [m/s] & $ c_{\textrm{s},i} $, [m/s] & $ \rho_i $, 
			  [kg/m$ ^3$] & $ \eta $,[Pa$\cdot$s] & $\tau_i $, [s] & $ n $, [-]\\
			 % \hline %inserts one horizontal lines
		Solid1 &  6000 & 3500 & 2500 & -- & $ \tau_2 = \infty $ &  --\\
	\rowcolor[HTML]{EFEFEF} 
		Solid2 &  5000 & 2800 & 2500 & -- & $ \tau_2 = \infty $ &  --\\
		Solid3 &  3000 & 1750 & 2000 & -- & $ \tau_2 = \infty $ &  --\\
	\rowcolor[HTML]{EFEFEF} 
		Fluid1 & 1500 & 0 & 1040 & -- & $ \tau_2 = \infty $ &  --\\
		Viscous Fluid2 & 1500 & 100 & 1040 & $ 10^{-2} $ & $ \tau_1 = \eta/\mu_1 $ & 8\\
	
		\end{tabular}
	\caption{Material parameters used in the layered model 
	%\textcolor{red}{Check Fluid1 and Viscous Fluid2 in text } 
	.}
	\label{tab.param.layered}
	\end{center}
\vspace{-0.75cm}
\end{table}

 %____________
\subsection{Model validation and verification} 

Note that the formulated model \eqref{stress.velocity} can be used for numerical simulations of 
wave propagation for a quite wide range of problems. It can serve for wavefields simulation in a 
porous 
medium saturated by viscous and inviscid fluid for all range of porosity $\phi=\alpha_1^0\in[0,1]$. 
In this section, we consider test problems illustrating the applicability of the model for wavefield simulation in the medium containing regions of pure isotropic elastic solid $\phi=0$ and pure fluid 
$\phi=1$ with the single PDE system \eqref{stress.velocity}.
We compare numerical results obtained by solving equations \eqref{stress.velocity} with those obtained by the classical approach based on the stress-velocity elasticity model with interface tracking between elastic medium and fluid, realized by the boundary conditions treatment in the numerical method. Numerical examples presented in this section include wavefields simulation in a finely layered medium composed of elastic isotropic solid/solid or solid/fluid alternate layers.

We study a periodic, two-layered system inside a square computational domain $[500m]\times[500m]$ 
centered at the origin with $5000^2$ grid mesh and uniform  grid spacing $0.1 m$. Layers of the 
thickness $0.5 m$ are settled horizontally. For linear elasticity model each layer is defined by a 
prescribed boundaries and the medium inside  is defined by the density  $ \rho_i (i=1,2) $, the 
compressional  velocity $c_{\textrm{p},i} (i=1,2)$ and the shear velocity $c_{\textrm{s},i} 
(i=1,2)$. For two-phase model \eqref{stress.velocity} the interface between layers is defined via 
the porosity function $\phi$. In the absence of the velosity and shear stress relaxation  ($\tau = 
\theta_2 = \infty$), $\phi = 0$ corresponds to the pure elastic medium and $\phi = 1$ corresponds 
to the pure fluid. 

We consider two cases of finely layered medium with parameters Solid1/Fluid1 and Solid1/Solid3 from Table \ref{tab.param.layered}. 

\begin{figure}
\centering
\begin{tabular}{cc}
  % \hline
   \pbox{8cm}{\vspace{0.0ex} \includegraphics[draft=false,width=9 cm]{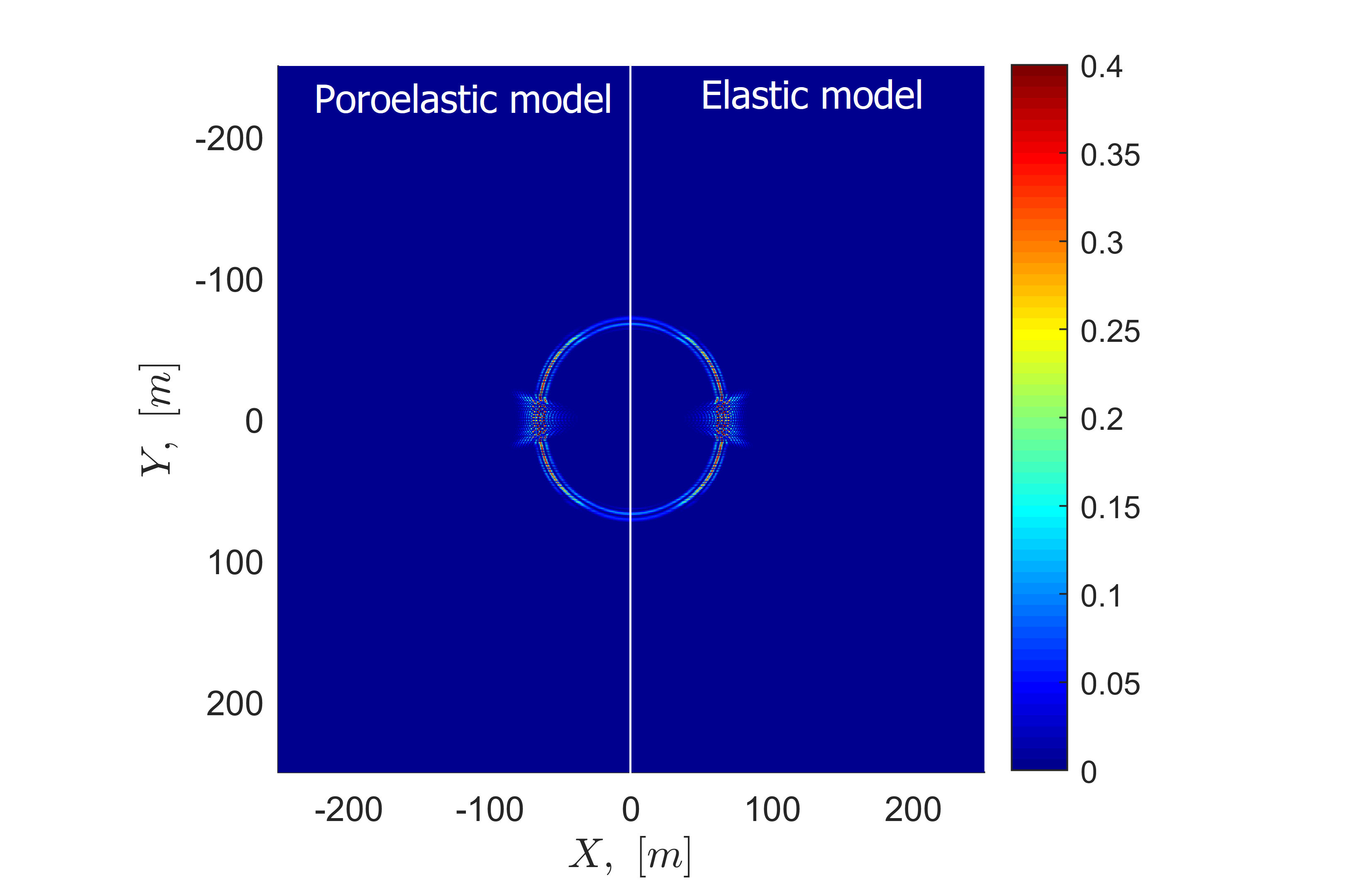}} &
   \pbox{8cm}{\vspace{0.0ex} \includegraphics[draft=false,width=9 cm]{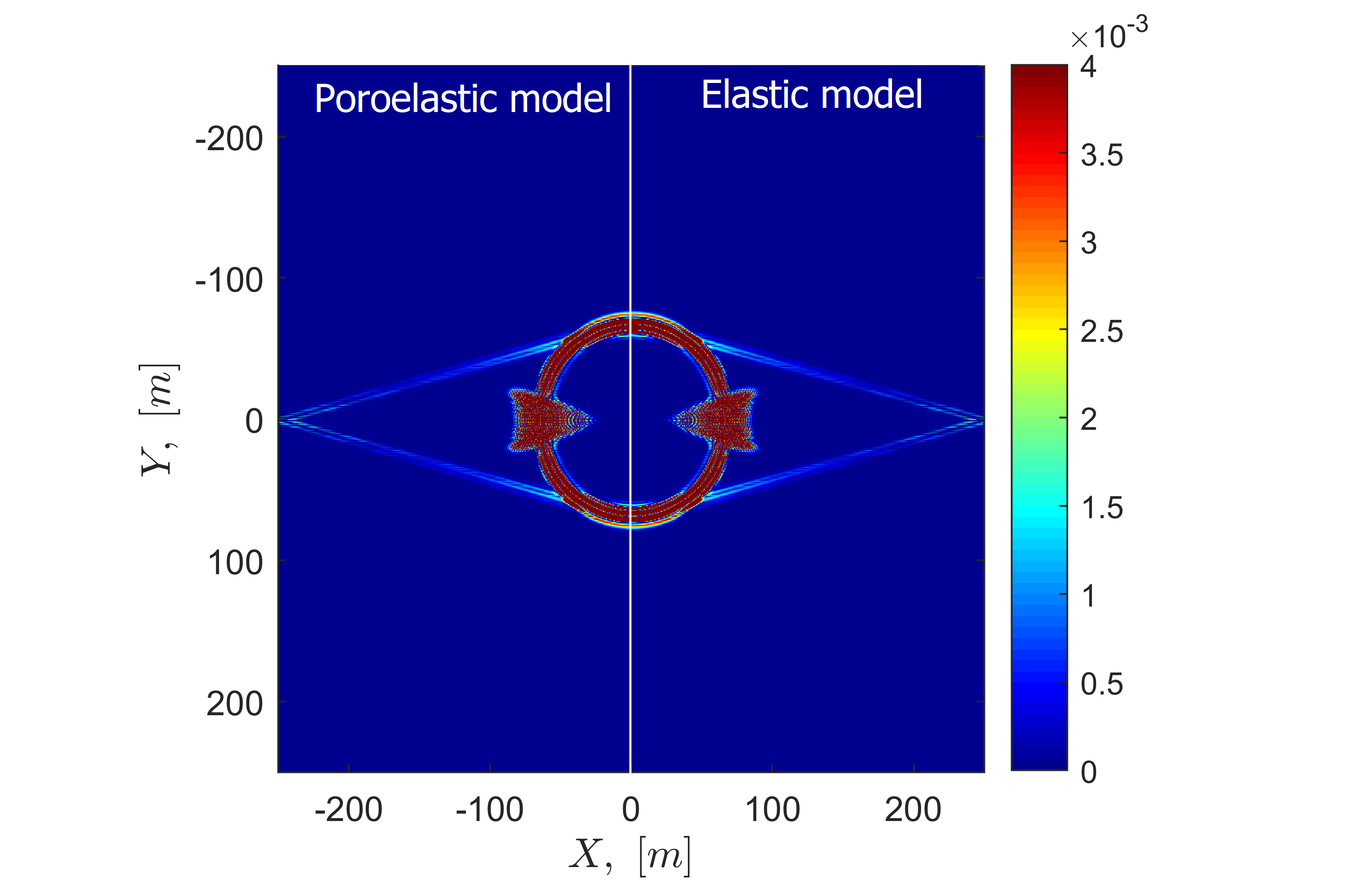}} \\
  {\pbox{9cm}{a   \vspace{1ex}}}     &       
  {\pbox{9cm}{b  \vspace{1ex}}} \\  
 % \hline
\end{tabular}
          \caption{ (a) Two-piece snapshot of the total velocity vector computed by model 
          \eqref{stress.velocity} (left) and by stress-velocity model of the dynamic linear 
          elasticity (rigth) for the finely layered model with Solid1/Fluid1 parameters from Table 
          \ref{tab.param.layered} at time $t =6.2\cdot10^{-2} $\,$ s $ and $\omega=10^{3}$.
          (b) Amplitude magnification of snapshot (a).}
          \label{fig:Verification_water_solid}
\end{figure}

Fig.\,\ref{fig:Verification_water_solid} shows snapshot of the total velocity vector computed  for 
the Solid1/Fluid1 layered medium at time $t =6.2\cdot10^{-2} $\,$ s $. The volumetric-type source 
with dominant frequency $f_0=\omega/2\pi=10^{3}/2\pi$ $Hz$ is located in a fluid layer at the 
center of the domain. For selected frequency and medium parameters, the thickness of the layers 
corresponds to $1/75$ of the dominant P-wavelength in the solid. 
To illustrate the agreement between the solutions obtained by the linear elasticity model and the two-phase model \eqref{stress.velocity}, Fig.\,\ref{fig:Verification_water_solid} is split to show both numerical results. On the left side, the snapshot obtained by the two-phase model \eqref{stress.velocity} is depicted, whereas on the right side the snapshot obtained by the classical stress-velocity formulation of the linear elasticity equations is shown.  It can be seen that they are practically identical, which is confirmed by semblance value 99,9\% computed by the formula
\begin{equation}
S=\frac{\sum_i(a_i+b_i)^2}{2\sum_i(a_i^2+b_i^2)}\cdot100\%
\end{equation}
where $a_i$ and $b_i$ are the wavefield components changing with time in both simulations. 
Good agreement of the results is achieved by a practically identical finite-difference schemes approximating the selected model for both calculations. 

\begin{figure}
\centering
\begin{tabular}{cc}
  % \hline
   \pbox{8cm}{\vspace{0.0ex} \includegraphics[draft=false,width=7cm]{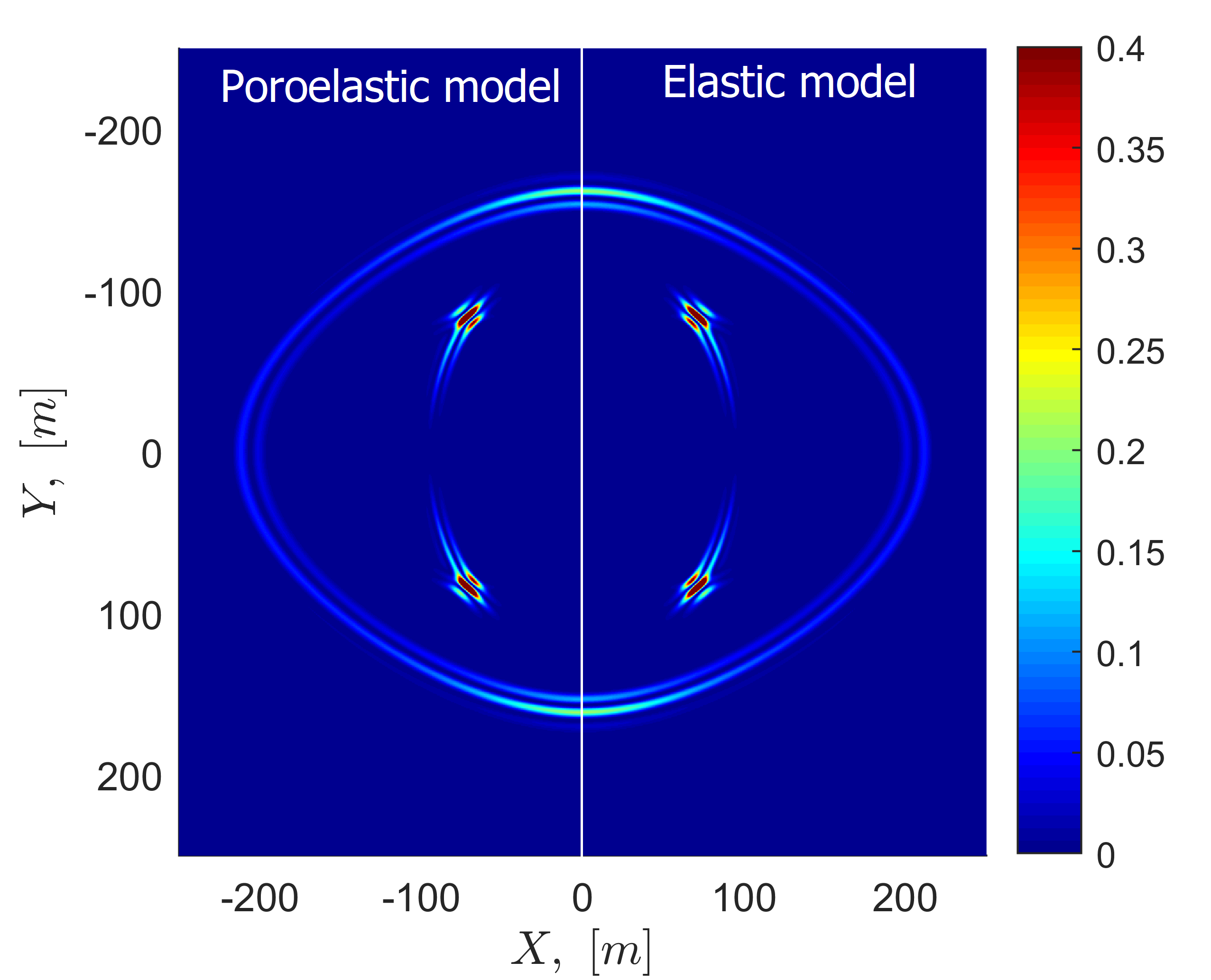}} &
   \pbox{8cm}{\vspace{0.0ex} \includegraphics[draft=false,width=9.2cm]{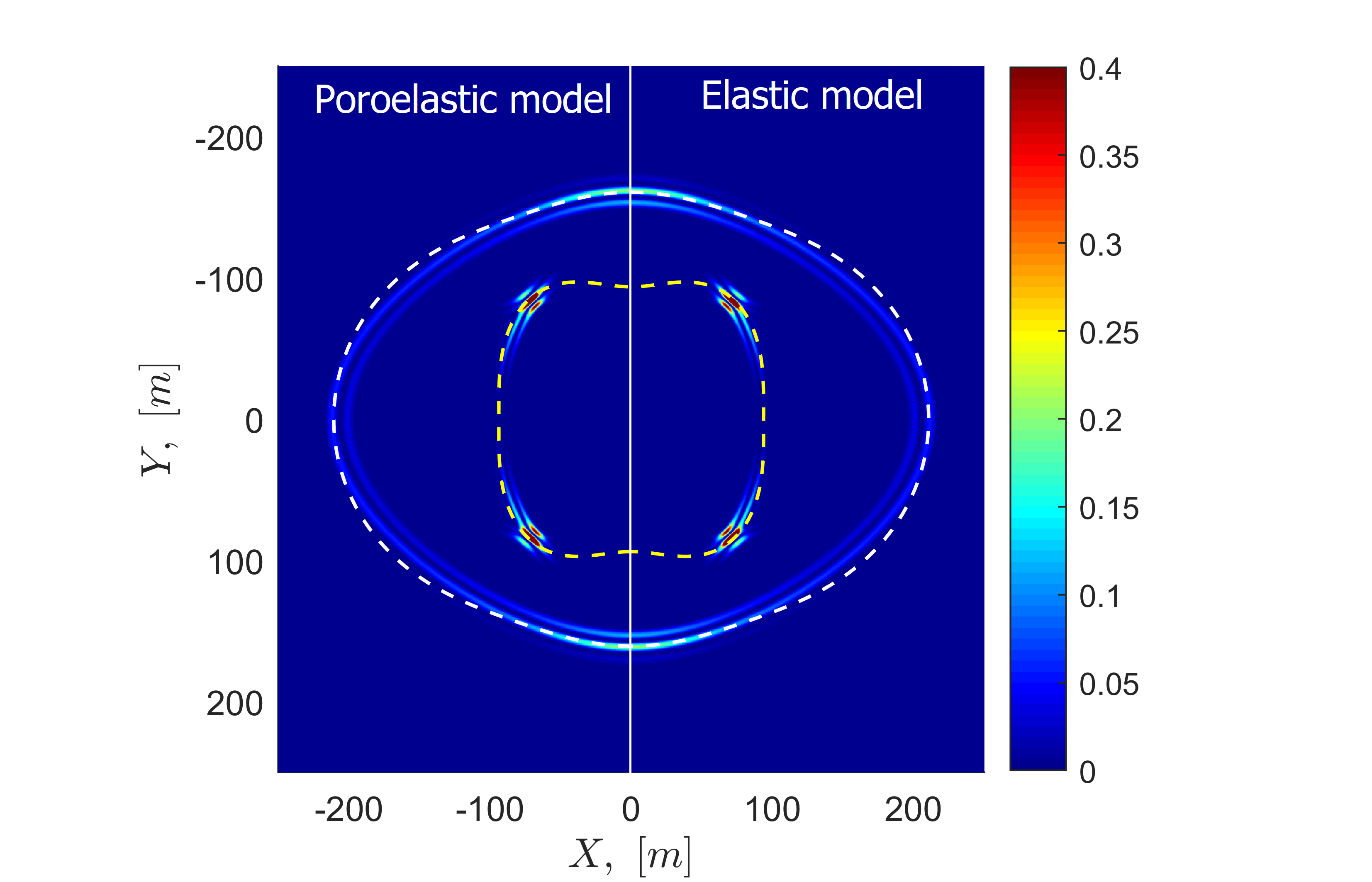}} \\
  {\pbox{9cm}{a   \vspace{1ex}}}     &       
  {\pbox{9cm}{b  \vspace{1ex}}} \\  
 % \hline
\end{tabular}
          \caption{ (a) Two-piece snapshot of the total velocity vector computed by model 
          \eqref{stress.velocity} (left) and by stress-velocity model of the dynamic linear 
          elasticity (rigth) for the finely layered model with Solid1/Solid3 parameters from Table 
          \ref{tab.param.layered} at time $t =6.2\cdot10^{-2} $\,$ s $ and $\omega=10^{3}$.
          (b) The position of the peak amplitude of quasi-P (white dashed curve) and quasi-S 
          (yellow dashed curve) wavefronts obtained by Backus averaging procedure at time $t 
          =5\cdot10^{-2} $\,$ s $.}
          \label{fig:Verification_solid_solid}
\end{figure}

The similar excellent agreement between solutions obtained by the linear elasticity model with 
interface tracking and the two-phase model can be obtained if to consider the case of elastic 
finely layered medium with layers parameters Solid1/Solid3. In this case for wavefields simulation 
we just replace Fluid1 layers in the previous consideration with the Solid3 layers. 
Fig.\,\ref{fig:Verification_solid_solid} (a) shows a two-piece snapshot of the total velocity 
vector computed by the model \eqref{stress.velocity} and linear elasticity equations for the 
layered medium Solid1/Solid3  for the source frequency $\omega=10^{3}$ at time $t =6.2\cdot10^{-2} 
$\,$ s $. 

A significant difference of wavefield in Fig.\,\ref{fig:Verification_water_solid}  and 
Fig.\,\ref{fig:Verification_solid_solid} can be explained by the homogenization theory for a finely 
layered medium composed of periodic elastic layers. The stack of the horizontally parallel thin 
isotropic layers acts as a VTI (transverse isotropy with a vertical axis of symmetry) anisotropic 
medium in the long-wavelength limit. We use Backus averaging procedure \cite{Backus1962} to 
estimate the so-called quasi-P wave and quasi-S wave phase velocities in order to compare with our 
numerical results. To this end,  the position of the peak amplitude of quasi-P (white dashed curve) 
and quasi-S (yellow dashed curve) wavefronts is imposed on the wavefield snapshot in 
Fig.\,\ref{fig:Verification_solid_solid}(right).   We remind that in the numerical modeling, the 
source wavelet is shifted in time by $2/f_0=1.2\cdot10^{-2} $\,$ s $, that is why the phase 
velocity curves are calculated at $t =5\cdot10^{-2} $\,s after the source excitation. We can 
observe that these curves are in good consistency with the finite-difference results, thus 
demonstrating the correctness and accuracy of our model.

\begin{figure}
\centering
\begin{tabular}{cc}
  % \hline
   \pbox{8cm}{\vspace{0.0ex} \includegraphics[draft=false,width=10cm]{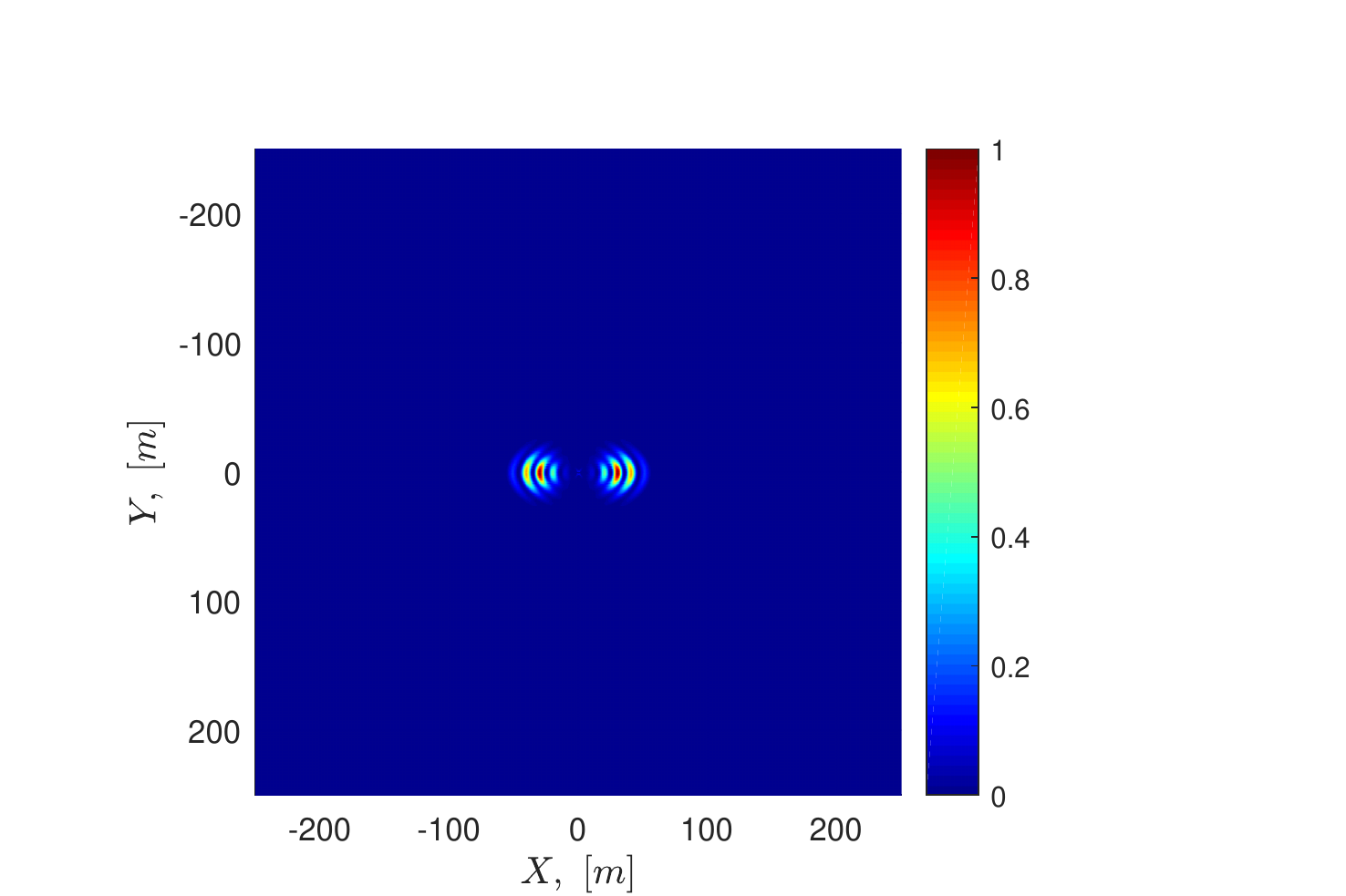}} &
   \pbox{8cm}{\vspace{0.0ex} \includegraphics[draft=false,width=10cm]{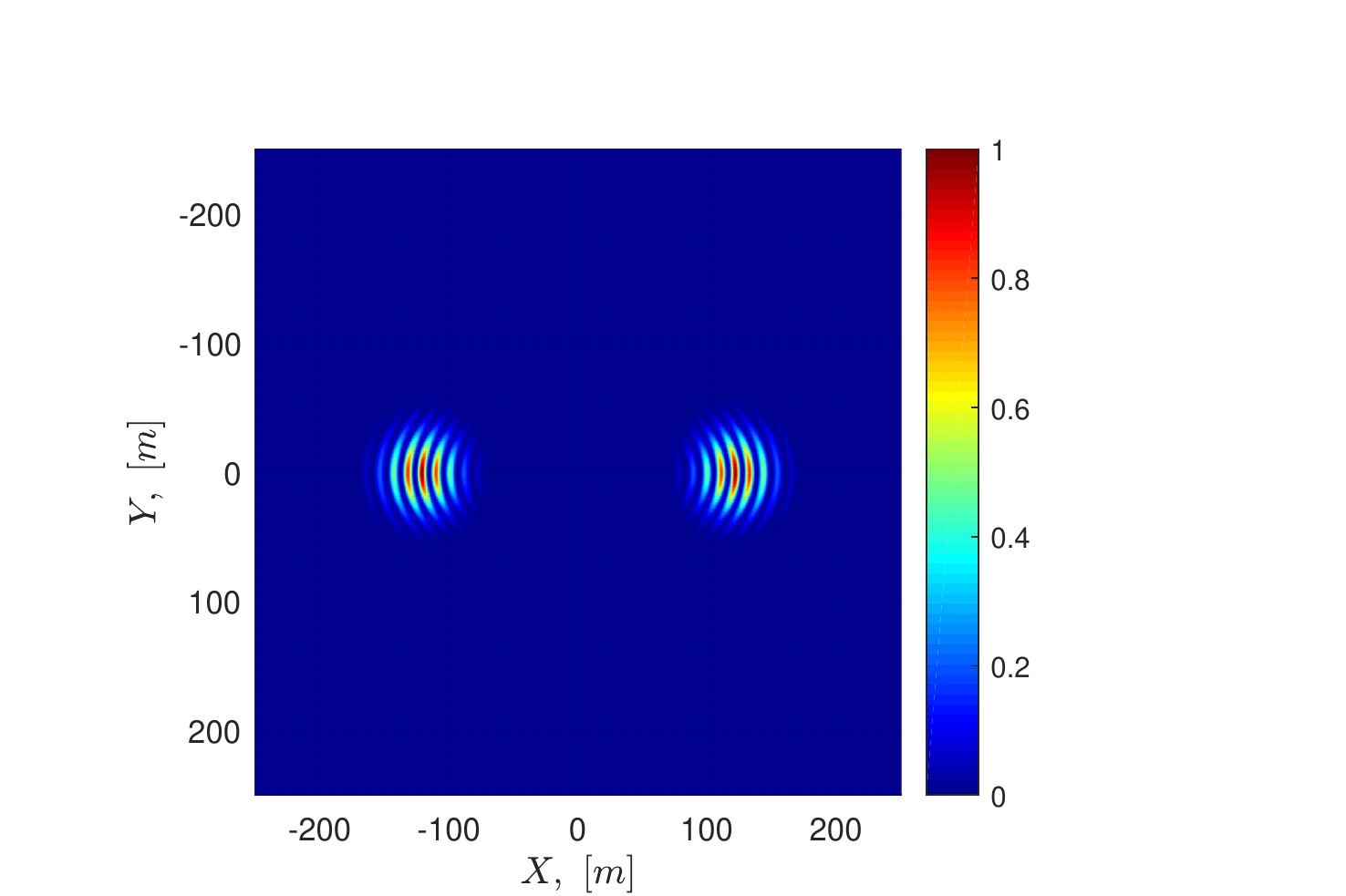}} \\
  {\pbox{9cm}{a   \vspace{1ex}}}     &       
  {\pbox{9cm}{b  \vspace{1ex}}} \\  
 % \hline
\end{tabular}
          \caption{ Snapshot of the normalized total velocity vector computed for a periodic 
          finely two-layered model consisting of alternating elastic layer Solid1 and poroelastic 
          layer Solid3/Viscous Fluid2 with porosity $\phi = 0.2$ at time $t=1.5\cdot10^{-2} $\,s 
          (a) and $t =6.2\cdot10^{-2} $\,s (b).}
          \label{fig:Verification_porous_solid}
\end{figure}

At the end of this section, we present one more calculation for a finely layered medium containing 
poroelastic layers with a viscous fluid in order to investigate the influence of the fluid 
viscosity on the wavefield behavior. For comparison purposes, we slightly modify the previous 
experiment with elastic layers  Solid1/Solid3, and substitute the pure elastic layer Solid3 with a 
poroelastic layer Solid3/Viscous Fluid2 from Table \ref{tab.param.layered} with porosity $\phi = 
0.2$. All other parameters are the same. Fig.\,\ref{fig:Verification_porous_solid} shows snapshot 
of the normalized total velocity vector computed  for this medium at the times $t =1.5\cdot10^{-2} 
$\,$ s $ and $t =6.2\cdot10^{-2} $\,$ s $. Comparing Fig.\,\ref{fig:Verification_solid_solid} and  
Fig.\,\ref{fig:Verification_porous_solid} demonstrates a significant change in wavefield behaviour 
produced by adding porosity and viscosity into the model. We observe strong seismic attenuation of 
all waves in vertical direction caused by the combined effect of multiple inter-layer 
reflection/refraction scattering and viscous mechanism. Also, note that wavefield front doesn't 
gather in the horizontal direction.

%____________
\subsection{Strongly heterogeneous medium} 

\begin{figure}[htbp]
\centering
          \includegraphics[draft=false,width=10cm]{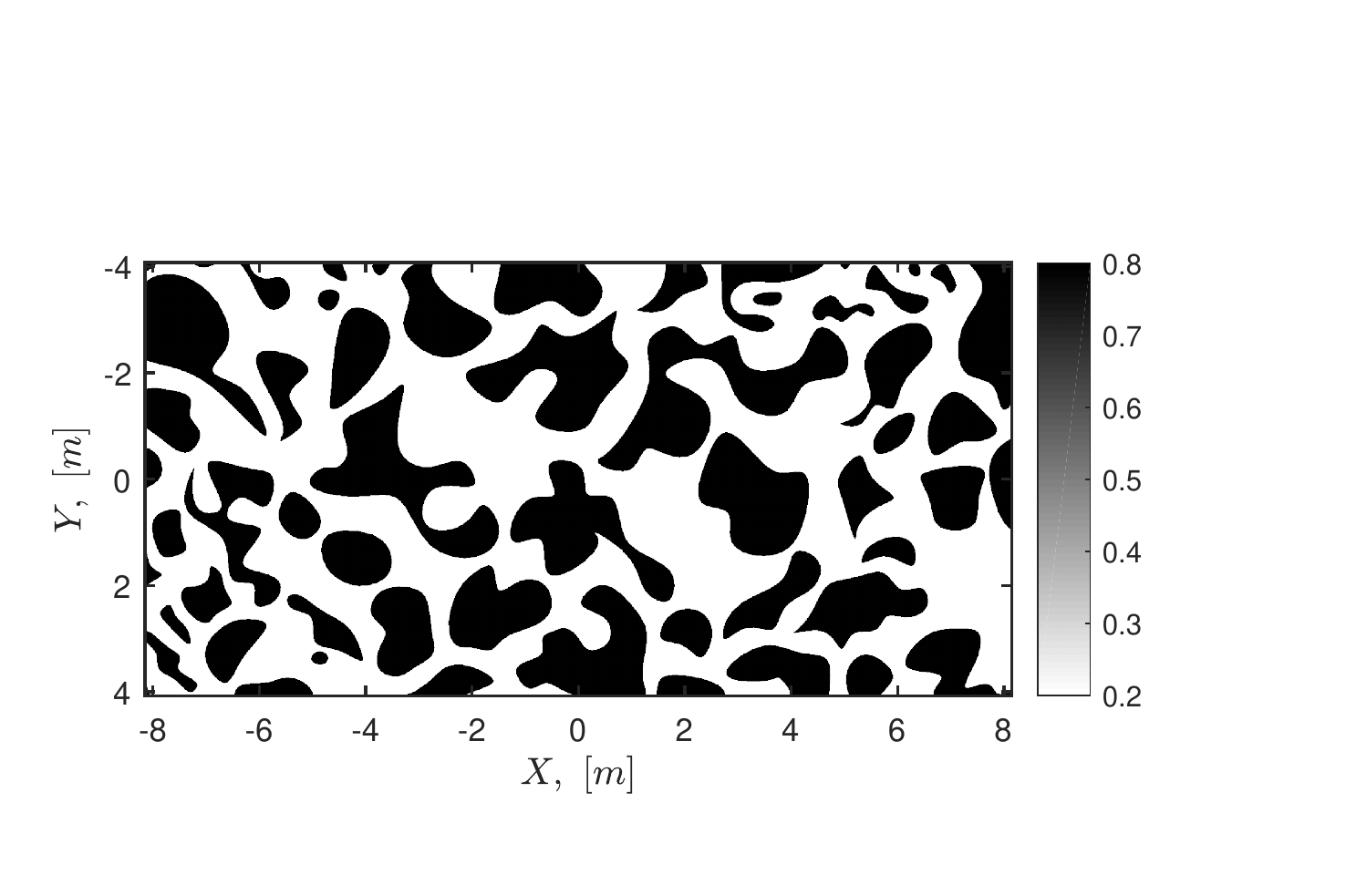}
          \caption{  Comsol model. Geometry of fluid channels (white). }
          \label{fig:Test4_geometry_fluid}
\end{figure}

This experiment is carried out for the porous media model downloaded from Comsol Application Gallery, Pore-Scale Flow, produced by scanning the electron microscope images during the pore-scale flow experiments and schematically presented in Figure \ref{fig:Test4_geometry_fluid}. We use this geometry to define porous medium with the material parameters from Table \ref{tab.param} and porosity $\phi = 0.8$, containing less %more 
permeable porous channels with porosity $\phi = 0.2$ ( white color in Fig. \ref{fig:Test4_geometry_fluid}). For numerical simulations the physical domain $X = 16.3$ $m$, $Y = 8.15$ $m$ centering at the origin of the Cartesian coordinate system is discredited by $N_x = 1630$, $N_y = 815$ grid points with a spatial sampling step $dx = dy = 0.01$ $m$. The wavefield is excited by a vertically-type source located in the permeable channel at the position $(x,y) = (-1.2; -0.5)$ $m$.

\begin{figure} [hbt!]
\centering
\begin{tabular}{cc}
  % \hline
   \pbox{9cm}{\vspace{0.0ex} \includegraphics[draft=false,width=8cm]{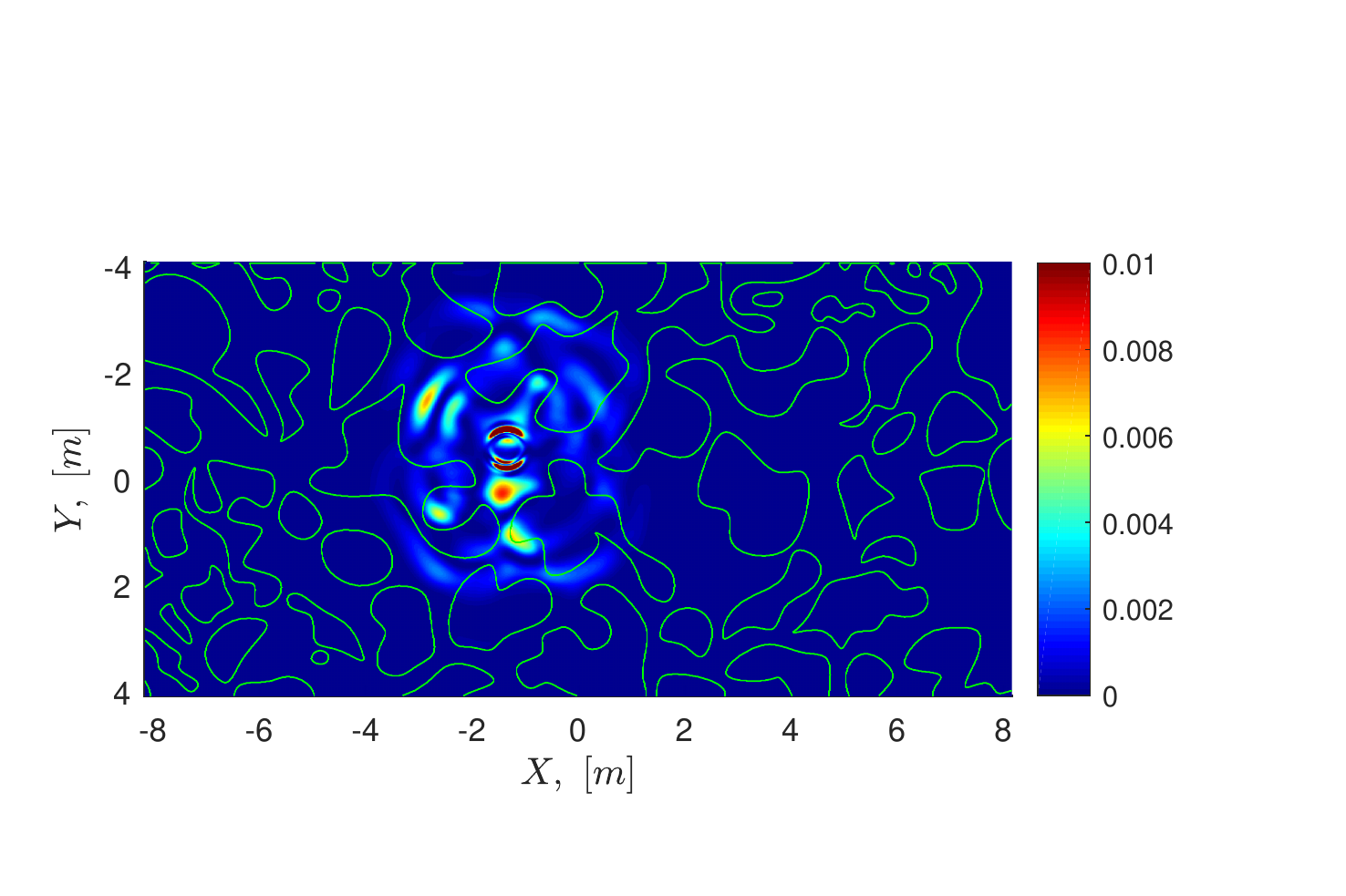}} &
   \pbox{9cm}{\vspace{0.0ex} \includegraphics[draft=false,width=8cm]{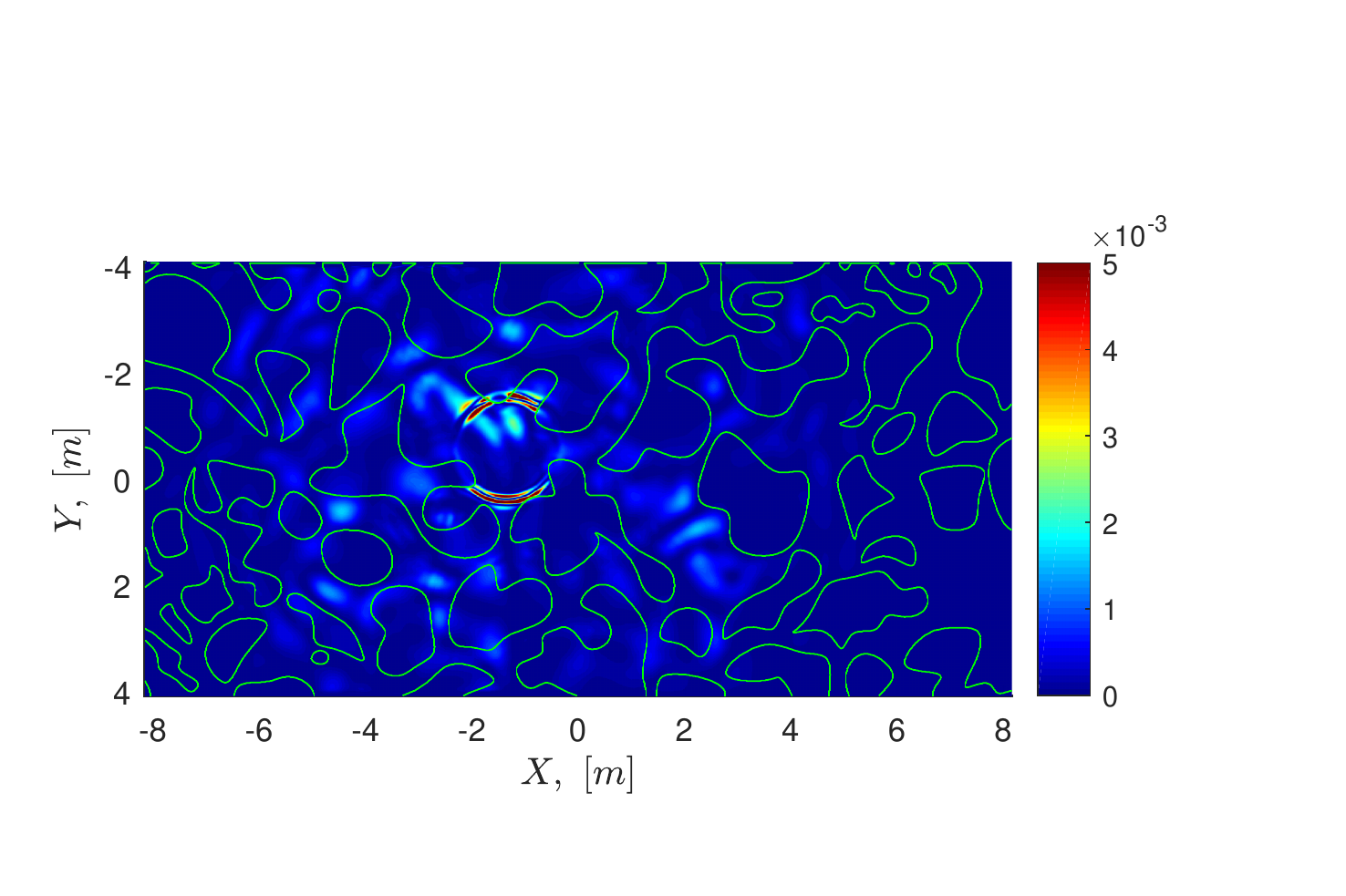}} \\
  {\pbox{9cm}{\vspace{1ex} $\tau = \theta_2 = \infty,\quad  t=1.2\cdot10^{-3}$             \vspace{1ex}}}     &       
  {\pbox{9cm}{\vspace{1ex} $\tau = \theta_2 = \infty,\quad  t=2.4\cdot10^{-3}$  \vspace{1ex}}} \\
  &  \\
    % \hline
   \pbox{9cm}{\vspace{0.0ex} \includegraphics[draft=false,width=8cm]{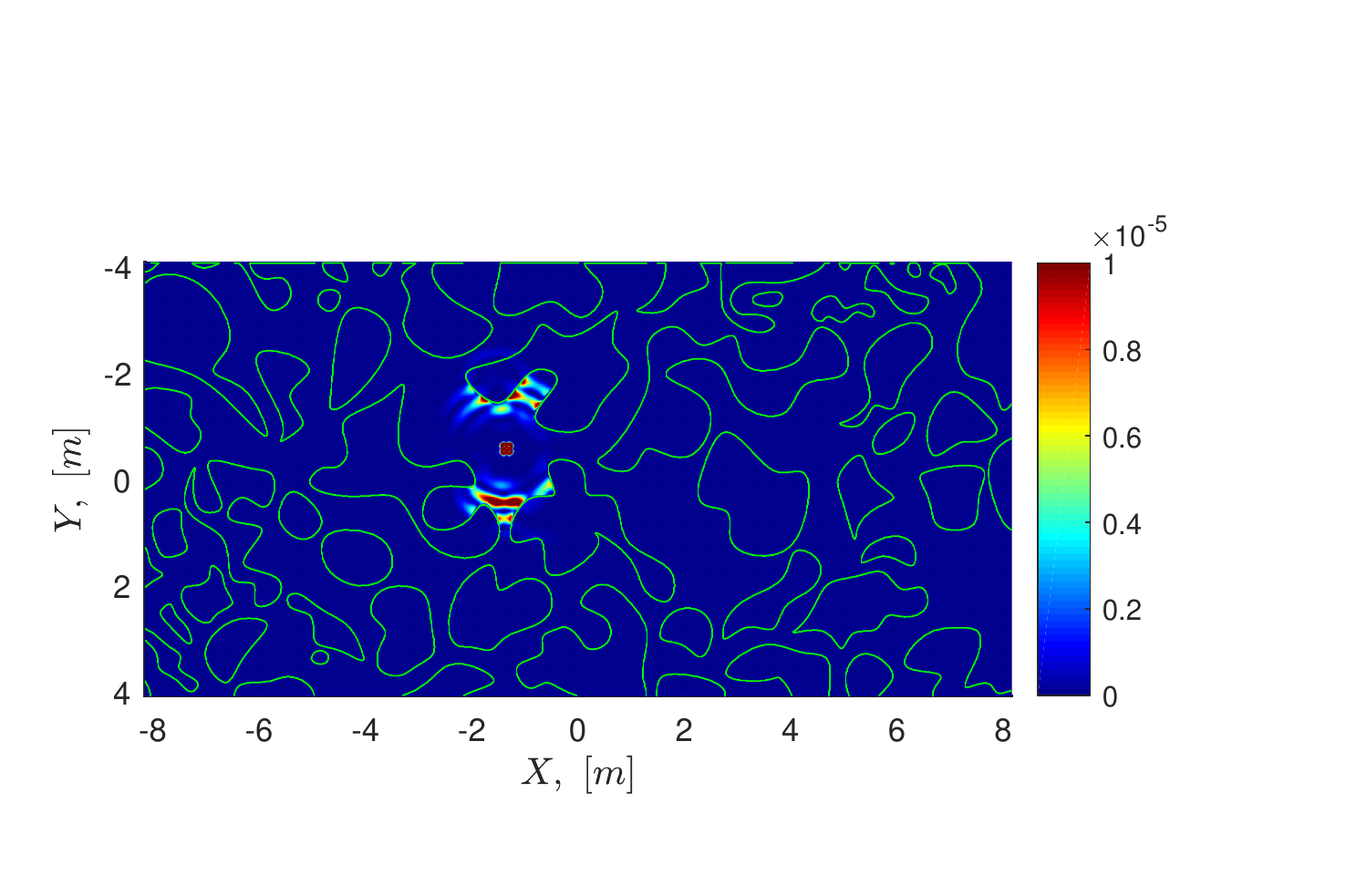}} &
   \pbox{9cm}{\vspace{0.0ex} \includegraphics[draft=false,width=8cm]{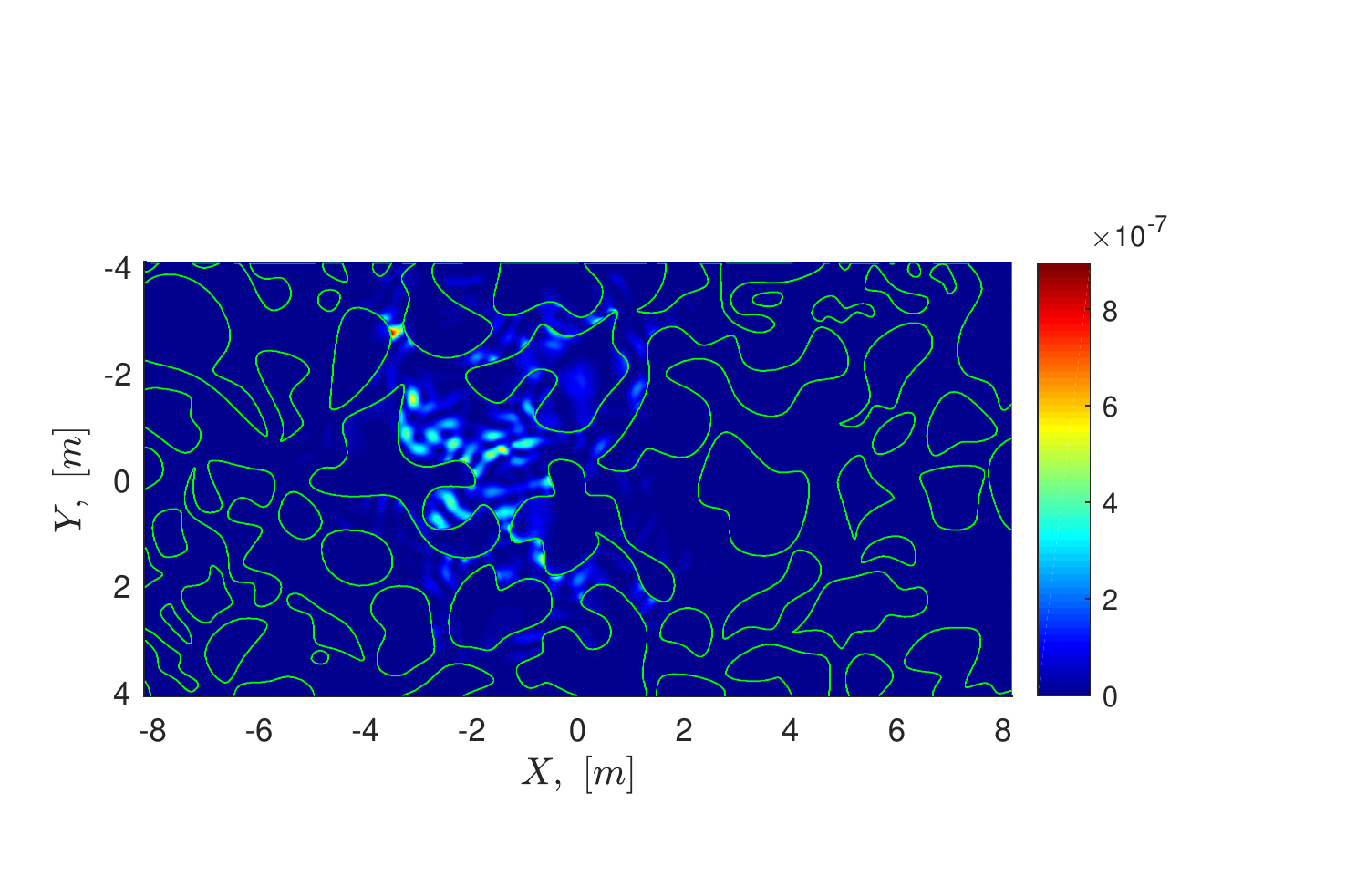}} \\
  {\pbox{9cm}{\vspace{1ex} $\tau = \theta_2 = 3.75\cdot10^{-4},\quad t=1.2\cdot10^{-3}$           \vspace{1ex}}}     &       
  {\pbox{9cm}{\vspace{1ex} $\tau = \theta_2 = 3.75\cdot10^{-4},\quad  t=2.4\cdot10^{-3}$   \vspace{1ex}}} \\
 % \hline
\end{tabular}
          \caption{ Snapshot of the total velocity vector for $\omega=10^{4}$  at  times $t = 1.2\cdot10^{-3} $\,s and $t = 2.4\cdot10^{-3} $\,s computed with parameters  $\tau = \theta_2 = \infty$ (upper line) and $\tau = \theta_2 = 3.75\cdot10^{-4}$ (down line) for Comsol model. }
          \label{fig:Comsol_compare_snap}
\end{figure}

For numerical simulations let us consider two cases of  relaxation mechanisms with $\tau = \theta_2 
= \infty$ and $\tau = \theta_2 = 3.75\cdot10^{-4}$. Fig.\,\ref{fig:Comsol_compare_snap} show 
wavefield snapshots of the total velocity vector computed for $\omega=10^{4}$  at  times $t = 
1.2\cdot10^{-3} $\,$ s $ and $t = 2.4\cdot10^{-3} $\,$ s $.
 At time $t = 1.2\cdot10^{-3} $\,$ s $, we see the propagation of the P wave and the initiation of 
 the Bio wave for both cases. But in the case of $\tau = \theta_2 = 3.75\cdot10^{-4}$, the 
 wavefield's amplitude is strongly attenuates from the beginning and makes it impossible to form 
 the Biot mode ( see time $t = 2.4\cdot10^{-3} $\,$ s $ ) if the relative velocity and shear stress 
 relaxations are taken into account.

\section{Conclusions} 

We have presented a new {\em hyperbolic} two-phase model of a porous medium saturated by a viscous 
fluid. 
The governing equations of the model are derived by generalizing the unified model of continuum \cite{HPR2016,DPRZ2016}, which in turn is derived using the theory of Symmetric Hyperbolic Thermodynamically Compatible (SHTC) systems \cite{GodRom2003,Rom1998,Peshkov2018}. 
The unified model of continuum is based on the first principles, mathematically well-posed, satisfies thermodynamics laws and describes heat-conducting solid and liquid states of a medium by a single hyperbolic PDE system.
Consideration of the saturated porous medium as a two-phase solid-fluid mixture and coupling of the unified model with the two-phase SHTC model of compressible fluid \cite{RomDrikToro2010} allows us to design the SHTC nonlinear model for compressible fluid flow in a deformed porous medium. 
The presented model, similarly to the unified model, can take into account the viscosity of a saturating fluid in a hyperbolic formulation.
Thus, in addition to interfacial friction, there is a dissipation in the model caused by the viscosity of the saturating fluid which should be taken into account in the interpretation of seismic data. 
Since there are no available experimental or well-established theoretical data on the properties of 
shear waves in the literature, we choose some empirical functions of state parameters for closing 
relations such as the equation of state (specific internal energy), interfacial friction 
coefficient and shear stress relaxation function.

Using the presented nonlinear finite-strain SHTC model, the governing equations for the propagation 
of small-amplitude waves in a porous medium saturated with a viscous fluid are derived. 
As in the well-known theories of porous media, three types of waves can be found: fast and slow compression waves and shear wave.
It turns out that the shear wave attenuates rapidly due to the viscosity of the saturating fluid, and this wave is difficult to see in typical test cases.
However, some test cases are presented in which shear waves can be observed in the vicinity of interfaces between regions with different porosity.

The presented model can be validated by comparison with experimental data on the propagation of waves in porous media and, in particular, on the propagation of shear waves, or by modeling flows in an elastic skeleton on a pore scale, which is one of the goals of our forthcoming studies.

%=============================================================================
%==========    A C K N O W L E D G M E N T S
\section*{Acknowledgments}
The work of E.R. and G.R. is supported by Mathematical Center in Akademgorodok, the agreement with Ministry of Science and
High Education of the Russian Federation number 075-15-2019-1613.
I.P. gratefully acknowledge funding from the Italian Ministry of Education, 
University and Research (MIUR) under the Departments of Excellence Initiative 2018--2022 attributed 
to DICAM of the University of Trento (grant L. 232/2016), as well as financial support from the 
University of Trento 
under the  \textit{Strategic Initiative Modeling and Simulation}. I.P. has further received funding 
from the University of Trento via the \textit{UniTN Starting Grant initiative}.

\appendix
\section{Asymptotic analysis for small shear stress relaxation time}\label{app.asympt}
Consider governing equations for a pure fluid presented in Section 3.
\begin{subequations}\label{APeqn.IsentropicFluid}
	\begin{eqnarray}
&&\displaystyle\frac{\partial 
\rho_1v^i_1}{\partial t}+
\frac{\partial(\rho_1v^i_1v^k_1 
+p_1\delta_{ik}-\sigma_{ik})}{\partial x_k}=0, 
\label{APeqn.momentumPVsFluid}\\[2mm]
&&\displaystyle\frac{\partial A_{i k}}{\partial t}+\frac{\partial A_{ij} 
	v^j_1}{\partial x_k}+v^j_1\left(\frac{\partial A_{ik}}{\partial 
	x_j}-\frac{\partial A_{ij}}{\partial x_k}\right)
=-\dfrac{ \psi_{ik} }{\theta},\label{APeqn.deformationPVsFluid}\\[2mm]
&& \frac{\partial \rho_1}{\partial t}+\frac{\partial \rho_1 
v^k_1}{\partial x_k}=0,\label{APeqn.contiPV2Fluid}
\end{eqnarray}
\end{subequations}
where $p_1=\rho_1^2\partial e_1/\partial \rho_1$ is the pressure of fluid, $\sigma_{ij} = 
-\frac{\rho_1 \cs{1}^2}{2}\left({g_{ik}g_{kj}-
	\frac{1}{3}{g_{mn}g_{nm}}\delta_{ij}}\right)$ is 
the the shear stress and the source term in \eqref{APeqn.deformationPVsFluid} contains 
$\Psi=[\psi_{ik}]=
\frac{\cs{1}^2}{2} \AA^{-T} \left(\g^2-\frac{\tr({\g^2})}{3} \II \right)$,
where $\g=|\GG|^{-1/3} \GG$, $|\GG|=\det\GG$.

Here
\begin{equation} \label{Theta}
    \theta= \theta_0 \tau, \quad \theta_0= \frac{2 \cs{1}^2}{\rho_1 |\GG|^{1/3}}.
\end{equation}
The starting point of the asymptotic analysis for small shear stress relaxation time 
$\tau$ is the equation for distortion in the form equivalent to \eqref{APeqn.deformationPVsFluid}
\begin{equation} \label{A_eqn}
\displaystyle\frac{\partial A_{i k}}{\partial t}+
v^j_1 \frac{\partial A_{ik}}{\partial x_j}
+A_{ij}\frac{\partial v^j_1}{\partial x_k}
=-\dfrac{ \psi_{ik} }{\theta}. 
\end{equation}
Since the stress tensor $\sigma$ is a function of metric (or Finger) tensor $\GG=\AA^T\AA$, 
one should use the equation for $\GG$ for asymptotic expansion, which is a direct consequence of 
\eqref{A_eqn} and in matrix form reads as
\begin{equation} \label{G_eqn}
\displaystyle\frac{\partial \GG}{\partial t}+
v^j_1 \frac{\partial \GG}{\partial x_j}
+\GG (\nabla {\mathbf v}_1) + (\nabla {\mathbf v}_1)^T \GG
=-\dfrac{ 1 }{\tau}\dfrac{ \cs{1}^2 }{\theta_0}\left(\g^2-\dfrac{\tr(\g^2)}{3} I \right), 
\end{equation}
where $\nabla {\mathbf v}_1$ is the velocity gradient matrix.

Note that in terms of metric tensor $\GG$, the stress tensor reads as 
\begin{equation} \label{Stress}
\bm{\sigma} = -\frac{\rho_1 \cs{1}^2}{2}(\text{det}\GG)^{-2/3}
\left(\GG^2-\frac{1}{3} \tr (\GG^2) \II\right)
\end{equation}

In order to define a small parameter, we introduce the time scale $t_0$ and denote $t^\prime=t/t_0$. After implementing the above scaling in \eqref{G_eqn}, we arrive at the following equation:
\begin{equation} \label{G_eqn_scale}
\displaystyle\frac{\partial \GG}{\partial t^\prime}+
t_0 v^j_1 \frac{\partial \GG}{\partial x_j}
+t_0\left(\GG (\nabla {\mathbf v}_1) + (\nabla {\mathbf v}_1)^T \GG \right)
=-\dfrac{ t_0 }{\tau}\dfrac{ \cs{1}^2 }{\theta_0}(\text{det}\GG)^{-2/3}
\left(\GG^2-\dfrac{\tr(\GG^2)}{3} \II \right), 
\end{equation}
Now assuming $t_0 \gg \tau$, we can introduce the small parameter $\varepsilon=\tau/t_0$.

Our goal is to construct a solution in the form of asymptotic expansion with respect to small 
parameter $\varepsilon$. Note that stress tensor in \eqref{Stress} depends on $\GG^2$, that is why 
we are looking for the solution in the form
\begin{equation} \label{Gexpansion}
\GG^2=\GG^2_0+\varepsilon \GG^2_1+... .
\end{equation}
From \eqref{G_eqn_scale} one can see that the zero term $\GG^2_0$ should satisfy the uniform 
volumetric deformation, i.e.
$\GG^2_0-\frac{1}{3} \tr (\GG^2_0) \II=\GG^2_0-{\Lambda} \II=0$,
where ${\Lambda}=(\det\GG_0)^{2/3}$ (the latter follows from the chain of equalities
$\det \GG^2_0=\det (\GG_0)^2=\Lambda^3$). Note that the density $\rho_1$ is a product of the 
reference density $\rho_{10}$ and square root of $\det \GG$: 
$\rho_1=\rho_{10}\sqrt{\det \GG}$, but the two first terms of its expansion (zero and first ones) 
has no influence on the final result and we use the notation $\rho_1$ everywhere below.

Now, with the use of equation \eqref{G_eqn_scale} we can derive equation for $\GG^2$:
\begin{equation} \label{G_eqn_scale}
\displaystyle\frac{\partial \GG^2}{\partial t^\prime}+
t_0 v^j_1 \frac{\partial \GG^2}{\partial x_j}
+t_0\left(\GG^2 (\nabla {\mathbf v}_1) 
+\GG\left(\nabla {\mathbf v}_1 + \nabla {\mathbf v}_1^T\right)\GG+
(\nabla {\mathbf v}_1)^T \GG^2 \right)
=-\dfrac{ 1 }{\varepsilon}\dfrac{2 \cs{1}^2 }{\theta_0}(\text{det}\GG)^{-2/3}
\GG \left(\GG^2-\dfrac{\tr(\GG^2)}{3} \II \right), 
\end{equation}

Substituting \eqref{Gexpansion} into \eqref{G_eqn_scale}, we obtain for the first order terms in $\varepsilon$
\begin{equation} \label{G_eqn_scale1}
\displaystyle\frac{\partial \GG^2_0}{\partial t^\prime}+
t_0 v^j_1 \frac{\partial \GG^2_0}{\partial x_j}
+t_0\left(\GG^2_0 (\nabla {\mathbf v}_1) 
+\GG_0\left(\nabla {\mathbf v}_1 + \nabla {\mathbf v}_1^T\right)\GG_0+
(\nabla {\mathbf v}_1)^T \GG^2_0 \right)
=-\dfrac{2 \cs{1}^2 }{\theta_0}(\text{det}\GG_0)^{-2/3}
\GG_0 \left(\GG^2_1-\dfrac{\tr(\GG^2_1)}{3} \II \right), 
\end{equation}
Then, taking into account that for zero terms $\GG_0=\Lambda^{1/2}\II$ and
$$
\left(\frac{\partial }{\partial t^\prime}+
t_0 v^j_1 \frac{\partial }{\partial x_j}\right)\left(\GG^2_0-\dfrac{\tr(\GG^2_0)}{3} \II\right)=0
$$
we obtain from \eqref{G_eqn_scale1}
$$
\left(\frac{\partial }{\partial t^\prime}+
t_0 v^j_1 \frac{\partial }{\partial x_j}\right)\left(\GG^2_0 \right) =
\left(\frac{\partial }{\partial t^\prime}+
t_0 v^j_1 \frac{\partial }{\partial x_j}\right)\left(\dfrac{\tr(\GG^2_0)}{3} \II\right)= 
-t_0 \frac{1}{3}\tr \left(\GG^2_0 (\nabla {\mathbf v}_1) 
+\GG_0\left(\nabla {\mathbf v}_1 + \nabla {\mathbf v}_1^T\right)\GG_0+
(\nabla {\mathbf v}_1)^T \GG^2_0 \right) \II.
$$
Substituting the above formula and $\GG_0=\Lambda^{1/2}\II$ in \eqref{G_eqn_scale1}, 
we obtain
$$
2t_0 \Lambda \left( \left( \nabla {\mathbf v}_1 + \nabla {\mathbf v}_1^T\right)
-\frac{1}{3}\tr \left(\nabla {\mathbf v}_1 + \nabla {\mathbf v}_1^T \right)\right)=
\dfrac{2 \cs{1}^2 }{\theta_0}\Lambda^{-1/2}\left(\GG^2_1-\dfrac{\tr(\GG^2_1)}{3} \II \right).
$$
Finally, for the first term of the asymptotic expansion, we obtain
\begin{equation}\label{TTT}
\left(\GG^2_1-\dfrac{\tr(\GG^2_1)}{3}\II \right)=
-t_0 \dfrac{\theta_0}{\cs{1}^2}\Lambda^{3/2}
\left( \left( \nabla {\mathbf v}_1 + \nabla {\mathbf v}_1^T\right)
-\frac{1}{3}\tr \left(\nabla {\mathbf v}_1 + \nabla {\mathbf v}_1^T \right)\right)
\end{equation}
and asymptotic expansion of expression \eqref{Stress} of stress tensor and
definition \eqref{Theta} of $\theta_0$ gives us 
\begin{equation} \label{StressNS}
\bm{\sigma} = -\varepsilon \frac{\rho_1 \cs{1}^2}{2}(\text{det}\bm{G_0})^{-2/3}
\left(\GG^2_1-\frac{1}{3} \tr (\GG_1^2) \II\right)=
\tau c^2_{s1} \left( \left( \nabla {\mathbf v}_1 + \nabla {\mathbf v}_1^T\right)
-\frac{1}{3}\tr \left(\nabla {\mathbf v}_1 + \nabla {\mathbf v}_1^T \right)\right)
\end{equation}
This formula \eqref{StressNS} is exactly the definition of viscous Navier-Stokes stress and can be written as 
$$
\sigma_{ik} = 2\eta (\dot \epsilon_{ik}-\delta_{ik}{\dot\epsilon_{jj}}/{3}),
$$
where $\dot \epsilon_{ik}=\left( \nabla {\mathbf v}_1 + \nabla {\mathbf v}_1^T\right)$ is the 
strain rate tensor and $\eta=\tau \cs{1}^2$ is the dynamic viscosity.

\printbibliography
%%=============================================================================
%%==========  B I B L I O G R A P H Y
%%\input{Bib}
%%\section*{References}
%\bibliographystyle{plain}
%\bibliography{./library}
%%=============================================================================

\end{document}